\documentclass[sigconf]{acmart}
\usepackage{subfigure}
\usepackage[norelsize,ruled,vlined,linesnumbered]{algorithm2e}
\usepackage{algorithmic}
\usepackage{url}

\AtBeginDocument{%
  \providecommand\BibTeX{{%
    \normalfont B\kern-0.5em{\scshape i\kern-0.25em b}\kern-0.8em\TeX}}}

\newcommand{\wsq}{\hspace{\fill}$\square$}
\newcommand{\vs}{\vspace{1.5mm}}

\newtheoremstyle{mystyle}
    {1.5mm}
    {1.5mm}
    {\it}
    {0mm}
    {\scshape}
    {.}
    { }
    {}
\theoremstyle{mystyle}
\newtheorem{defi}{Definition}
\newtheorem{exa}{Example}
\newtheorem{lemm}{Lemma}
\newtheorem{theo}{Theorem}

\setcopyright{acmcopyright}
\copyrightyear{2021}
\acmYear{2021}
\acmDOI{10.1145/1122445.1122456}

\acmConference[SIGMOD '21]{SIGMOD '21: ACM SIGMOD International Conference of Management of Data}{June 21--26, 2021}{Xi'an, Shaanxi, China}
\acmBooktitle{SIGMOD '21: ACM SIGMOD International Conference of Management of Data, June 21--26, 2021, Xi'an, Shaanxi, China}
\acmPrice{15.00}
\acmISBN{978-1-4503-XXXX-X/18/06}

\setcopyright{none}
\renewcommand\footnotetextcopyrightpermission[1]{}
\begin{document}

\title{Fast and Exact Outlier Detection in Metric Spaces: A Proximity Graph-based Approach}

\author{Daichi Amagata}
\affiliation{%
\institution{Osaka University, PRESTO}
\country{Japan}}
\email{amagata.daichi@ist.osaka-u.ac.jp}

\author{Makoto Onizuka}
\affiliation{%
\institution{Osaka University}
\country{Japan}}
\email{onizuka@ist.osaka-u.ac.jp}

\author{Takahiro Hara}
\affiliation{%
\institution{Osaka University}
\country{Japan}}
\email{hara@ist.osaka-u.ac.jp}

\begin{abstract}
Distance-based outlier detection is widely adopted in many fields, e.g., data mining and machine learning, because it is unsupervised, can be employed in a generic metric space, and does not have any assumptions of data distributions.
Data mining and machine learning applications face a challenge of dealing with large datasets, which requires efficient distance-based outlier detection algorithms.
Due to the popularization of computational environments with large memory, it is possible to build a main-memory index and detect outliers based on it, which is a promising solution for fast distance-based outlier detection.

Motivated by this observation, we propose a novel approach that exploits a proximity graph.
Our approach can employ an arbitrary proximity graph and obtains a significant speed-up against state-of-the-art.
However, designing an effective proximity graph raises a challenge, because existing proximity graphs do not consider efficient traversal for distance-based outlier detection.
To overcome this challenge, we propose a novel proximity graph, MRPG.
Our empirical study using real datasets demonstrates that MRPG detects outliers significantly faster than the state-of-the-art algorithms.
\end{abstract}

\maketitle

\section{Introduction}	\label{section_introduction}
Outlier detection is a fundamental task in many applications, such as fraud detection, health check, and noise data removal \cite{aggarwal2015outlier, wang2019progress}.
These applications often employ distance-based outlier detection (DOD) \cite{knorr1998algorithms} (as described later), because DOD is unsupervised, can be employed in any metric spaces, and does not have any assumptions of data distributions.
This paper addresses the DOD problem.

\vs
\noindent
\underline{\textbf{Motivation.}}
DOD requires a range threshold $r$ and a count threshold $k$ as input parameters.
Given a set $P$ of objects, an object $p \in P$ is an outlier if there are less than $k$ objects $p' \in P$ such that $dist(p,p') \leq r$, where $dist(p,p')$ evaluates the distance between $p$ and $p'$ in a data space.
(Density-based clustering also employs this definition to identify noises \cite{amagata2021dpc, ester1996density}.) 
Motivated by a recent trend of machine learning-related applications, we are interested in an efficient solution that can be employed in any metric spaces.

Classification, prediction, and regression utilize machine learning techniques, because they can provide high accuracy.
To train high performance models, noises (i.e., outliers) should be removed from training datasets, because the performances of models tend to be affected by outliers \cite{aggarwal2015outlier, lerman2018overview, wang2018iterative}.
It is now a common practice for many applications to remove noises as a pre-processing of training \cite{batista2004study, hodge2004survey}, and DOD can contribute to this noise removal.
Besides this, natural language processing, medical diagnostics, and image analysis also receive benefits from DOD.
For example, DOD is utilized to make datasets clean and diverse by finding error or unique sentences from sentence embedding vectors \cite{larson2019outlier}.
Campos et al. tested Euclidean DOD on medical and image datasets and confirmed that DOD successfully finds unhealthy people and irregular images \cite{campos2016evaluation}.

To cover these applications, a DOD technique needs to deal with many distance functions.
This is because the above noise removal application can have many data types (e.g., multi-dimensional points, strings, and time-series) and word (sentence) embedding vectors usually exist in angular distance spaces \cite{pennington2014glove}.
In addition, they need to deal with large datasets \cite{ilyas2019data}, thereby a scalable solution for metric spaces is required.
Due to the popularization of main-memory databases \cite{zhang2015memory}, in-memory processing of DOD on a large dataset is possible.
Fast DOD would be achieved by building an efficient main-memory index offline.
Some studies proposed metric DOD techniques \cite{angiulli2009dolphin, knorr1998algorithms, tao2006mining}, but they miss this observation.

\vs
\noindent
\underline{\textbf{Challenge.}}
To design an efficient index-based solution for any metric spaces, we address the following challenges: 
(i) general and effective index to any $r$ and $k$, (ii) space efficiency, and (iii) robustness to any metric spaces.

\vs
\noindent
(i) General and effective index to any $r$ and $k$.
Because we do not know $r$ and $k$ in advance, an index has to deal with any $r$ and $k$.
Building an index that is general to $r$ and $k$ and effectively supports fast DOD is not trivial.
The state-of-the-art algorithms \cite{angiulli2009dolphin, tao2006mining} build a simple data structure in an online fashion after $r$ and $k$ are specified.
The pruning efficiency of this index built online is limited, so they need long time to detect outliers.

\vs
\noindent
(ii) Space efficiency.
Let $p'$ be the $k$-th nearest neighbor of an object $p$.
If $dist(p,p') \leq r$, $p$ is not outlier.
Therefore, if $p$ stores a sorted array that maintains the distance to each object in $P$, whether $p$ is an outlier or not can be evaluated in $O(1)$ time.
However, this approach requires $O(n^2)$ space, where $n = |P|$, so is not practical.

\vs
\noindent
(iii) Robustness to any metric spaces.
Since we consider metric spaces and recent applications usually deal with middle or large dimensional data, robustness to any data types and dimensionality is important.
Notice that we can employ range queries to evaluate whether given objects are outliers or not.
A simple and practical solution is to build a tree-based index offline and iteratively conduct a range query on the index for each object.
However, space-partitioning approaches like tree structures are efficient only for low-dimensional data.
That is, the computational performances of existing techniques \cite{angiulli2009dolphin, tao2006mining} degrade on high-dimensional data.

\vs
\noindent
\underline{\textbf{Our Contributions.}}
We overcome the above challenges and make the following contributions\footnote{This is the full version of \cite{amagata2021dod}.}.

\begin{figure}[!t]
	\centering
	\includegraphics[width=0.40\linewidth]{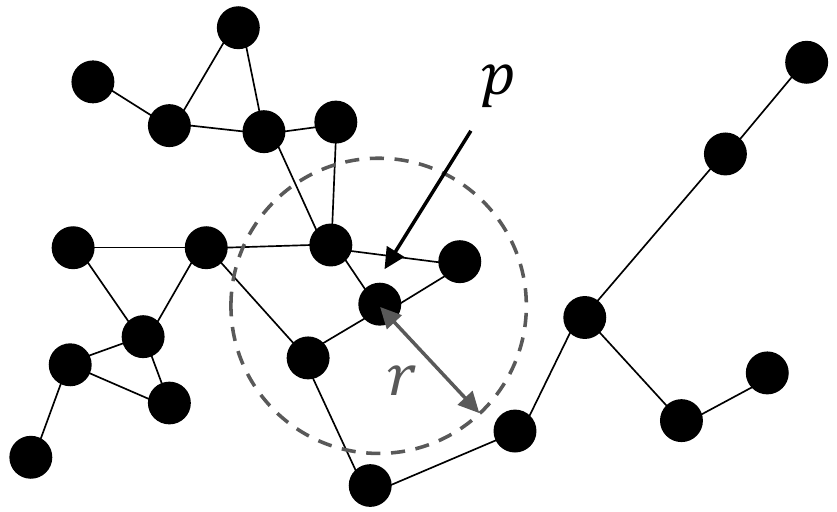}
    \caption{Example of a proximity-graph. Each object (black vertex) has links to its similar (nearby) objects.}
    \label{figure_proximity-graph}
\end{figure}

\vs
\noindent
\,$\bullet$\,\textit{Novel DOD algorithm that exploits a proximity graph} (Section \ref{section_overview}).
We propose a new technique for the DOD problem that filters non-outliers efficiently while guaranteeing the correctness by exploiting a proximity graph.
In a proximity graph, an object $p$ is a vertex, and each object has links to some of its similar objects, as shown in Figure \ref{figure_proximity-graph}, which assumes a Euclidean space.
The following example intuitively explains the filtering power of a proximity graph (a non-outlier can be filtered in $O(k)$ time).

\begin{exa}
\textit{Let $p$ be the center of the gray circle with radius $r$ in Figure \ref{figure_proximity-graph}.
Assume $k = 3$, and we can see that $p$ is not an outlier by traversing its links.}
\end{exa}

\noindent
This novel idea of graph-based filtering yields a significant improvement, because it avoids the impact of the curse of dimensionality and we need to verify only non-filtered objects.
Note that our algorithm (i) is orthogonal to any metric proximity graphs, (ii) is parallel-friendly, and (iii) detects all outliers correctly.

Furthermore, the above idea provides a new result: the time complexity of our solution is $O((f+t)n)$, where $f$ is the number of false positives incurred by the filtering and $t$ is the number of outliers.
This result states that, if $f + t = o(n)$ in the worst case, our solution does not need $O(n^2)$ time.
Real datasets usually have this case, whereas the existing DOD algorithms \cite{angiulli2009dolphin,knorr1998algorithms,tao2006mining} essentially incur $O(n^{2})$ time.
(Empirically, our solution scales almost linearly to $n$ on real datasets.)

\vs
\noindent
\,$\bullet$\,\textit{Novel metric proximity graph} (Section \ref{section_mrpg}).
To maximize the performance of our solution, $f$ should be minimized, and high reachability of neighbors (objects within distance being not larger than $r$) achieves this.
Motivated by this observation, as our second contribution, we devise MRPG (Metric Randomized Proximity Graph), a new proximity graph specific to the DOD problem.
When $r$ or $k$ is large, to evaluate whether $p$ is not an outlier, we may need to traverse objects existing in more than 1-hop from $p$ in a proximity graph.
However, existing proximity graphs are not designed to consider reachability to neighbors, which increases $f$.
The novelty of MRPG is that MRPG improves the reachability of neighbors by making pivot-based monotonic paths between objects with small distances, so that, for a given $p$, \textit{we can greedily traverse $p$'s neighbors from $p$}.
The space of an MRPG is reasonable, i.e., linear to $n$.

How to build a MRPG efficiently is not trivial, so we also propose an efficient algorithm that builds a MRPG in linear time to $n$.
We show that simply improving reachability between objects incurs $\Omega(n^{2})$ time, which clarifies that our algorithm is much faster.
Our MRPG building algorithm improves the reachability of neighbors while keeping a theoretically comparable efficiency with the state-of-the-art algorithm that builds an approximate $K$ nearest neighbor graph \cite{dong2011efficient} (and our algorithm is empirically faster).

\vs
\noindent
\,$\bullet$\,\textit{Extensive experiments} (Section \ref{section_experiment}).
We conduct experiments using various real datasets and distance functions.
The results demonstrate that our algorithm significantly outperforms the state-of-the-art.
Besides, MRPG provides faster response time than existing metric proximity graphs.

\vs
\noindent
Besides the above contents, Section \ref{section_problem-definition} defines the problem, Section \ref{section_related-work} reviews related work, and Section \ref{section_conclusion} concludes this paper.


\section{Problem Definition}    \label{section_problem-definition}
Let $P$ be a set of $n$ objects (i.e., $n = |P|$).
The neighbors of an object $p \in P$ are defined as follows:

\begin{defi}[\textsc{Neighbor}]
Given a distance threshold $r$ and an object $p \in P$, $p' \in P\backslash\{p\}$ is a neighbor of $p$ if $dist(p,p') \leq r$.
\end{defi}

\noindent
We consider that $dist(\cdot,\cdot)$ satisfies metric, i.e., triangle inequality.
We next define distance-based outlier and our problem.

\begin{defi}[\textsc{Distance-based outlier}]
Given a distance threshold $r$, a count threshold $k$, and a set of objects $P$, an object $p \in P$ is a distance-based outlier if $p$ has less than $k$ neighbors.
\end{defi}

\noindent
\textsc{Problem statement.}
\textit{Given a distance threshold $r$, a count threshold $k$, and a set of objects $P$, the distance-based outlier detection problem finds all distance-based outliers.}

\vs
Hereinafter, a distance-based outlier is called an outlier.
We use inliers to denote objects that are not outliers.
As with recent works \cite{fu2019fast, li2019approximate, perdacher2019cache, zois19efficient}, we focus on a single machine and static and memory-resident $P$.
(If $P$ is dynamic, we can use one of the state-of-the-art algorithms, e.g., \cite{kontaki2011continuous, tran2020real}.)

\section{Related Work}	\label{section_related-work}
\textbf{Outlier detection in metric spaces.}
A nested-loop algorithm \cite{knorr1998algorithms} is a straightforward solution for our problem.
Given an object $p \in P$, this algorithm counts the number of neighbors of $p$ by scanning $P$ and terminates the scan when the count reaches $k$.
This algorithm incurs $O(n^2)$ time, so does not scale to large datasets.

Given $r$, SNIF \cite{tao2006mining} forms clusters with radius $r/2$ (cluster centers are randomly chosen).
If the distance between an object $p$ and a cluster center is within $r/2$, $p$ belongs to the corresponding cluster.
From triangle inequality, the distances between any objects in the same cluster are within $r$.
Therefore, if a cluster has more than $k$ objects, they are not outliers.
Even if a cluster has less than $k + 1$ objects, objects in the cluster do not have to access the whole $P$.
This is because each object $p$ can avoid accessing objects $p'$ such that $dist(p,p') > r$ by using clusters.

DOLPHIN \cite{angiulli2009dolphin} is also a scan-based algorithm.
This algorithm indexes already accessed objects to investigate whether the next objects are inliers.
DOLPHIN can know how many objects exist within a distance from the current object $p$.
If there are at least $k$ objects within the distance, DOLPHIN does not need to evaluate the number of neighbors of $p$ any more.

The main issue of the above algorithms is their time complexity.
They rely on the (group-based) nested-loop approach and incur $O(n^2)$ time.
Besides, they lose distance bounds for high-dimensional data due to the curse of dimensionality, rendering degraded performance.
In addition to these solutions, an algorithm that exploits range search can also solve the DOD problem, as can be seen from Definition 1.
As one of baselines, we employ VP-tree \cite{yianilos1993data}, because \cite{chen2017pivot} demonstrated that VP-tree is the most efficient solution for the range search problem in metric spaces.
Each node of a VP-tree stores a subset $P'$ of $P$, the centroid object in $P'$, and the maximum value among the distances from the centroid to the objects in $P'$.
A range search on VP-tree is conducted as follows.
The lower-bound distance between a query and any node can be obtained by using the maximal value and triangle inequality.
If this lower-bound distance is larger than $r$, the sub-tree rooted at this node is pruned (otherwise, its child nodes are accessed).
How to build a VP-tree is introduced in Section \ref{section_nndescent-p}.

\vs
\noindent
\textbf{Proximity graphs} have been demonstrated to be the most promising solution to the $k$-NN search problem \cite{li2019approximate}.
If the distance between an object $p$ and its $k$-th NN is within $r$, $p$ is not an outlier.
From this observation, we see that proximity graphs have a potential to solve the DOD problem efficiently.
Some proximity graphs \cite{arya1993approximate, fu2019fast, harwood2016fanng} are dependent on $L_{2}$ space, so we review only proximity graphs that can be built in metric spaces.

One of the most famous proximity graphs is KGraph.
In this graph, each object is considered as a vertex and has links to its approximate $K$-NN (AKNN) objects (i.e., $K$ is the degree of the graph).
This graph is built by \textsc{NNDescent} algorithm \cite{dong2011efficient}.
Our proximity graph is also based on an AKNN graph, and we extend \textsc{NNDescent} to build an AKNN graph more efficiently in Section \ref{section_nndescent-p}.
Actually, simply employing KGraph may incur some problems.
For example, its reachability to neighbors can be low if $k > K$.

\begin{figure}[!t]
	\begin{center}
		\subfigure[ANN problem]{%
		\includegraphics[width=0.4\linewidth]{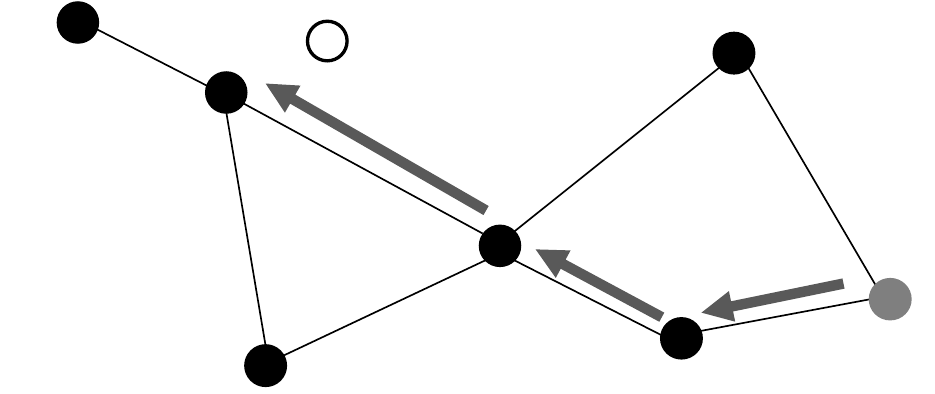}	\label{fig_ann}}
        \subfigure[Our problem]{%
		\includegraphics[width=0.4\linewidth]{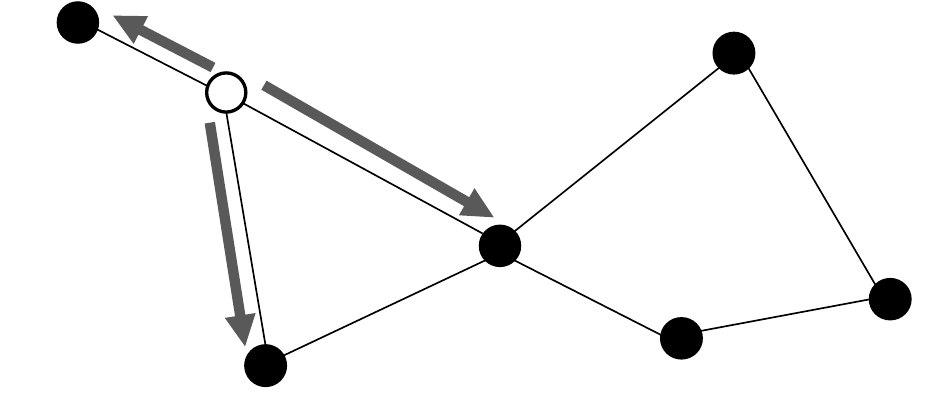}	\label{fig_outlier}}
        \vspace{-2.0mm}
        \caption{Difference between our and ANN problems w.r.t. graph traversal which is represented by arrows. Grey and white vertices (objects) respectively represent a starting and a query vertex.}
        \label{figure_related-work}
	\end{center}
\end{figure}

Another famous proximity graph is based on navigable small-world network models \cite{boguna2009navigability}.
In a graph based on this model, the number of hops between two arbitrary nodes is proportional to $\log n$.
Building a graph based on this model incurs $O(n^2)$ time, thereby an approximate solution, NSW, was proposed in \cite{malkov2014approximate}.
To accelerate approximate nearest neighbor (ANN) search, its hierarchical version, HNSW, was proposed in \cite{malkov2018efficient}.
The upper layers of HNSW are built by sampling objects in their lower layers.
This structure aims at skipping redundant vertices to quickly reach vertices with small distances to a query.
When we evaluate the number of neighbors of $p \in P$, $p$ can be considered as a query object.
Figure \ref{fig_ann} depicts the search process in ANN problem: it starts from a random vertex (the grey one) and traverses the proximity graph so that the next vertex is closer to the query object (white one) than the former one.
On the other hand, in our problem, a query object is one of the objects in $P$.
It is clearly better to traverse the graph from the query object for finding its neighbors, as shown in Figure \ref{fig_outlier}.
Therefore, we do not need the skipping structure of HNSW (thus do not consider it as a baseline).

Although these proximity graphs can be employed in our solution proposed in Section \ref{section_overview}, they cannot optimize the performance of our solution.
This is because they are not designed for the DOD problem and do not consider reachability to neighbors.
Therefore, we propose a new proximity graph for the DOD problem that takes the reachability into account in Section \ref{section_mrpg}.

\section{Our DOD Algorithm}	\label{section_overview}
 Let $t$ be the number of outliers in $P$.
A range search in metric spaces with arbitrary dimensionality needs $O(n)$ time.
Therefore, the DOD problem needs $\Omega(tn)$ time (because we have to evaluate not only outliers but also inliers).
To scale well to large datasets, it is desirable that the time complexity of a solution nearly matches the lower-bound.
Designing such a solution is however not straightforward.
Our new technique for the DOD problem overcomes this non-trivial challenge.

\vs
\noindent
\textbf{Main idea.}
Given $P$, the ratio of outliers in $P$ is small (usually less than one percent) \cite{yoon2019nets}.
That is, most objects in $P$ are inliers, so we should identify them as inliers quickly, to reduce computation time.
The evaluation of whether or not $p$ is an inlier can be converted to answering the problem of range counting with query object $p$ and radius $r$.
Therefore, to filter inliers quickly, we need an efficient solution for the problem of range counting, with early termination when the count reaches $k$.
Proximity graphs recently have shown high potential for solving the approximate nearest neighbor search problem \cite{li2019approximate}, thanks to their property of the connections between similar objects.
This property is also promising for the range counting problem.
Because, each object $p$ has links to its similar objects in a proximity graph, we can efficiently count the number of neighbors of $p$ by traversing the graph from $p$, \textit{regardless of the dimensionality of the dataset}.
Figures \ref{figure_proximity-graph} and \ref{fig_outlier} depict its intuition.

To implement this idea, we propose a proximity graph-based solution, a novel approach for the DOD problem.
Algorithm \ref{algo_framework} describes its overview.
This solution consists of a filtering phase (lines \ref{algo_framework_filter_b}--\ref{algo_framework_filter_e}) and a verification phase (lines \ref{algo_framework_verification_b}--\ref{algo_framework_verification_e}).

\vs
\noindent
\textbf{\underline{Filtering phase.}}
In this phase, we filter inliers by exploiting a proximity graph $G$, which is built in one-time pre-processing phase.
More specifically, we propose \textsc{Greedy-Counting} (Algorithm \ref{algo_greedy-counting}) to count the number of neighbors of an object $p$ on $G$.
Consider that a vertex $v$ in $G$ corresponds to an object $p$ ($v$ and $p$ are hereinafter used interchangeably in the context of $G$).
Let $v.E$ be the set of links between $v$ and some other vertices.
Given an object $p$, $r$, and $k$, \textsc{Greedy-Counting} greedily traverses $G$ from $v$, as long as a visited vertex $v'$ satisfies $dist(p,p') \leq r$, so as to count the number of neighbors of $p$.
In other words, we first check $v.E$: for each $(v,v') \in v.E$ where $v'$ has not been visited, we increment the count by one and insert $p'$ into a queue $Q$, iff $dist(p,p') \leq r$.
We next pop the front of $Q$, say $v'$, check $v'.E$, and do the same as $v$.
One exception appears in line \ref{algo_greedy-counting_pivot_b}, and this is necessary for MRPG, which is explained in Section \ref{section_link}.
\textsc{Greedy-Counting} is terminated when the count reaches $k$ or $Q$ becomes empty.
It is important to see that:

\begin{lemm}    \label{lemma_false-negative}
Our filtering does not incur false negatives.
\end{lemm}

\noindent
\textsc{Proof.}
All proofs appear in Appendix.   \wsq

\begin{algorithm}[!t]
	\caption{Proximity Graph-based DOD}	\label{algo_framework}
	\DontPrintSemicolon
    {\small
        \KwIn {$P$, $r$, $k$, and a proximity graph $G$}
		\vspace{1.0mm}
		/* Filtering phase */\;
        $P' \leftarrow \varnothing$\;									\label{algo_framework_filter_b}
		\For {each $p \in P$}{
            \If {\rm{\textsc{Greedy-Counting}}$(p,r,k,G)$ $< k$}{       \label{algo_framework_filter_b_}
            	$P' \leftarrow P' \cup \{p\}$							\label{algo_framework_filter_e}
            }
        }
        \vspace{1.0mm}
        /* Verification phase */\;
        $P_{out} \leftarrow \varnothing$\;								\label{algo_framework_verification_b}
        \For {each $p \in P'$}{											\label{algo_framework_verification_b_}
            \If {\rm{\textsc{Exact-Counting}}$(p,r,k)$ $< k$}{
            	$P_{out} \leftarrow P_{out} \cup \{p\}$					\label{algo_framework_verification_e}
            }
        }
        \textbf{return} $P_{out}$	\label{algo_framework_verification_e_}
	}
\end{algorithm}

\vs
\noindent
\textbf{\underline{Verification phase.}}
Let $P'$ be the set of objects whose counts returned by \textsc{Greedy-Counting} are less than $k$.
From Lemma \ref{lemma_false-negative}, $P'$ contains all outliers but does false positives (i.e., inliers but not filtered).
We therefore have to verify whether or not objects in $P'$ are really outliers.
\textsc{Exact-Counting} in Algorithm \ref{algo_framework} verifies them in the following way:
\begin{itemize}
	\setlength{\leftskip}{-5.0mm}
    \setlength{\itemsep}{0.0mm}
	\item	For data with low intrinsic dimensionality\footnote{This is the minimum number of parameters (variables) that are needed to represent a given dataset.
	        For example, when this is less than 5, it can be considered as low.}, we conduct a range counting on a VP-tree \cite{yianilos1993data} for each object in $P'$.
	\item	For the other data, we use a linear scan, because this is more efficient than any indexing methods for high-dimensional data.
\end{itemize}
We terminate range counting or sequential scan for $p$ when the count of $p$ reaches $k$.
Since this phase counts the exact number of neighbors for all outliers in $P'$ and $P'$ contains all outliers, \textit{Algorithm \ref{algo_framework} returns the exact answer}.

\vs
\noindent
\textbf{Time analysis.}
Hereinafter, we assume that the dimensionality is fixed.
We analyze the time complexity of Algorithm \ref{algo_framework}.

\begin{theo}    \label{theorem_online-time}
Algorithm \ref{algo_framework} requires $O((f + t)n)$ time, where $f$ and $t$ are respectively the numbers of false positives and outliers.
\end{theo}

\noindent
\textbf{Remark.}
From the above result, we see that our solution theoretically does not need $O(n^2)$ time, if $f + t = o(n)$ in the worst case.
This holds in practice, so our result supports a significant speed-up over the existing $(r,k)$-DOD algorithms.
We here note that (1) real datasets have $t \ll n$ \cite{kontaki2011continuous,yoon2019nets}, and (2) $t$ is usually dependent not on $n$ but on data distributions.
These and Theorem \ref{theorem_online-time} suggest that our solution with a proximity graph yielding a small $f$ can be (almost) linear to $n$ in practice.

\vs
\noindent
\textbf{Multi-threading.}
Algorithm \ref{algo_framework} iteratively evaluates objects independently, thereby can be parallelized.
(Given multi-threads, each thread independently evaluates assigned objects in both the filtering and verification phases.)
However, to exploit multi-threading, balancing the load of each thread is important.
The early termination in the verification phase cannot function for outliers, since they do not have $k$ neighbors.
The filtering and verification costs of outliers are hence larger than those of inliers.
That is, keeping load balance is hard in our problem theoretically, as we do not know outliers in advance.
To relieve this, we employ a random partitioning approach for assigning objects into each thread.

\begin{algorithm}[!t]
	\caption{\textsc{Greedy-Counting}}	\label{algo_greedy-counting}
	\DontPrintSemicolon
    {\small
        \KwIn {$p_{i}$, $r$, $k$, and a proximity graph $G$}
		\vspace{1.0mm}
        count $\leftarrow 0$, $Q \leftarrow \{v_{i}\}$, check $v_{i}$ as visited\;
        \While {$Q \neq \varnothing$}{
        	$v \leftarrow$ the front of $Q$\;
        	$Q \leftarrow Q\backslash\{v\}$\;
            \For {each $v' \in v.E$ where $v'$ has not been checked as visited}{
                	Check $v'$ as visited\;
                    \eIf {$dist(p,p') \leq r$}{
                    	count $\leftarrow$ count $+ 1$\;
                        \If {count $= k$} {
                        	\textbf{break}
                        }
                        $Q \leftarrow Q \cup \{v'\}$
                    }
                    {
                    	\If {$p'$ is a pivot}{				\label{algo_greedy-counting_pivot_b}
                        	$Q \leftarrow Q \cup \{v'\}$	\label{algo_greedy-counting_pivot_e}
                        }
                    }
            }
            \If {count $= k$} {
            	\textbf{break}
            }
        }
        \textbf{return} count
	}
\end{algorithm}

\section{MRPG}	\label{section_mrpg}
hile our DOD algorithm is orthogonal to any proximity graphs, its performance (i.e., $f$) depends on a given proximity graph.
To maximize the performance, we have to minimize $f$ in the filtering phase.
Therefore, the main challenge of this section is to reduce $f$.
To overcome this, in a proximity graph, neighbors of an arbitrary object $p$ should be reachable from $p$ for \textsc{Greedy-Counting}.

Consider an inlier $p$.
To accurately identify $p$ as an inlier in the filtering phase (i.e., to reduce $f$), a proximity graph $G$ should have paths from $p$ to its neighbors that can be traversed by \textsc{Greedy-Counting}.
Our idea that achieves this is to introduce \textit{monotonic path}, a path from $p$ such that \textsc{Greedy-Counting} can traverse its neighbors in non-decreasing order w.r.t. distance.

\begin{defi}[\textsc{Monotonic path}]
Consider two objects $p_{i}$ and $p_{i+x}$ in $P$.
Let $v_{i}$, $v_{i+1}$, ..., $v_{i+x}$ be a path from $p_{i}$ to $p_{i+x}$ in a proximity graph (that is, $v_{i+j}$ has a link to $v_{i+j+1}$ for all $j \in [0,x-1]$).
Iff $dist(v_{i},v_{i+j}) \leq dist(v_{i},v_{i+j+1})$ for all $j \in [0,x-1]$, this path is a monotonic path.
\end{defi}

If $G$ has at least one monotonic path between any two objects, $G$ is a monotonic search graph (MSG) \cite{dearholt1988monotonic}.
Although a MSG can reduce $f$, building it in metric spaces requires $\Omega(n^2)$ time (see Theorem \ref{theorem_msg-time}), meaning that reducing $f$ with a proximity graph that can be built in a reasonable time is not trivial.
To solve this challenge, we propose MRPG (Metric Randomized Proximity Graph), an approximate version of MSG.
MRPG incorporates the following properties to reduce $f$.
\begin{description}
    \setlength{\leftskip}{-4.0mm}
    \item[Property 1:]  each object has links to its approximate $K$-NNs.
    \item[Property 2:]  monotonic paths are created based on pivots (a subset of $P$).
    \item[Property 3:]  candidates of outliers have their exact $K'$-NNs, where $K' \geq K$.
\end{description}
The benefits of these properties are as follows.
First, thanks to the first property, \textsc{Greedy-Counting} tends not to miss accessing similar objects.
Second, the graph traversal in Algorithm \ref{algo_greedy-counting} goes through pivots.
Assume that we now visit a pivot when counting the number of neighbors of $p$.
If the pivot has a monotonic path to the neighbors of $p$, reachability between $p$ and its neighbors is improved.
Now the challenge is how to choose pivots to receive this benefit for many objects.
Random sampling is clearly not effective because it produces biased samples from dense subspaces (objects in dense spaces are easy to reach their neighbors).
Our approach is that we choose pivots from each subspace of $P$, because this approach can choose pivots from (comparatively) sparse spaces and reachability between objects existing in such spaces is also improved.
(How to efficiently identify subspaces is introduced in Section \ref{section_nndescent-p}.)
Last, the third property is simple yet important.
If objects that would be outliers have links to their exact $K'$-NNs, we can efficiently know whether or not they are outliers, if $k \leq K'$.

This section presents a non-trivial MRPG building algorithm, which satisfies the above properties through the following steps:

\vs
\noindent
1. \textsc{NNDescent+}: this builds an AKNN graph.
We extend a state-of-the-art AKNN graph building algorithm \textsc{NNDescent}, to quickly build it.

\noindent
2. \textsc{Connect-SubGraphs}: this connects such sub-graphs to guarantee that MRPG is strongly connected, because an AKNN may have disjoint sub-graphs.

\noindent
3. \textsc{Remove-Detours}: this creates monotonic paths by removing detours.
We utilize a heuristic approximation.

\noindent
4. \textsc{Remove-Links}: this removes unnecessary links to avoid redundant graph traversal.

\vs
\noindent
The first step achieves the properties 1 and 3.
Then, we obtain the property 2 in the third step.
In Section \ref{section_discussion}, we show that this algorithm achieves \textit{linear time to} $n$.
That is, we achieve a reduction of $f$ by using a MRPG (i.e., the three properties) that can be obtained in a reasonable time.

\subsection{NNDescent+}	\label{section_nndescent-p}
MRPG is based on an AKNN graph, so we need an efficient algorithm for building an AKNN graph.
Building an exact $K$-NN graph needs $O(n^2)$ time, thereby we consider an AKNN graph.
\textsc{NNDescent} \cite{dong2011efficient} is a state-of-the-art algorithm that builds an AKNN graph in any metric spaces.
We first introduce it.
(Note that the AKNN graph obtained by \textsc{NNDescent} satisfies only property 1.)

\vs
\noindent
\textbf{NNDescent.}
This algorithm is based on the idea that, given an object $p$ and its similar object $p'$, similar objects of $p'$ would be similar to $p$.
That is, approximate $K$-NNs of $p$ can be obtained by accessing its similar objects and their similar ones iteratively.
Given $K$ and $P$, the specific operations of \textsc{NNDescent}\footnote{We consider the basic version of \textsc{NNDescent} in \cite{dong2011efficient}, because it is parallel-friendly (almost no synchronization).} are as follows:
\begin{enumerate}
	\setlength{\leftskip}{-3.0mm}
    \setlength{\itemsep}{0.0mm}
    \item	For each object $p \in P$, \textsc{NNDescent} first chooses $K$ random objects as its initial AKNNs.
    \item	For each object $p \in P$, \textsc{NNDescent} obtains a similar object list that contains its AKNNs and reverse AKNNs.
    		(If $p \in$ AKNNs of $p'$, $p'$ is a reverse AKNN of $p$, thereby how to obtain reverse AKNNs is trivial.)
    		Given $p \in P$ and the objects $p'$ in the similar object list of $p$, \textsc{NNDescent} accesses the similar object list of $p'$.
            If the list contains objects with smaller distances to $p$ than those to the current AKNN of $p$, \textsc{NNDescent} updates its AKNNs.
    \item	\textsc{NNDescent} iteratively conducts the above procedure until no updates occur (or a fixed iteration times).
\end{enumerate}


\begin{theo}    \label{theorem_nndescent}
\textit{\textsc{NNDescent} requires $O(nK^2\log K)$ time.}
\end{theo}

\noindent
\textbf{Drawbacks of NNDescent.}
The accuracy of the AKNN graph built by \textsc{NNDescent} is empirically high, but it has the following:
\begin{itemize}
	\setlength{\leftskip}{-5.0mm}
    \setlength{\itemsep}{0.0mm}
	\item	The initial completely random links incur many AKNN updates in the second operation.
	        Due to this initialization, each object cannot have links to its similar objects in an early stage, incurring unnecessary distance computations.
    \item	The similar object list of $p'$ is redundantly accessed even when the list has no updates from the previous iteration.
\end{itemize}
We overcome them by \textsc{NNDescent+}, an extension of \textsc{NNDescent}.
This is of independent interest for building an AKNN graph.

\vs
\noindent
\textbf{NNDescent+.}
We overcome the first drawback by utilizing data partitioning that clusters similar objects and do the second drawback by maintaining the update status of similar object lists.

\vs
\noindent
\textbf{\underline{Initialization by VP-tree based partitioning.}}
Each object needs to find its (approximate) $K$-NNs quickly, to reduce the number of update iterations.
We achieve this by utilizing a VP-tree based partitioning approach.

Given an object set $P$, a VP-tree for $P$ is built by recursive partitioning.
Specifically, consider that a node of the VP-tree has $P$.
If a node contains more objects than the capacity $c$, this node (or $P$) is partitioned into two nodes, left and right.
(Otherwise, this node is a leaf node.)
Let $p$ be a randomly chosen object from $P$.
The partitioning algorithm computes the distances between $p$ and the other objects in $P$, sorts the distances, and obtains the mean distance $\mu$.
If an object $p' \neq p$ has $dist(p,p') \leq \mu$, it is assigned to the left child of $p$.
Otherwise, it is assigned to the right one.
This partition is repeated until no nodes can be partitioned.

We set $c = O(K)$.
Consider a leaf node that is the left node of its parent.
Let $P'$ be the set of objects held by this leaf node.
Objects in $P'$ tend to be similar to each other, due to the ball-based partitioning property.
Therefore, for each $p \in P'$, we set its $K$-NNs in $P'$ as its initial AKNNs.
This approach can have much more accurate AKNNs at the initialization stage than the random-based one.
Besides, the efficiency of \textsc{NNDescent} is not lost.

\begin{lemm}    \label{lemma_init}
\textsc{NNDescent+} needs $O(nK^2\log K)$ time at its initialization.
\end{lemm}

\begin{algorithm}[!t]
	\caption{\textsc{Partition}}	\label{algo_partition}
	\DontPrintSemicolon
    {\small
        \KwIn {A set of objects $P' \subseteq P$}
		\vspace{1.0mm}
        \eIf {$|P'| > c$}{
        	$p \leftarrow$ a randomly chosen object from $P'$, $D \leftarrow \varnothing$	\;
        	\For {each $p' \in P'$}{
        		$D \leftarrow \langle dist(p,p'), p' \rangle$
        	}
        	$\mu \leftarrow$ the mean distance in $D$	\;
        	$L \leftarrow \varnothing$, $R \leftarrow \varnothing$	\;
        	\For {each $p' \in P'$}{
        		\eIf{$dist(p,p') \leq \mu$}{
               		$L \leftarrow L \cup \{p'\}$
               	}
               	{
               		$R \leftarrow R \cup \{p'\}$
               	}
        	}
            \textsc{Partition}$(L)$, \textsc{Partition}$(R)$	\;
            \If {$|L| \leq c$}{
            	Set $p$ as a pivot
            }
        }
        {
        	\If {$P' = L$}{
           		\For {each $p' \in P'$}{
           			Update AKNN of $p'$ from $P'$
           		}
           	}
        }
	}
\end{algorithm}

\vs
Because of the random nature, some objects cannot be contained in $P'$.
We hence do this partitioning a constant number of times.
(For objects that could not be contained in $P'$ after repeating the partitioning, random objects are set as their AKNNs.)
It is also important to note that nodes, whose left child is a leaf node, are set as \textit{pivots}, which are utilized in future steps.
The ball-based partitioning makes pivots being distributed in each subspace of the given data space.
This is also the reason why we use this partitioning approach.
Note that we have $o(n)$ pivots.
Algorithm \ref{algo_partition} summarizes our initialization approach, and \textsc{NNDescent+} replaces the first operation of \textsc{NNDescent} with Algorithm \ref{algo_partition}.

\vs
\noindent
\textbf{\underline{Skipping similar object lists with no updates.}}
When obtaining the similar object list of an object $p$, \textsc{NNDescent+} adds objects $p'$, which are AKNNs or reverse AKNNs of $p$, to the similar object list iff AKNNs of $p'$ have been updated in the previous iteration.
We employ a hash table to maintain the AKNN update status of each object.
The space complexity of this hash table is thus $O(n)$, and confirmation of the update status needs $O(1)$ amortized time for each object.
Therefore, \textsc{NNDescent+} reduces the cost of the second operation of \textsc{NNDescent}.

\vs
\noindent
\textbf{\underline{Exact K'-NN Retrieval.}}
The above initialization and skipping approaches respectively reduce the number of iterations and unnecessary distance computations.
However, for objects $p$ such that their $K$-NNs are relatively far from $p$, the initialization may provide inaccurate results.
The initialization approach clusters objects with small distances and is difficult to cluster objects with $K$-NNs that have large distances to them.
This may generate biased clusters, which derives biased similar object lists in the second procedure.
If this occurs, reachability to accurate $K$-NNs is degraded, and $p$ may suffer from this.
To alleviate this, \textsc{NNDescent+} computes the exact $K$-NNs for such objects.

After the iterative AKNN updates (the third procedure in \textsc{NNDescent}), \textsc{NNDescent+} sorts objects in $P$ in descending order of the sum of the distances to their (approximate) $K$-NNs.
If the sum is large, it is perhaps inaccurate.
\textsc{NNDescent+} picks the first $m$ objects and retrieves their exact $K'$-NNs, where $K' \geq K$ is sufficiently large (but $K' \ll n$).
We present why we use $K'$ in Section \ref{section_discussion}.
Note that $m$ is a constant and $m \ll n$.
Therefore, this approach incurs $O(n(K + \log n))$ time.

\vs
Now, we see the time complexity of \textsc{NNDescent+}.

\begin{lemm}    \label{lemma_nndescent-plus}
\textsc{NNDescent+} requires $O(nK^2\log K)$ time.
\end{lemm}

\noindent
Although \textsc{NNDescent+} theoretically requires the same time as \textsc{NNDescent}, \textsc{NNDescent+} is empirically faster (in most cases), because of reducing the number of iterations and pruning unnecessary similar object lists of neighbors.
In addition, the procedure of \textsc{NNDescent+} (except for obtaining reverse AKNNs) can exploit multi-threading (by using \textsf{parallel for} and \textsf{parallel sort}).

\subsection{Connecting Sub-Graphs}	\label{section_connect}
Since $K \ll n$, an AKNN graph may have some disjoint sub-graphs.
If this holds for the AKNN graph built by \textsc{NNDescent+}, \textsc{Greedy-Counting} may not be able to traverse some of neighbors.
We therefore make MRPG strongly connected\footnote{A similar idea has been proposed in \cite{fu2019fast}, but how to add links to make a proximity graph strongly connected is different from our approach. In addition, \cite{fu2019fast} does not have a theoretical time bound to achieve it.}.
Algorithm \ref{algo_connect} details our approach \textsc{Connect-SubGraphs} that consists of two phases.

\vs
\noindent
\textbf{\underline{Reverse AKNN phase}} (lines \ref{algo_connect_first_b}--\ref{algo_connect_first_e}).
In the first phase, we consider reverse AKNNs.
Specifically, if an object $p$ is included in AKNNs of $p'$, $p$ creates a link to $p'$ (if $p$ does not have it).
The AKNN graph built by \textsc{NNDescent+} is a directed graph.
This phase converts it to an undirected graph.
Although this is simple, reachability between objects and their neighbors can be improved, because reverse AKNNs of each object are (probably) similar to it.

\vs
\noindent
\textbf{\underline{BFS with ANN phase}} (lines \ref{algo_connect_second_b}--\ref{algo_connect_second_e}).
In the second phase, we propose a randomized approach that exploits breadth-first search (BFS) and ANN search on an AKNN graph.
We confirm the connection between any two objects through BFS (from a random object).
If BFS did not traverse some objects (line \ref{algo_connect_ann_b}), the AKNN graph has some disjoint sub-graphs.

Let $P'$ be a set of objects that have not been traversed by BFS.
We make a path between a pivot in $P'$ and a pivot in $P\backslash P'$.
Let $v'_{piv}$ be a random pivot in $P'$.
Also, let $V_{piv}$ be a set of random pivots in $P\backslash P'$ (note that $|V_{piv}|$ is a small constant).
We search for an ANN object for $v'_{piv}$ among $P\backslash P'$ and create links between them (lines \ref{algo_connect_ann_m}--\ref{algo_connect_second_e}).
Since pivots are distributed uniformly in each subspace, this approach creates links between objects with small distances as much as possible, which is the behind idea of this phase.

To find an ANN, we employ the greedy algorithm proposed in \cite{malkov2014approximate}.
The inputs of this algorithm are, a query object ($v'_{piv}$), a starting object ($v \in V_{piv}$), and a proximity graph.
Given $v$, this algorithm traverses objects in $v.E$, computes the object $v'$ with the minimum distance to $v'_{piv}$, goes to $v'$, and repeats this until we cannot get closer to $v'_{piv}$.
Let $v_{ann}$ be the answer to this algorithm.
We conduct this search for each $v \in V_{piv}$, select the object $v_{res}$ with the minimum distance to $v'_{piv}$, and create links between $v'_{piv}$ and $v_{res}$.
Then, we re-start BFS from a random object in $P'$ (already traversed objects are skipped).
The above operations are repeated until BFS traverses all objects.

\begin{algorithm}[!t]
	\caption{\textsc{Connect-SubGraphs}}	\label{algo_connect}
	\DontPrintSemicolon
    {\small
        \KwIn {$G$}
        \For {each $p \in P$}{						\label{algo_connect_first_b}
        	\For {each $(v,v') \in v.E$ such that $v' \notin K'$-NN}{
            	$v'.E \leftarrow v'.E \cup \{v\}$	\label{algo_connect_first_e}
            }
        }
        
        $P' \leftarrow P$	\;						\label{algo_connect_second_b}
        \While {$P' \neq \varnothing$}{
        	$Q \leftarrow$ a random node (object) $v$ ($p$) in $P'$	\;
        	$P' \leftarrow P'\backslash\{p\}$	\;
        	\While {$Q \neq \varnothing$}{
        		$v \leftarrow$ the front of $Q$	\;
            	$Q \leftarrow Q\backslash\{v\}$	\;
            	\For {each $v' \in v.E$}{
            		\If {$p' \in P'$}{
                		$P' \leftarrow P'\backslash\{p'\}$, $Q \leftarrow Q \cup \{v'\}$
                	}
           		}
            }
            \If {$P' \neq \varnothing$}{	\label{algo_connect_ann_b}
            	$v'_{piv} \leftarrow$ a random pivot in $P'$	\;
                $V_{piv} \leftarrow$ a set of random pivots in $P\backslash P'$	\;
                $dist_{min} \leftarrow \infty$, $v_{res} \leftarrow v'_{piv}$	\;
                \For {each $v \in V_{piv}$}{															\label{algo_connect_ann_m}
                	$v_{ann} \leftarrow$ \textsc{ANN-Search}$(v,v'_{piv},G)$	\;
                    \If {$dist(v_{ann},v'_{piv}) < dist_{min}$}{
                    	$dist_{min} \leftarrow dist(v_{ann},v'_{piv})$, $v_{res} \leftarrow v_{ann}$
                    }
                }
                $v'_{piv}.E \leftarrow v'_{piv}.E \cup \{v_{res}\}$, $v_{res}.E \leftarrow v_{res}.E \cup \{v'_{piv}\}$	\label{algo_connect_second_e}
            }
        }
	}
\end{algorithm}

\begin{figure}[!t]
	\begin{center}
		\subfigure[ANN search]{%
		\includegraphics[width=0.485\linewidth]{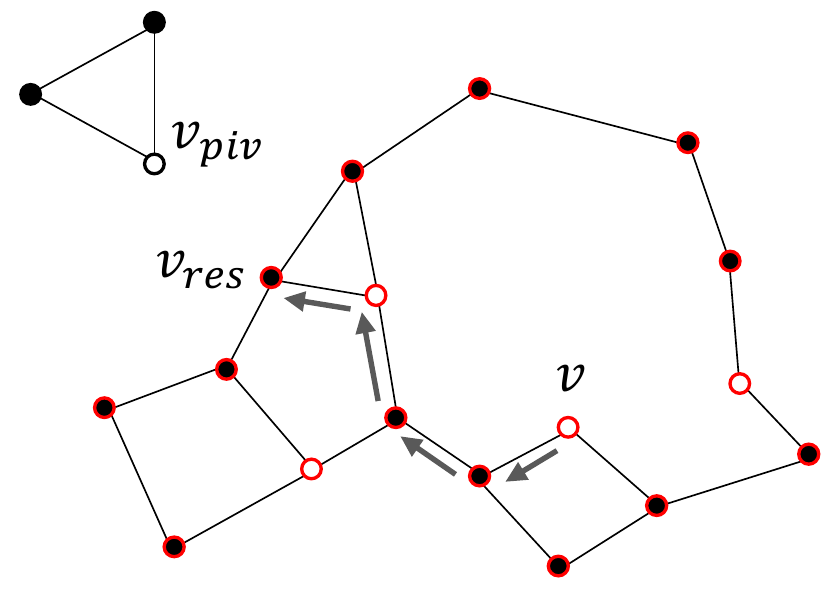}	\label{fig_connect-a}}
        \subfigure[Re-starting BFS]{%
		\includegraphics[width=0.485\linewidth]{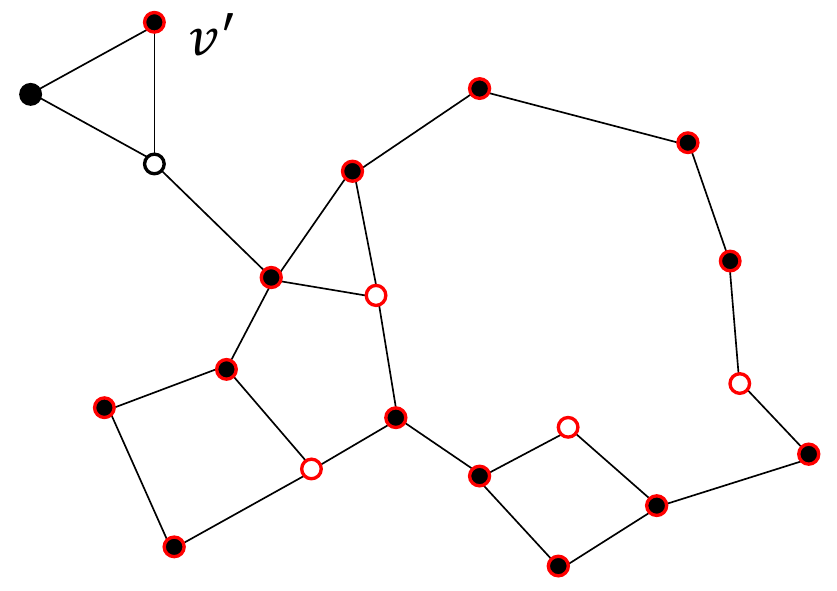}	\label{fig_connect-b}}
        \vspace{-3.0mm}
        \caption{Example of \textsc{Connect-SubGraphs}. White vertices represent pivots. BFS traversed red-marked vertices.}
        \label{figure_connect-subgraphs}
        \vspace{3.0mm}
	\end{center}
\end{figure}

\begin{exa} \label{example_connect}
Figure \ref{figure_connect-subgraphs} illustrates an example of \textsc{Connect-SubGraphs}.
Figure \ref{fig_connect-a} shows the AKNN graph obtained by \textsf{NNDescent+} ($K' = K$ for ease of presentation).
BFS has traversed the red-marked vertices, and now we conduct an ANN search, where the query and starting objects are respectively $v_{piv}$ and $v$.
The ANN search traverses the grey arrows (each traversed vertex selects the vertex that is the closest to $v_{piv}$) and obtains $v_{res}$.
We then create a link between $v_{piv}$ and $v_{res}$, as illustrated in Figure \ref{fig_connect-b}.
After that, we re-start BFS from a random vertex, e.g., $v'$, in Figure \ref{fig_connect-b}, that has not been traversed yet.
\end{exa}

We set the maximum hop count for the ANN search.
(It should be sufficiently large to make MRPG strongly connected, and is 10 in our implementation).
This yields that the time complexity of this algorithm is $O(K)$ (since $|V_{piv}| =O(1)$).
Then we have:

\begin{lemm}    \label{lemma_connect-subgraph}
\textsc{Connect-SubGraphs} requires $O(nK)$ time.
\end{lemm}

\subsection{Removing Detours}	\label{section_detour}
If a path from an object $p$ to its neighbor $p'$ is not monotonic (i.e., it is a detour), \textsc{Greedy-Counting} may not be able to access $p'$.
For example, consider two objects $p_{1}$ and $p_{2}$ where $dist(p_{1},p_{2}) \leq r$.
Assume that there is only a single path between $p_{1}$ and $p_{2}$, e.g., $p_{1} \rightarrow p_{3} \rightarrow p_{2}$.
If $dist(p_{1},p_{3}) > r$, \textsc{Greedy-Counting} cannot reach $p_{2}$ from $p_{1}$.
This increases the number of false positives, so we consider making monotonic paths.
We first demonstrate that making a monotonic search graph (MSG) is not practical.
Then, we propose a pivot-based approximation.

\vs
\noindent
\textbf{Building a MSG.}
Theoretically, building a MSG needs $\Omega(n^2)$ time, because we have to check a path between all object pairs in $P$.
We propose \textsc{Get-Non-Monotonic}$()$, which is based on BFS and searches for objects with no monotonic paths for a given object, to make a MSG.

\vs
\noindent
\textsc{Get-Non-Monotonic}$()$.
Given $p_{1}$, this function conducts BFS from $p_{1}$.
Assume that we now access $p_{3}$ during BFS and BFS traversed a path $p_{1} \rightarrow p_{2} \rightarrow p_{3}$.
If $dist(p_{1},p_{2}) > dist(p_{1},p_{3})$, this path is a detour, so we need a monotonic path.
We maintain objects, such that a monotonic path from $p_{1}$ to them could not be confirmed by BFS, and distances to them in an array $A_{1}$.
After all objects are traversed, we sort $A_{1}$ in ascending order of distance.

\vs
\noindent
We conduct this function for each object.
Now we have an array $A_{i}$ for each object.
We add a link between $A_{i}[j]$ and $A_{i}[j+1]$ for each $j \in [1,s-1]$, where $s$ is the size of $A_{i}$.
($A_{i}[1]$ is linked to $p_{i}$.)
This approach guarantees that a given proximity graph becomes a MSG.
However, a huge cost is incurred.

\begin{theo}    \label{theorem_msg-time}
Building a MSG needs $O(n^2(K + \log n))$ time.
\end{theo}

\noindent
\textbf{Approximation by heuristic.}
This theorem proves that building a MSG is not practical.
Note that it is not necessary to make monotonic paths between any two objects, because $r$ and $k$ are generally small \cite{kontaki2011continuous, tran2016distance, yoon2019nets}.
It is thus important to retain monotonic paths to objects with small distances in practice.
From this observation, we propose a heuristic that creates links between similar objects.
In addition to the observation, our heuristic utilizes the following observations:
(i) an AKNN graph has a property that similar objects of an object $p$ tend to exist within a small hop count from $p$, and
(ii) given $p$ and its similar object $p'$, similar objects of $p'$ tend to be similar to $p$ (i.e., the idea of \textsc{NNDescent}).
That is, our heuristic is based on the idea: we can create necessary links for $p$ if we traverse such objects appearing in observations (i) and (ii).

Algorithm \ref{algo_remove} describes our heuristic.
Line \ref{algo_remove_sample} samples $|P'|$ objects as target for making monotonic paths (we do not choose object with links to exact $K'$-NNs).
Pivots are weighted for this sampling, since \textsc{Greedy-Counting} traverses pivots.
For each $p \in P'$, we do the following:
\begin{enumerate}
	\setlength{\leftskip}{-3.0mm}
    \setlength{\itemsep}{0.0mm}
	\item	We conduct 3-hop BFS from $p$ (which terminates traversal when the hop count of the current object from $p$ is 3), to obtain objects with no monotonic path from $p$ (line \ref{algo_remove_3hop}).
    		This corresponds to \textsc{Get-Non-Monotonic}$()$ with a hop count constraint, and the objects obtained are maintained similarly.
	\item	We sample $|P_{piv}|$ pivots with small distances to $p$ (pivots existing within one hop from $p$ and/or having their exact $K'$-NNs are not sampled).
			Then, for each $p' \in P_{piv}$, 2-hop BFS from $p'$ is done, and we obtain objects with no monotonic path from $p$ (lines \ref{algo_remove_sample_}--\ref{algo_remove_2hop_e}).
    		That is, BFS starts from $p'$ but computes distances between $p$ and the objects within two hops from $p'$.
\end{enumerate}
After that, we create necessary links, similar to MSG building (lines \ref{algo_remove_add_b}--\ref{algo_remove_add_e}).
(We can increase the above hop counts to improve the accuracy, but the computational cost becomes significantly larger, which can be observed from Lemma \ref{lemma_remove-detours}.)

\begin{algorithm}[!t]
	\caption{\textsc{Remove-Detours}}	\label{algo_remove}
	\DontPrintSemicolon
    {\small
        \KwIn {$G$}
		\vspace{1.0mm}
		$P' \leftarrow$ a set of randomly chosen objects	\;	\label{algo_remove_sample}
        $P_{non} \leftarrow \varnothing$	\;
        \For {each $p \in P'$}{																	\label{algo_remove_b}
        	$P_{non} \leftarrow P_{non} \,\cup$ \textsc{Get-Non-Monotonic}$(p,p,3,G)$	\;		\label{algo_remove_3hop}
            $P_{piv} \leftarrow$ a set of randomly chosen pivots	\;							\label{algo_remove_sample_}
            \For {each $p' \in P_{piv}$}{														\label{algo_remove_2hop_b}
            	$P_{non} \leftarrow P_{non} \,\cup$ \textsc{Get-Non-Monotonic}$(p,p',2,G)$	\;	\label{algo_remove_2hop_e}
            }
        }
        \For {each $\langle p,p'\rangle \in P_{non}$}{	\label{algo_remove_add_b}
        	Create links between $p$ and $p'$			\label{algo_remove_add_e}
        }
	}
\end{algorithm}

\begin{figure}[!t]
	\begin{center}
		\subfigure[3-hop BFS]{%
		\includegraphics[width=0.485\linewidth]{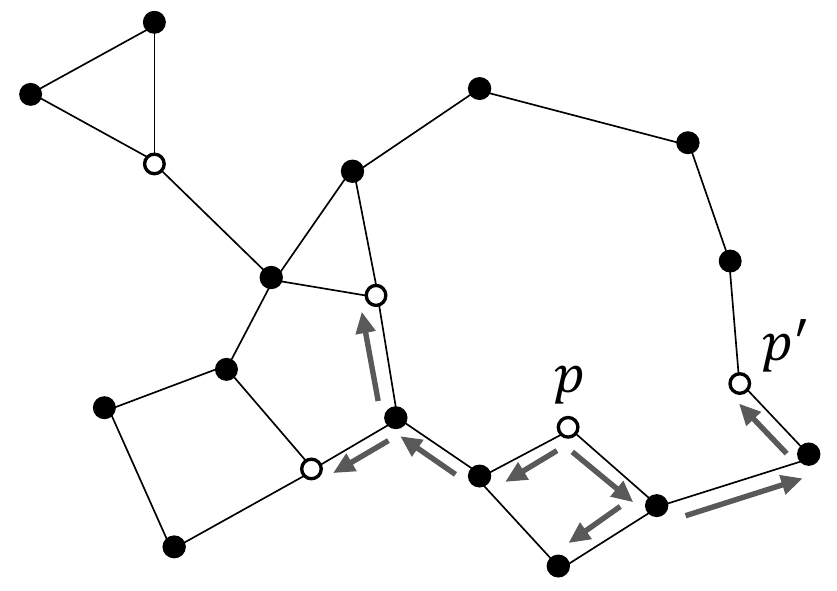}	\label{fig_detour-a}}
        \subfigure[Removing a detour]{%
		\includegraphics[width=0.485\linewidth]{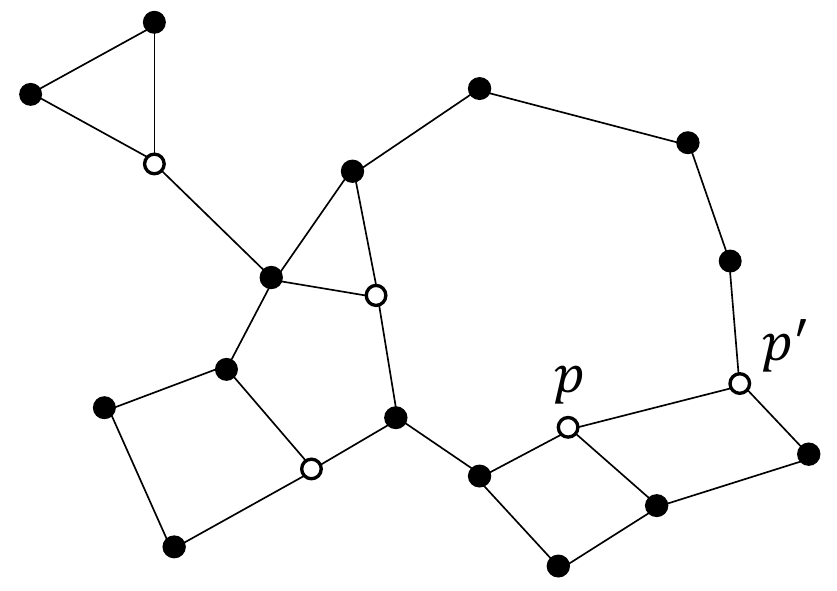}	\label{fig_detour-b}}
        \vspace{-4.0mm}
        \caption{Example of \textsc{Remove-Detours}}
        \label{figure_detour}
	\end{center}
\end{figure}

\begin{exa} \label{example_detour}
We present an example of Algorithm \ref{algo_remove}.
Figure \ref{fig_detour-a}, which shows the proximity graph obtained in Example \ref{example_connect}, depicts 3-hop BFS.
For ease of presentation, assume $P' = \{p\}$, and 3-hop BFS is conducted from $p$.
We see that the path from $p$ to $p'$ is a detour, i.e., is not a monotonic path.
After sampling pivots near $p$ and 2-hop BFS from them (not described here), we have $A = \{p'\}$.
Hence we add a link between $p$ and $p'$, as shown in Figure \ref{fig_detour-b}.
\end{exa}

Note that we set $|P'| = O(\frac{n}{K})$ and $|P_{piv}| = O(K)$.
Recall that \textsc{Get-Non-Monotonic}$()$ maintains $A$, and we limit the size of $A$ so that $|A|$ is at most $O(K^2)$ by maintaining only objects with the smallest distances to $p$.
Then, we have:

\begin{lemm}    \label{lemma_remove-detours}
\textsc{Remove-Detours} needs $O(nK^2\log K)$ time.
\end{lemm}

\subsection{Removing Links}	\label{section_link}
Since each object has links to its similar objects, $p_{1}$ and $p_{2}$, which is connected to $p_{1}$, may have links to other common objects, say $p_{3}$.
If $p_{1}$ and $p_{2}$ are traversed by \textsc{Greedy-Counting}, $p_{3}$ is accessed at least two times.
If there are many common links between objects within one hop, redundant accesses are incurred many times.
To reduce them, \textsc{Remove-Links} removes links based on pivots.

If a non-pivot object $p$ has a link to a pivot $p'$, we remove links to common objects between $p$ and $p'$.
We do this link removal for each non-pivot object.
(Because of this removal, lines \ref{algo_greedy-counting_pivot_b}--\ref{algo_greedy-counting_pivot_e} of Algorithm \ref{algo_greedy-counting} are necessary.)

\begin{figure}[!t]
	\begin{center}
		\subfigure[Checking common pivots]{%
		\includegraphics[width=0.485\linewidth]{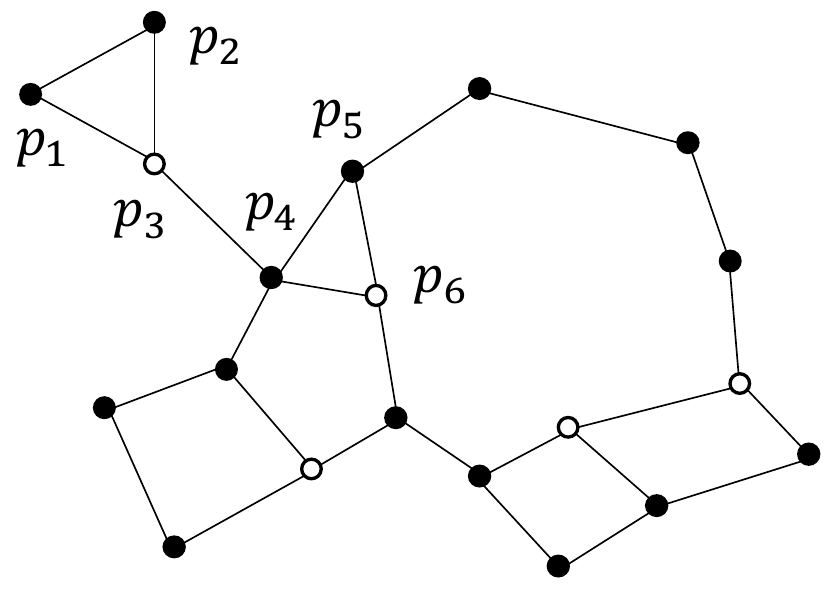}	\label{fig_link-a}}
        \subfigure[Removing unnecessary links]{%
		\includegraphics[width=0.485\linewidth]{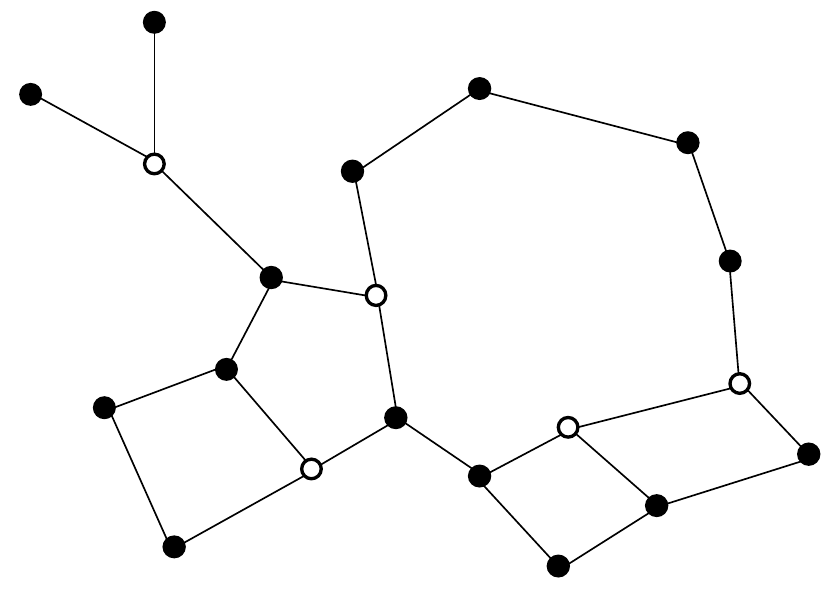}	\label{fig_link-b}}
        \vspace{-4.0mm}
        \caption{Example of \textsc{Remove-Links}}
        \label{figure_link}
	\end{center}
\end{figure}

\begin{exa} \label{example_link}
Figure \ref{figure_link} presents an example of \textsc{Remove-Links}.
We use the graph obtained in Example \ref{example_detour}.
Two non-pivot objects $p_{1}$ and $p_{2}$ in Figure \ref{fig_link-a} respectively have a link to a common pivot $p_{3}$.
Objects $p_{4}$, $p_{5}$, and $p_{6}$ have the same case.
Therefore, links $(p_{1},p_{2})$ and $(p_{4},p_{5})$ are removed, then we have an MRPG shown in Figure \ref{fig_link-b}.
\end{exa}

By using hash-based link management, it is trivial to see that

\begin{lemm}    \label{lemma_remove-links}
\textsc{Remove-Links} incurs $O(nK)$ time.
\end{lemm}

\subsection{Discussion} \label{section_discussion}
From Lemmas \ref{lemma_nndescent-plus}--\ref{lemma_remove-links}, we see that:

\begin{theo}    \label{theorem_mrpg-time}
We need $O(nK^2\log K)$ time to build a MRPG.
\end{theo}

\noindent
In addition,

\begin{theo}    \label{theorem_mrpg-space}
The space complexity of a MRPG is $O(nK)$.
\end{theo}

In MRPG, there are objects that have links to their exact $K'$-NNs, and these objects have a larger distance to their approximate $K$-NNs compared with the other objects in $P$.
It can be intuitively seen that these objects tend to be outliers for any (reasonable) $r$.
Assume that $p$ has links to its exact $K'$-NNs.
If $K' \geq k$, we can evaluate whether $p$ is outlier or not in $O(k)$ time, by traversing only its links.
That is, if the count does not reach $k$, we can accurately determine that $p$ is an outlier without verification, which reduces $t$ in Theorem \ref{theorem_online-time}.
For such objects $p$, we replace lines \ref{algo_framework_filter_b_}--\ref{algo_framework_filter_e} of Algorithm \ref{algo_framework} with the above operation.
By setting a sufficiently large integer as $K'$, when $k$ is reasonable, MRPG detects outliers very quickly.
(If $k > K'$, MRPG utilizes the original Algorithm \ref{algo_framework} to keep correctness, so it does not lose generality).
As analyzed in Section \ref{section_overview}, the main cost of online processing is the verification cost.
Therefore, reducing this cost from $O(n)$ to $O(k)$ yields significant efficiency improvement.

\section{Experiments}	\label{section_experiment}
This section reports our experimental results.
Our experiments were conducted on a machine with dual 12-core Intel Xeon E5-2687w v4 processors (3.0GHz) that share a 512GB RAM.
This machine can run at most 48 threads by using hyper-threading.
All evaluated algorithms were implemented in C++ and compiled by g++ 7.4.0 with -O3 flag.
We used OpenMP for multi-threading.

\vs
\noindent
\textbf{Datasets.}
We used seven real datasets, Deep \cite{babenko2016efficient}, Glove\footnote{\url{https://nlp.stanford.edu/projects/glove/}}, HEPMASS\footnote{\url{https://archive.ics.uci.edu/ml/datasets/HEPMASS}}, MNIST\footnote{\url{https://www.csie.ntu.edu.tw/~cjlin/libsvmtools/datasets/multiclass.html}}, PAMAP2\footnote{\url{https://archive.ics.uci.edu/ml/datasets/PAMAP2+Physical+Activity+Monitoring}}, SIFT\footnote{\url{http://corpus-texmex.irisa.fr/}}, and Words\footnote{\url{https://github.com/dwyl/english-words}}.
(For MNIST, we randomly sampled 3 million objects from the original dataset.)
Table \ref{table_dataset} summarizes their statistics and distance functions we used\footnote{We have updated the implementation for computing $L_{4}$-norm, and the experimental results on MNIST have also been updated from \cite{amagata2021dod}.}.
We normalized PAMAP2, so that the domain of each dimension is $[0,10^5]$.
We observed that the distance distribution of SIFT follows Gaussian mixture distribution and that of the other datasets follows Gaussian distribution.

\vs
\noindent
\textbf{Algorithms.}
We evaluated the following exact algorithms:
\begin{itemize}
	\setlength{\leftskip}{-4.0mm}
    \setlength{\itemsep}{0.0mm}
	\item	State-of-the-art: \textit{Nested-loop} \cite{bay2003mining}, \textit{SNIF} \cite{tao2006mining}, \textit{DOLPHIN} \cite{angiulli2009dolphin}, and \textit{VP-tree} \cite{yianilos1993data}, which are introduced in Section \ref{section_related-work}.
    \item	Proximity graph-based algorithm: \textit{NSW} \cite{malkov2014approximate}, \textit{KGraph} \cite{dong2011efficient}, \textit{MRPG-basic}, and \textit{MRPG}.
    		MRPG-basic is a variant of MRPG, and, in \textsc{NNDescent+}, we compute the exact $K$-NNs for some objects, instead of $K'$-NNs.
    		Therefore, by comparing MRPG with MRPG-basic, the efficiency of optimizing the verification is understandable. 
    		For outlier detection with NSW and KGraph, we used Algorithms \ref{algo_framework} and \ref{algo_greedy-counting} without lines \ref{algo_greedy-counting_pivot_b}--\ref{algo_greedy-counting_pivot_e} of Algorithm \ref{algo_greedy-counting}.
    		They used the same verification phase as MRPG.
            We employed a VP-tree in the verification phase, i.e., \textsc{Exact-Counting}, on HEPMASS, PAMAP2, and Words.
\end{itemize}
We followed the original papers to set the system parameters in the state-of-the-art.
For KGraph, MRPG-basic, and MRPG on PAMAP2, we set $K = 40$, and we set $K = 25$ for the other datasets.
The number of links for each object in NSW is set so that its memory is almost the same as that of KGraph.
For MRPG, we set $K' = 4 \times K$.
Codes are available in a GitHub repository\footnote{\url{https://github.com/amgt-d1/DOD}}.

We set 12 and 8 hours as time limit for pre-processing (offline time) and outlier detection (online time), respectively.
In cases that algorithms could not terminate pre-processing or detect all outliers within the time limit, we represent NA as the result.

\vs
\noindent
\textbf{Parameters.}
Table \ref{table_parameter} shows the default parameters.
They were specified so that the outlier ratio is small \cite{cao2016sharing, yoon2019nets} or clear outliers are identified\footnote{If objects have a small number of neighbors for a reasonable $r$ such that (most of) the other objects have enough or many neighbors, they are clear outliers.
We provided such $r$ and $k$.}.
We confirmed that the number of neighbors in each dataset follows power law and most objects have many neighbors.
We used 12 (48) threads as the default number of threads for outlier detection (pre-processing).
For outlier detection on Deep and MNIST, we used 48 threads, because they need time to be processed (due to large $n$ and dimensionality).

\begin{table}[!t]
\begin{center}
	\caption{Datasets}
    \label{table_dataset}
    \vspace{-3.0mm}
	\begin{tabular}{c||c|c|c} \hline
                Dataset		& Cardinality	& Dim.	& Distance function	\\ \hline \hline
                Deep		& 10,000,000	& 96	& $L_{2}$-norm		\\ \hline
                Glove		& 1,193,514		& 25	& Angular distance	\\ \hline
                HEPMASS		& 7,000,000		& 27	& $L_{1}$-norm		\\ \hline
                MNIST		& 3,000,000		& 784	& $L_{4}$-norm		\\ \hline
                PAMAP2		& 2,844,868		& 51	& $L_{2}$-norm		\\ \hline
                SIFT		& 1,000,000		& 128	& $L_{2}$-norm		\\ \hline
                Words		& 466,551		& 1--45	& Edit distance		\\ \hline
	\end{tabular}
	\vspace{-3.0mm}
\end{center}
\end{table}

\begin{table}[!t]
\begin{center}
	\caption{Default parameters}
    \label{table_parameter}
    \vspace{-3.0mm}
	\begin{tabular}{c||c|c|c} \hline
                Dataset	& $r$		& $k$	& Outlier ratio	\\ \hline \hline
                Deep	& 0.93		& 50	& 0.62\%		\\ \hline
                Glove	& 0.25		& 20	& 0.55\%		\\ \hline
                HEPMASS	& 15		& 50	& 0.65\%		\\ \hline
                MNIST	& 600		& 50	& 0.34\%		\\ \hline
                PAMAP2	& 50,000	& 100	& 0.61\%		\\ \hline
                SIFT	& 320		& 40	& 1.04\%		\\ \hline
                Words	& 5			& 15	& 4.16\%		\\ \hline
	\end{tabular}
\end{center}
\end{table}

\subsection{Evaluation of Pre-processing}
We first evaluate the pre-processing efficiencies of NSW, KGraph, MRPG-basic, and MRPG.
Nested-loop, SNIF, and DOLPHIN do not have a pre-processing phase, whereas building a VP-tree took less than 310 seconds for each dataset.

\begin{table}[!t]
\begin{center}
	\caption{Pre-processing time [sec]}
    \label{table_pretime}
    \vspace{-3.0mm}
	\begin{tabular}{c||c|c|c|c} \hline
                Dataset		& NSW		& KGraph	& MRPG-basic	& MRPG  	\\ \hline \hline
                Deep		& NA		& 20079.80	& 13417.40	    & 17230.30	\\ \hline
                Glove		& 2333.47	& 923.83	& 755.54    	& 791.53	\\ \hline
                HEPMASS		& NA		& 7935.25	& 4345.63	    & 5221.86	\\ \hline
                MNIST		& 33368.0	& 2972.96	& 1566.05	    & 2281.55	\\ \hline
                PAMAP2		& 4522.14	& 1325.40	& 729.54	    & 888.61	\\ \hline
                SIFT		& 4910.94	& 929.48	& 723.75	    & 817.33	\\ \hline
                Words		& 871.27	& 455.15	& 707.08		& 868.62	\\ \hline
	\end{tabular}
    \vspace{3.0mm}
	\caption{Decomposed time of pre-processing on Glove [sec]}
    \label{table_pre_decompose}
    \vspace{-3.0mm}
	\begin{tabular}{c||c|c|c} \hline
                Algorithm					& KGraph	& MRPG-basic    & MRPG  	\\ \hline \hline
        		\textsc{NNDescent(+)}		& 923.83	& 464.34	    & 474.20	\\ \hline
                \textsc{Connect-SubGraphs}	& -			& 20.36		    & 24.28		\\ \hline
                \textsc{Remove-Detours}		& -			& 278.21	    & 271.41	\\ \hline
                \textsc{Remove-Links}		& -			& 19.44		    & 19.61		\\ \hline
	\end{tabular}
	\vspace{5.0mm}
\end{center}
\end{table}

\begin{figure*}[!t]
	\begin{center}
    \includegraphics[width=0.40\linewidth]{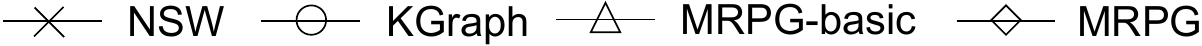}
    \vspace{-1.5mm}
    
	\subfigure[Deep]{%
		\includegraphics[width=0.235\linewidth]{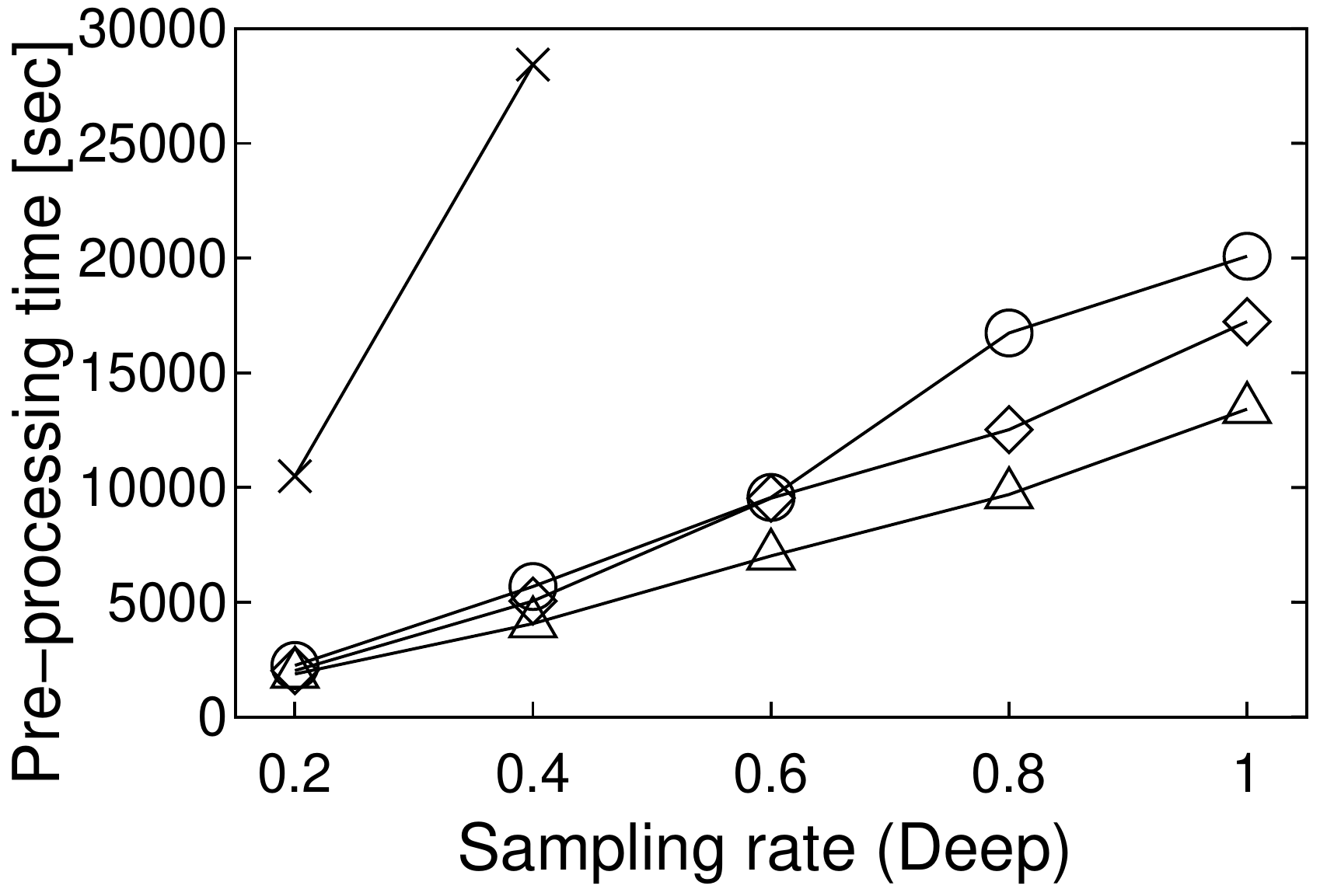}		\label{fig_deep_n_pre}}
        \subfigure[Glove]{%
		\includegraphics[width=0.235\linewidth]{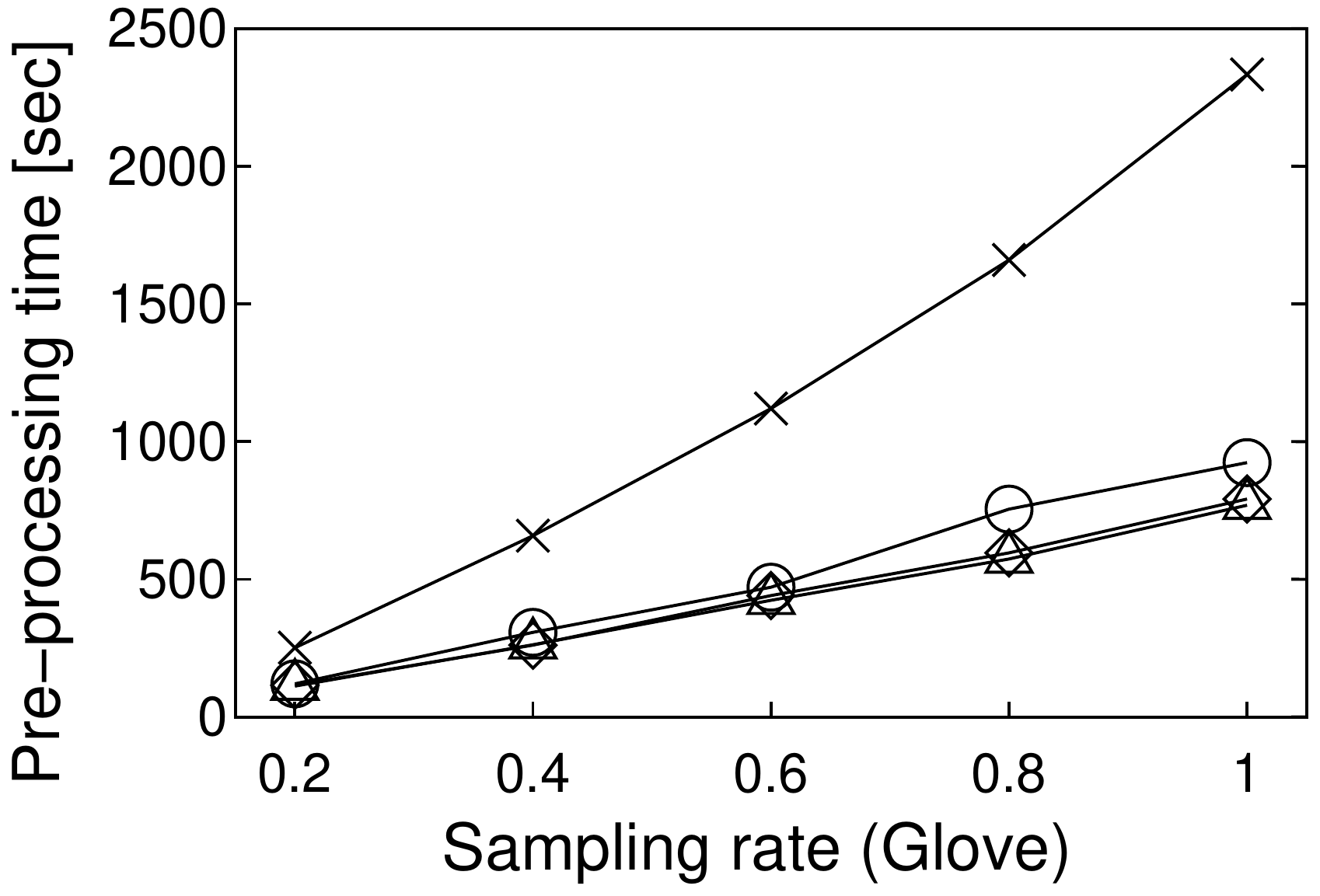}		\label{fig_glove_n_pre}}
        \subfigure[HEPMASS]{%
		\includegraphics[width=0.235\linewidth]{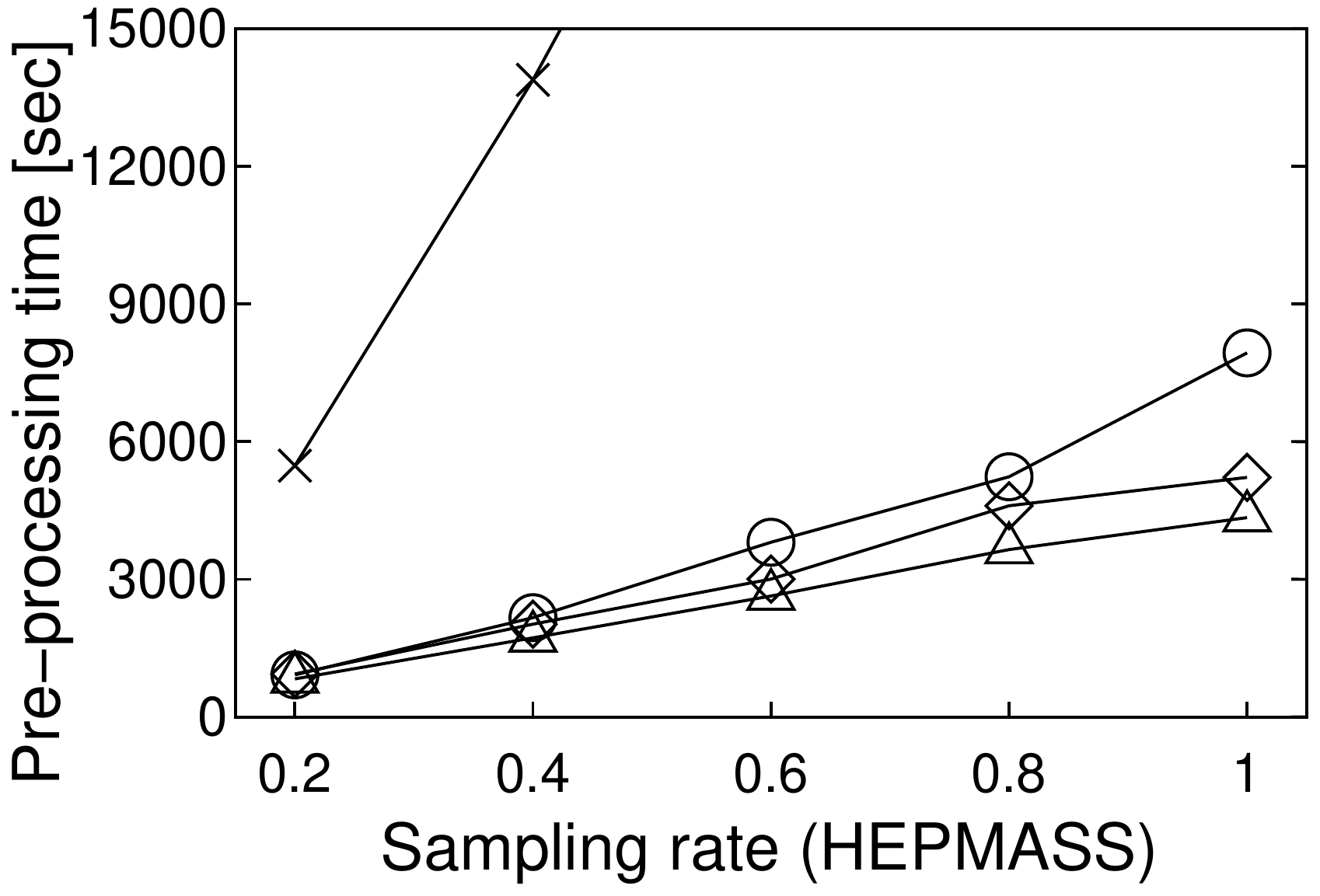}	\label{fig_hepmass_n_pre}}
        \subfigure[MNIST]{%
		\includegraphics[width=0.235\linewidth]{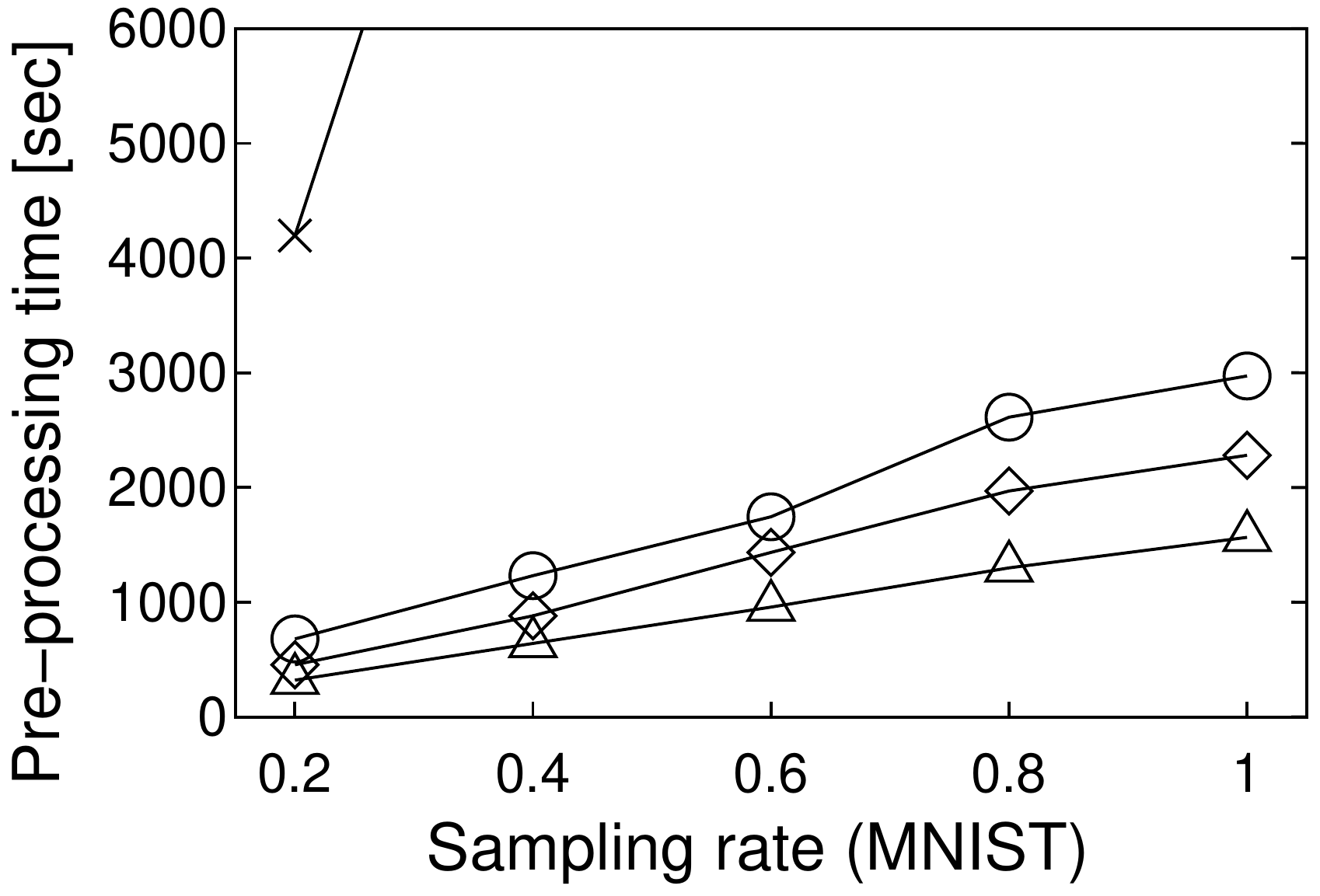}		\label{fig_mnist_n_pre}}
        \subfigure[PAMAP2]{%
		\includegraphics[width=0.235\linewidth]{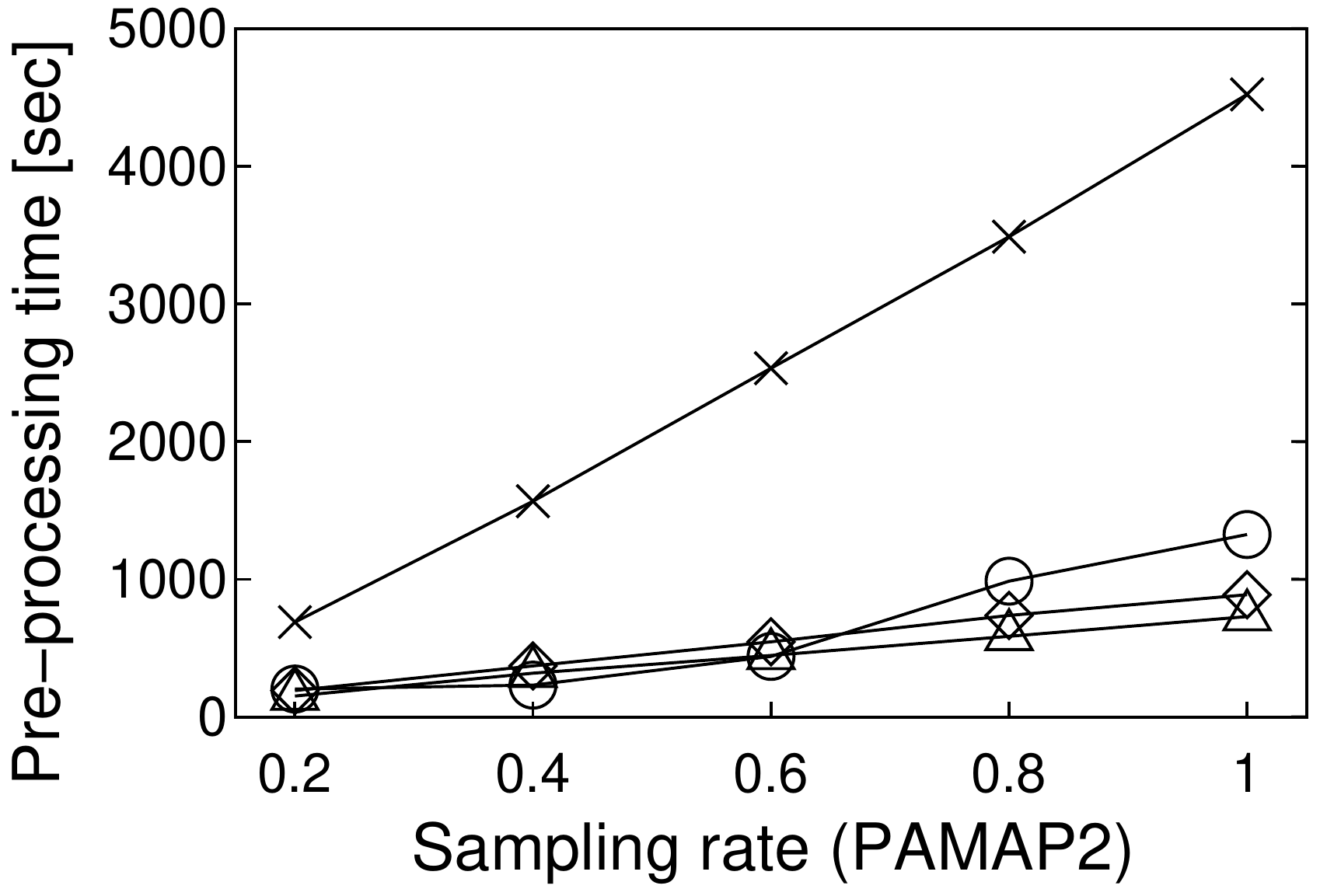}	\label{fig_pamap2_n_pre}}
        \subfigure[SIFT]{%
		\includegraphics[width=0.235\linewidth]{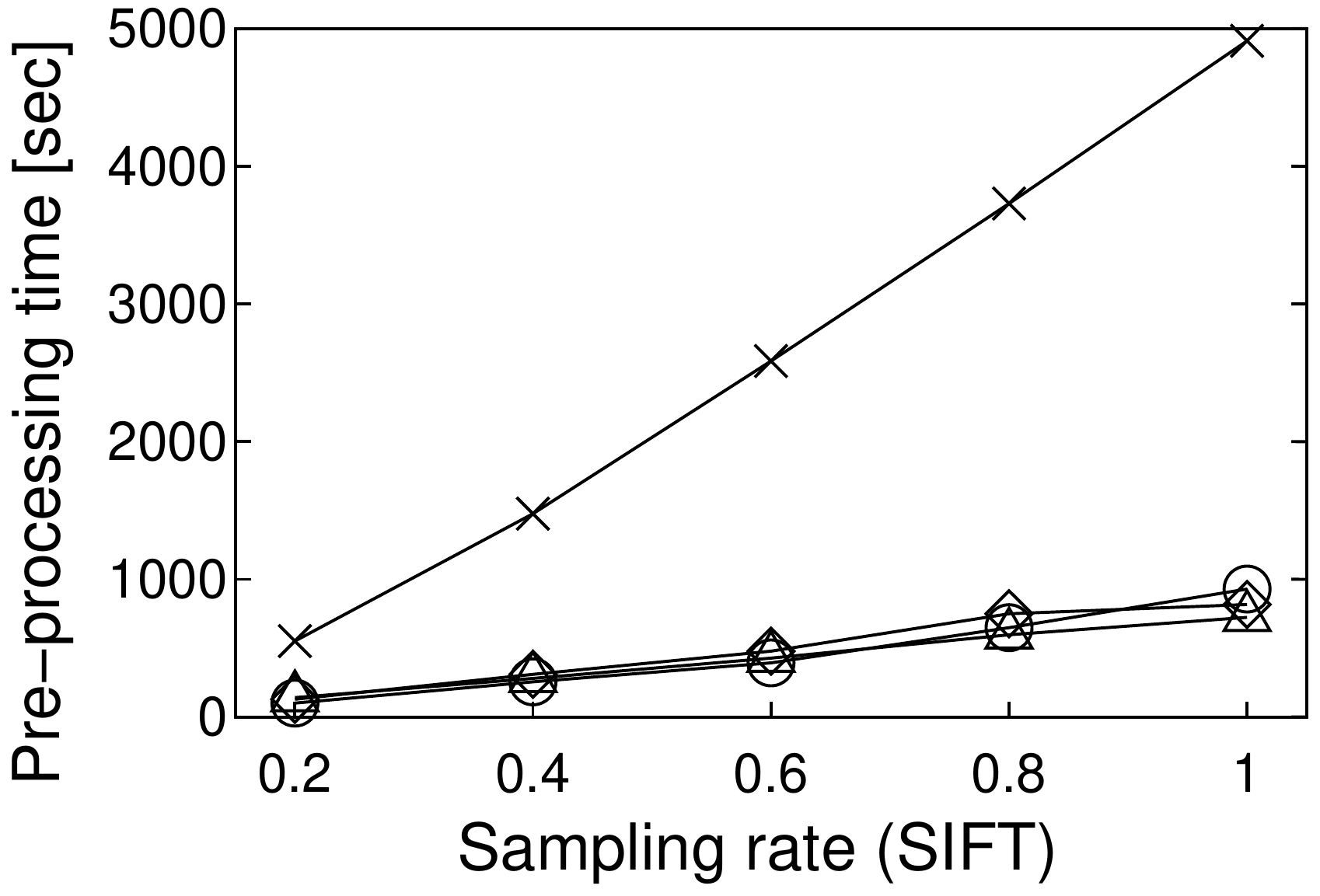}		\label{fig_sift_n_pre}}
        \subfigure[Words]{%
		\includegraphics[width=0.235\linewidth]{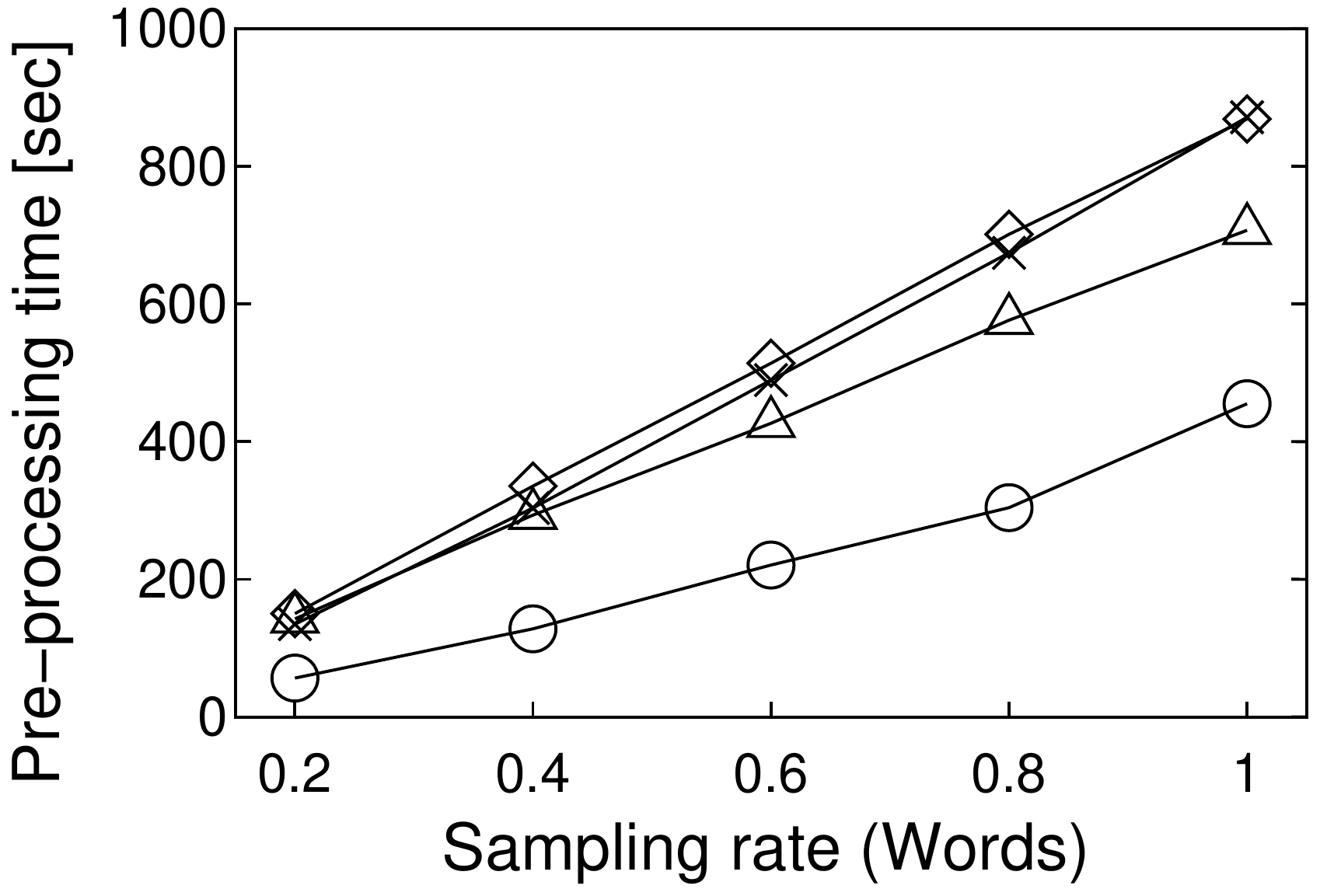}		\label{fig_words_n_pre}}
		\vspace{-5.0mm}
        \caption{Impact of $n$ (pre-processing time)}
        \label{figure_sample_pre}
        \vspace{-4.0mm}
	\end{center}
\end{figure*}
\begin{table*}[!t]
\begin{center}
	\caption{Running time [sec]. Bold shows the winner.}
    \label{table_time}
    \vspace{-3.0mm}
	\begin{tabular}{c||c|c|c|c|c|c|c|c} \hline
                Dataset		& Nested-loop	& SNIF		& DOLPHIN	& VP-tree		& NSW		& KGraph		& MRPG-basic	& MRPG  			\\ \hline \hline
        		Deep		& NA			& NA		& NA		& NA			& NA		& 8616.10		& 5474.10		& \textbf{1966.17}	\\ \hline
                Glove		& 1045.47		& 1222.43	& 9277.89	& 1398.92		& 147.00	& 83.82			& 56.80			& \textbf{2.63}		\\ \hline
                HEPMASS		& 17295.40		& 20360.80	& NA		& 8597.23		& NA		& 780.19		& 638.83		& \textbf{38.40}	\\ \hline
                MNIST		& 15494.00		& 22579.80	& NA		& 13836.60		& 1630.25	& 1702.57		& 1264.26		& \textbf{918.91}	\\ \hline
                PAMAP2		& 422.40		& 1213.56	& 1819.90	& 208.55		& 22.16		& 23.77			& 18.16			& \textbf{10.55}	\\ \hline
                SIFT		& 1427.74		& 1507.58	& 8684.08	& 2005.27		& 200.89	& 175.88		& 144.11		& \textbf{11.32}	\\ \hline
                Words		& 1844.65		& 2086.50	& 7061.50	& 1021.39		& 498.34	& 393.95		& 374.08		& \textbf{2.67}		\\ \hline
	\end{tabular}
\end{center}
\vspace{-3.0mm}
\end{table*}

\vs
\noindent
\textbf{MRPG(-basic) vs. KGraph.}
Table \ref{table_pretime} presents the pre-processing time of each proximity graph at the default parameters.
In most cases, building a MRPG-basic is the most efficient and building a MRPG is also more efficient than building a KGraph.
This result is derived from the efficiency of \textsc{NNDescent+}.
We depict the decomposed time of building a KGraph, MRPG-basic, and MRPG on Glove in Table \ref{table_pre_decompose}, as an example.
This table shows that \textsc{NNDescent+} is faster than \textsc{NNDescent}, demonstrating the effectiveness of the VP-tree based partitioning approach and the skipping approach.
Also, the other functions for building a MRPG do not incur significant costs.
These provide a high performance for building a MRPG.

One exception appears in the Words case.
We used edit distance for Words, and this distance function needs a large computational cost for objects with large dimensionality.
We observed that objects, whose exact $K'$-NNs are computed, have large dimensionality, thereby exact $K'$-NN computation incurs a long time.

\vs
\noindent
\textbf{MRPG vs. MRPG-basic.}
Building a MRPG needs longer time than building a MRPG-basic.
This is because, for some objects, we compute their $K'$-NNs where $K' > K$, during building a MRPG.
That is, \textsc{NNDescent+} for MRPG incurs longer time than that for MRPG-basic, as Table \ref{table_pre_decompose} presents.

\vs
\noindent
\textbf{NSW vs. the other proximity graphs.}
Table \ref{table_pretime} shows that building a NSW consistently needs longer time than building the other proximity graphs.
Because the NSW building algorithm is based on incremental object insertion, building a NSW cannot use multi-threads.
This property lacks the scalability to large datasets and ones with large dimensionality.
Therefore, NSW cannot be built on Deep and HEPMASS within a half day.

\vs
\noindent
\textbf{Scalability test.}
Figure \ref{figure_sample_pre} illustrates the scalability to $n$ when building the proximity graphs.
We varied the size of $P$ by random sampling (i.e., we varied sampling rate).

As discussed, NSW basically needs (much) larger time for building.
This algorithm is competitive only in the case of Words, because its cardinality is smaller than the other datasets.
KGraph, MRPG-basic, and MRPG have linear scalability to $n$, due to Theorems \ref{theorem_nndescent} and \ref{theorem_mrpg-time}.
Notice that MRPG scales better in most cases.

\subsection{Evaluation of DOD Algorithms}
We next evaluate outlier detection algorithms.
Tables \ref{table_time} and \ref{table_index} describe the running time and index size of each algorithm, respectively.
We did not test algorithms whose index could not be obtained within the time limit.

\begin{table*}[!t]
\begin{center}
	\caption{Index size [MB]}
    \label{table_index}
    \vspace{-3.0mm}
	\begin{tabular}{c||c|c|c|c|c|c|c|c} \hline
                Dataset		& Nested-loop	& SNIF		& DOLPHIN	& VP-tree	& NSW		& KGraph	& MRPG-basic	& MRPG  	\\ \hline \hline
                Deep		& 0				& NA		& NA		& 324.35	& NA		& 1405.94	& 5516.58	    & 7350.83	\\ \hline
                Glove		& 0				& 13.26		& 69.14		& 54.86		& 188.62	& 167.91	& 460.48	    & 438.76	\\ \hline
                HEPMASS		& 0				& 61.04		& NA		& 265.39	& NA		& 1195.35	& 2188.65	    & 2450.84	\\ \hline
                MNIST		& 0				& 27.75		& NA		& 117.80	& 417.95	& 404.29	& 589.08	    & 591.27	\\ \hline
                PAMAP2		& 0				& 18.36		& 65.12		& 128.00	& 819.17	& 528.26	& 553.87	    & 760.69	\\ \hline
                SIFT		& 0				& 8.10		& 47.00		& 39.99		& 157.58	& 140.54	& 433.48	    & 489.14	\\ \hline
                Words		& 0				& 4.41		& 26.86		& 27.79		& 102.20	& 93.92		& 191.73	    & 178.74	\\ \hline
	\end{tabular}
	\vspace{-2.0mm}
\end{center}
\end{table*}

\vs
\noindent
\textbf{Our approach vs. state-of-the-art.}
Let us compare our approach, proximity graph-based solution (NSW, KGraph, MRPG-basic, and MRPG), with state-of-the-art (Nested-loop, SNIF, DOLPHIN, and VP-tree).
Table \ref{table_time} shows that our approach is clearly faster than the state-of-the-art, demonstrating its robustness to the distance functions listed in Table \ref{table_dataset}.
This result is derived from the reduction of unnecessary distance computation.
Specifically, in our approach (or a proximity graph), each object has links (or paths) to its neighbors.
This yields efficient early termination, i.e., inliers are quickly identified.
For example, MRPG is 397.5, 223.9, 15.1, 19.8, 126.1, and 382.5 times faster than the best algorithm among the state-of-the-art on Glove, HEPMASS, MNIST, PAMAP2, SIFT, and Words, respectively.
The state-of-the-art could not detect outliers within the time limit on Deep (largest dataset), whereas MRPG and MRPG-basic successfully deal with it.
Also, we see that \textit{MRPG provides a significant speed-up} by sacrificing a bit longer pre-processing time than MRPG-basic.
This speed-up is derived from the reduction of the verification cost by detecting (some) outliers in the filtering phase (see Section \ref{section_discussion}).

Table \ref{table_index} shows that our approach needs a larger index size than the state-of-the-art (Nested-loop does not build an index, so its index size is 0).
However, its index size is not significant, and recent main-memory systems afford to retain the proximity graph, as its space requirement is $O(nK)$.

\begin{table}[!t]
\begin{center}
	\caption{Number of false positives after the filtering phase}
    \label{table_fp}
    \vspace{-3.0mm}
	\begin{tabular}{c||c|c|c|c} \hline
                Algorithm	& NSW		& KGraph	& MRPG-basic	& MRPG     	\\ \hline \hline
                Deep		& NA		& 81,140	& 33,180	    & 20,616	\\ \hline
        		Glove		& 19,970	& 3,356		& 40		    & 24		\\ \hline
                HEPMASS		& NA		& 11,133	& 2,363		    & 438		\\ \hline
                MNIST		& 7,079		& 4,698		& 2,509		    & 2,061		\\ \hline
                PAMAP2		& 18,346	& 22,543	& 4,290		    & 3,986		\\ \hline
                SIFT		& 4,899		& 2,513		& 585		    & 51		\\ \hline
                Words		& 9,569		& 989		& 120		    & 4 		\\ \hline
	\end{tabular}
\end{center}
\vspace{-4.0mm}
\end{table}
\begin{table}[!t]
\begin{center}
	\caption{Decomposed time of outlier detection on Glove [sec]}
    \label{table_decompose}
    \vspace{-3.0mm}
	\begin{tabular}{c||c|c|c|c} \hline
                Algorithm		& NSW		& KGraph	& MRPG-basic	& MRPG  	\\ \hline \hline
        		Filtering		& 1.28		& 0.86		& 2.43	        & 1.98		\\ \hline
                Verification	& 147.00	& 82.96		& 57.03         & 0.65		\\ \hline
	\end{tabular}
\end{center}
\end{table}

\begin{figure*}[!t]
	\begin{center}
		\subfigure[Glove]{%
		\includegraphics[width=0.235\linewidth]{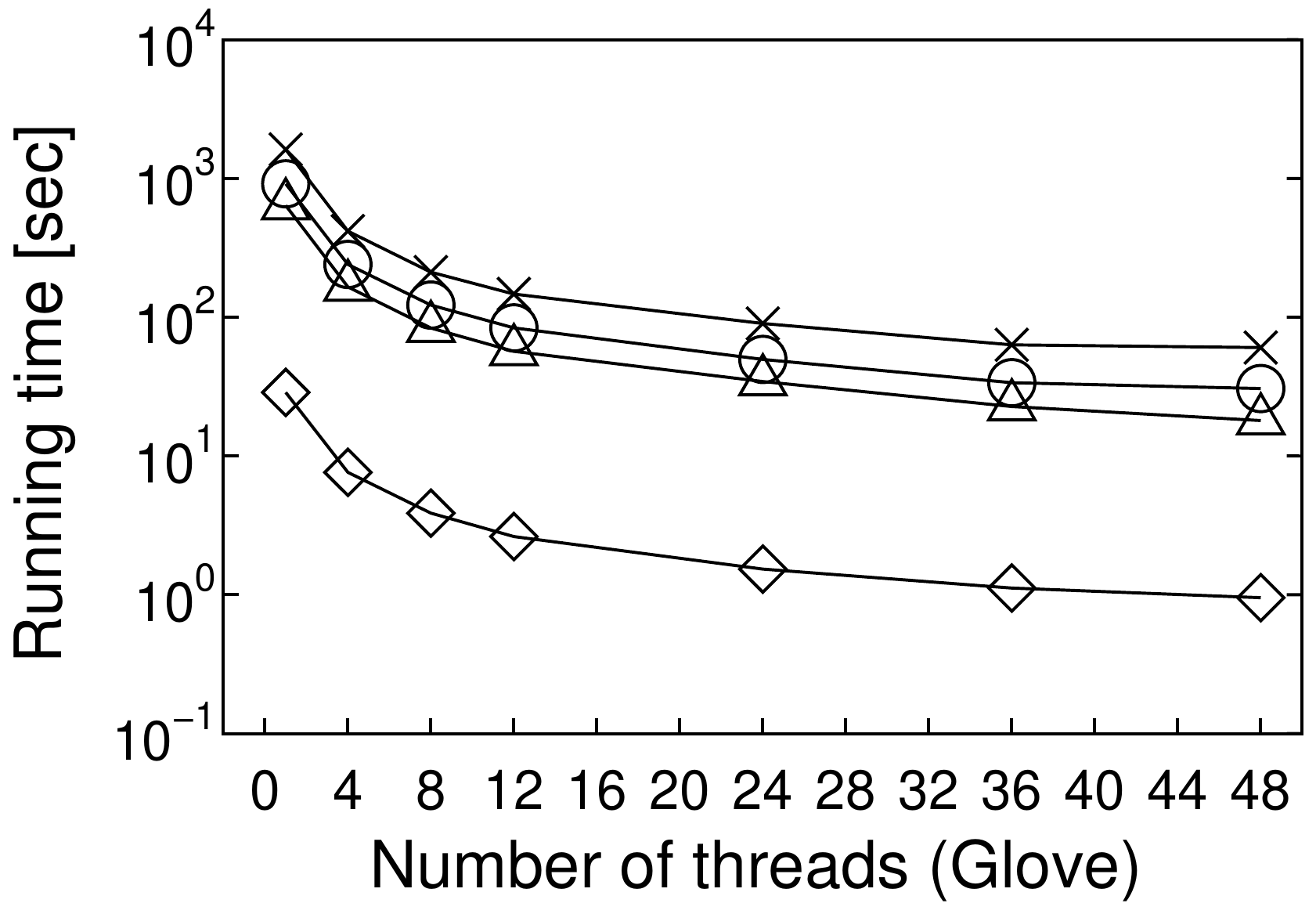}		\label{fig_glove_t}}
        \subfigure[HEPMASS]{%
		\includegraphics[width=0.235\linewidth]{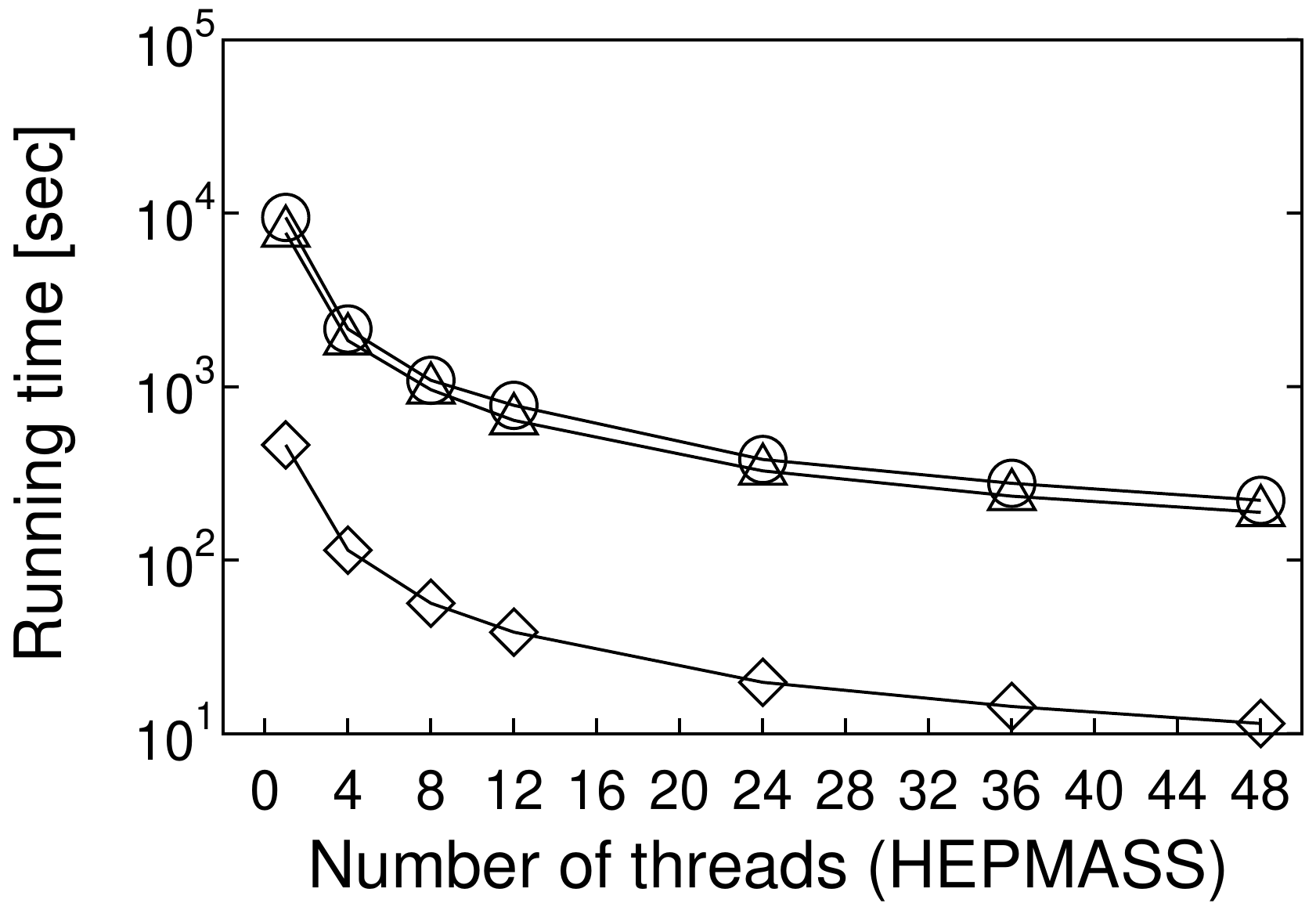}	\label{fig_hepmass_t}}
        \subfigure[PAMAP2]{%
		\includegraphics[width=0.235\linewidth]{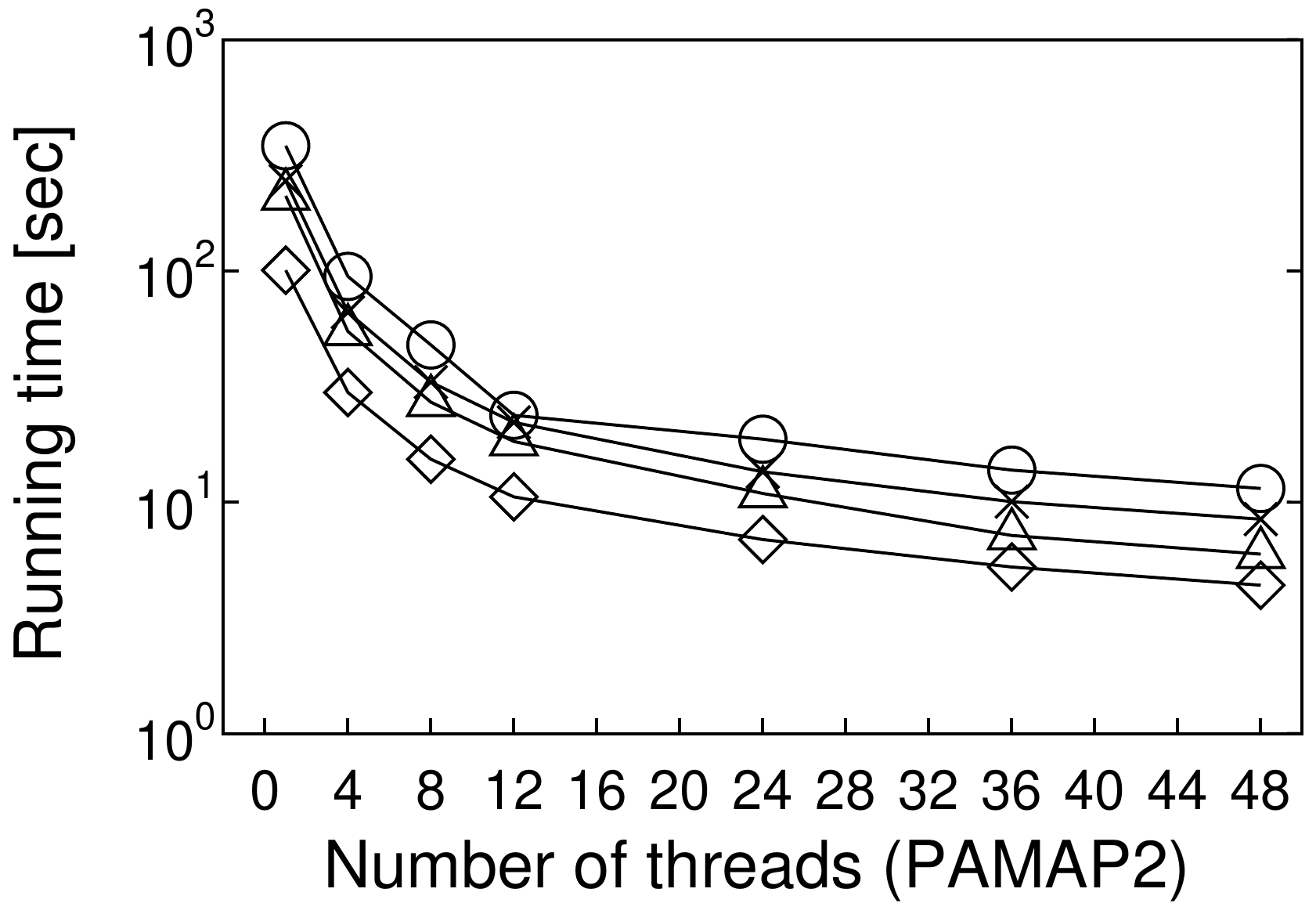}	\label{fig_pamap2_t}}
		
        \subfigure[SIFT]{%
		\includegraphics[width=0.235\linewidth]{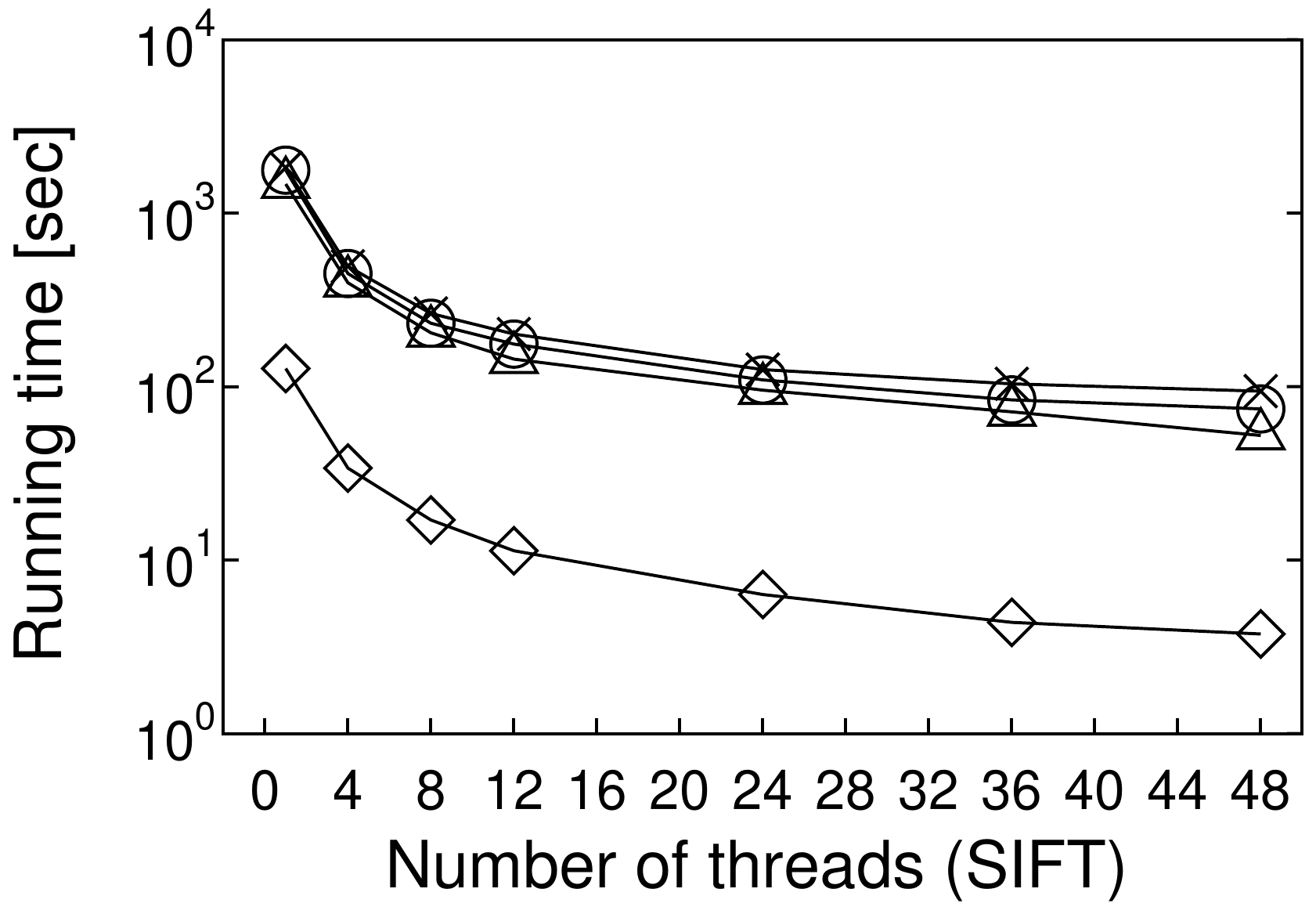}		\label{fig_sift_t}}
        \subfigure[Words]{%
		\includegraphics[width=0.235\linewidth]{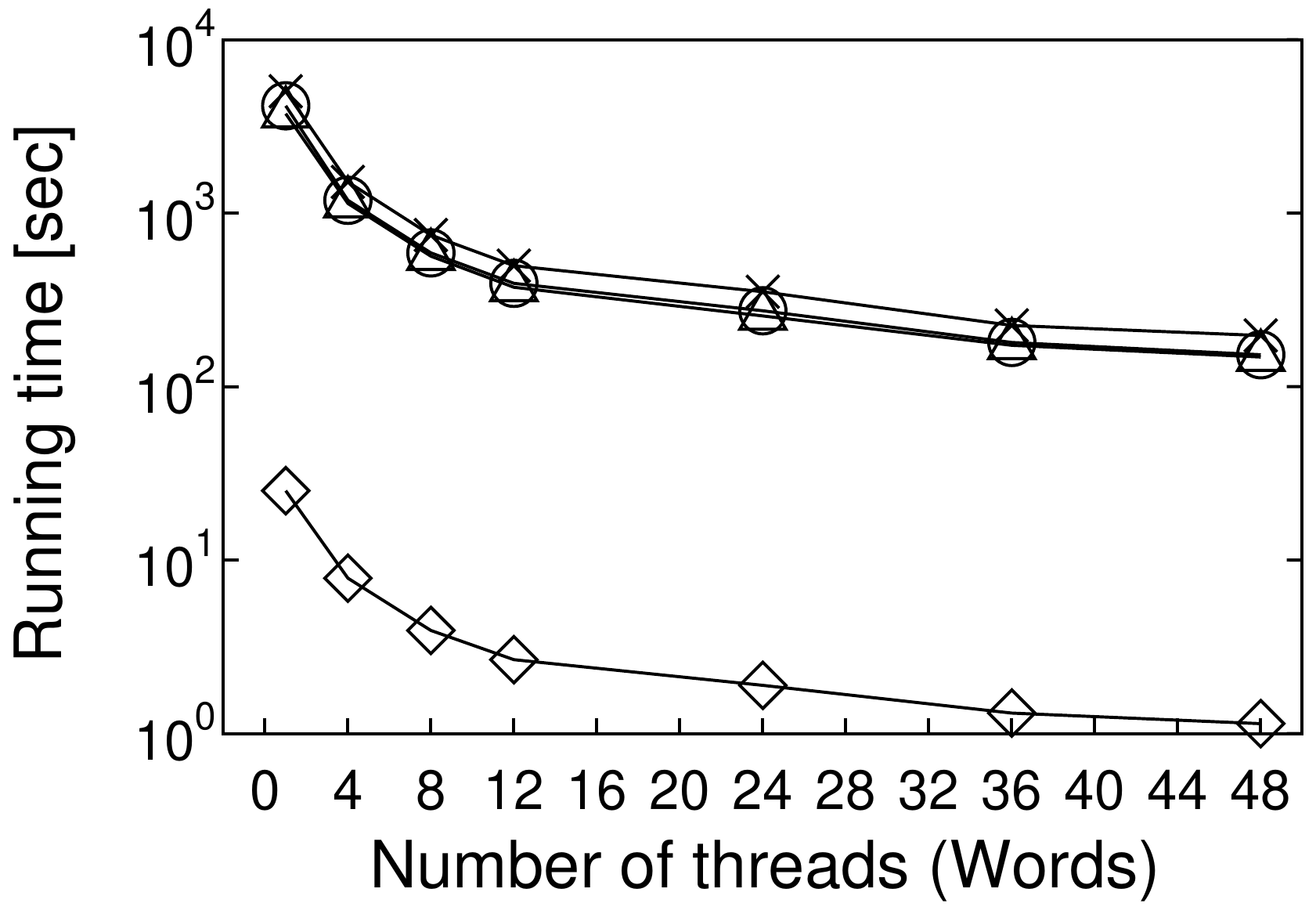}		\label{fig_words_t}}
        \vspace{-4.0mm}
        \caption{Impact of the number of threads}
        \label{figure_t}
	\end{center}
\end{figure*}

\vs
\noindent
\textbf{MRPG(-basic) vs. the other proximity graphs.}
We next focus on the performances of the proximity graphs.
Table \ref{table_time} reports that MRPG is clear winner.
Recall that, to make Algorithm \ref{algo_framework} faster, we have to reduce the number of false positives $f$, as demonstrated in Theorem \ref{theorem_online-time}.
Table \ref{table_fp} shows that MRPG and MRPG-basic reduce $f$ more compared with KGraph and NSW, so we obtain faster running time than those of KGraph and NSW\footnote{We have corrected some errors in Table \ref{table_fp} from \cite{amagata2021dod} (but this does not affect our claim in \cite{amagata2021dod}).}.
This fact demonstrates the effectiveness of monotonic paths, i.e., MRPG and MRPG-basic have a better reachability than the others.
We notice that the performance difference between MRPG and KGraph is not significant on Deep and MNIST compared with the other datasets.
In Deep, we observed that false positive objects of MRPG have only nearly $k$ neighbors, which makes the early termination not function. 
In MNIST, we found that some objects having links to their exact $K'$-NNs are inliers and false positive objects have the same observation as with Deep\footnote{Although NSW has more $f$ than that of KGraph, NSW is faster than KGraph on MNIST.
We found that the false positives of NSW have more neighbors than those of KGraph, thus, for NSW, the early termination in the sequential scan functions, rendering its faster time.}.
The verification cost of outliers and false positives therefore still remains on them, as with the other proximity graphs.
However, this can be alleviated by using additional CPUs (cores/threads), as shown in Figure \ref{figure_t}.

Recall that each dataset follows a power law distribution w.r.t. the number of neighbors.
If a dataset has many objects that are inliers but exist in sparse areas, $f$ of MRPG tends to be large.
This is because the reachability to their neighbors still tends to be lower than that to neighbors of dense objects.
The number of inliers in sparse areas is affected by data distributions, so $f$ between the datasets are different as in Table \ref{table_fp}.
For example, we observed that Deep is sparser than the other datasets\footnote{The reasonable $r$ of Deep is far from the mean of its distance distribution, compared with the other datasets.}, so its $f$ is large.

Table \ref{table_decompose} exhibits the time for filtering and verification on Glove.
Due to the reachability, MRPG and MRPG-basic incur longer filtering time but this reduces the verification time the most.
(This result is consistent for the other datasets.)
Besides, the running time of MRPG is shorter than those of the other proximity graphs.
This is due to the heuristic that objects, which would be outliers, have links to exact $K'$-NNs.
They are usually outliers in real datasets and are identified as outliers when 1-hop links are traversed from them, so verification is not needed for them.
This provides a (significant) speed-up, and MRPG is 1.3--140.1 times faster than MRPG-basic.
Recall that, in most cases, MRPG needs less pre-processing time than the others.
Therefore, in terms of computational performance, MRPG normally dominates the other proximity graphs.

As for index size, MRPG needs more memory than NSW and KGraph, because MRPG creates links to improve reachability.
However, MRPG removes unnecessary links, so its index size is competitive with those of NSW and KGraph for datasets with skew, such as PAMAP2.
The index size of MRPG is smaller than that of MRPG-basic on Glove and Words.
This is also derived from the unnecessary link removal.

\vs
\noindent
\textbf{Effectiveness of \textsc{Connect-SubGraphs} and \textsc{Remove-Detours}.}
We evaluated (i) MRPG without Algorithms \ref{algo_connect} and \ref{algo_remove}, (ii) MRPG without Algorithm \ref{algo_connect}, and (iii) MRPG without Algorithm \ref{algo_remove}, to investigate how they contribute to improving reachability.
We here report the numbers of false positives only on PAMAP2 for the three variants of MRPG, because the result is consistent with those on the other datasets.
The numbers of false positives provided by the first, second, and third variants are respectively 11937, 4712, and 9720.
They are less than those of NSW and KGraph, see Table \ref{table_fp}.
This result verifies that \textsc{Connect-SubGraphs} is useful and \textsc{Remove-Detours} is important to improve reachability, i.e., provide fewer false positives.
(Note that \textsc{Remove-Links} does not affect the number of false positives, since it does not improve reachability.)

\begin{figure*}[!t]
	\begin{center}
    \includegraphics[width=0.40\linewidth]{Figure/label.pdf}
    \vspace{-1.5mm}
    
    	\subfigure[Deep]{%
		\includegraphics[width=0.235\linewidth]{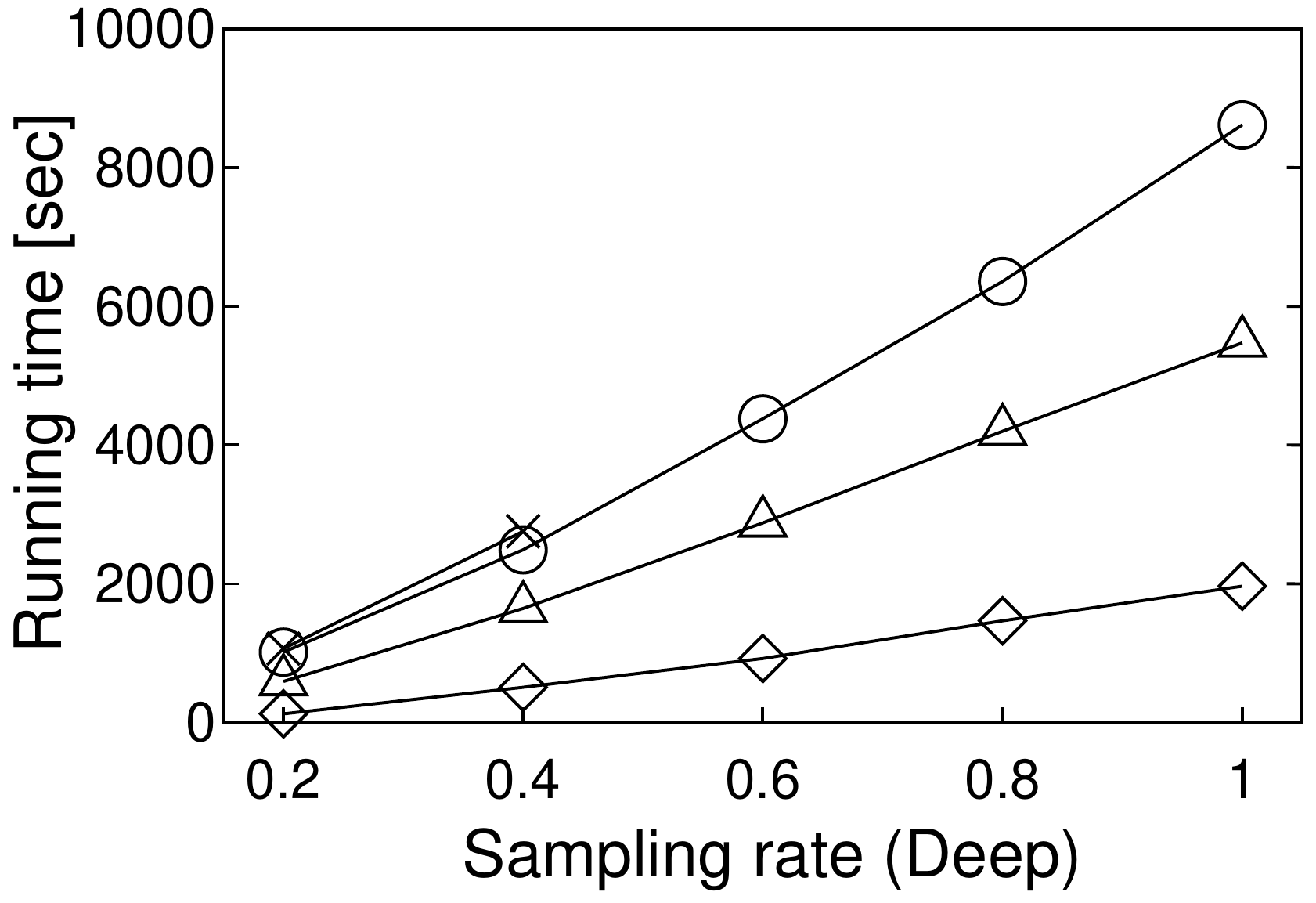}		\label{fig_deep_n}}
        \subfigure[Glove]{%
		\includegraphics[width=0.235\linewidth]{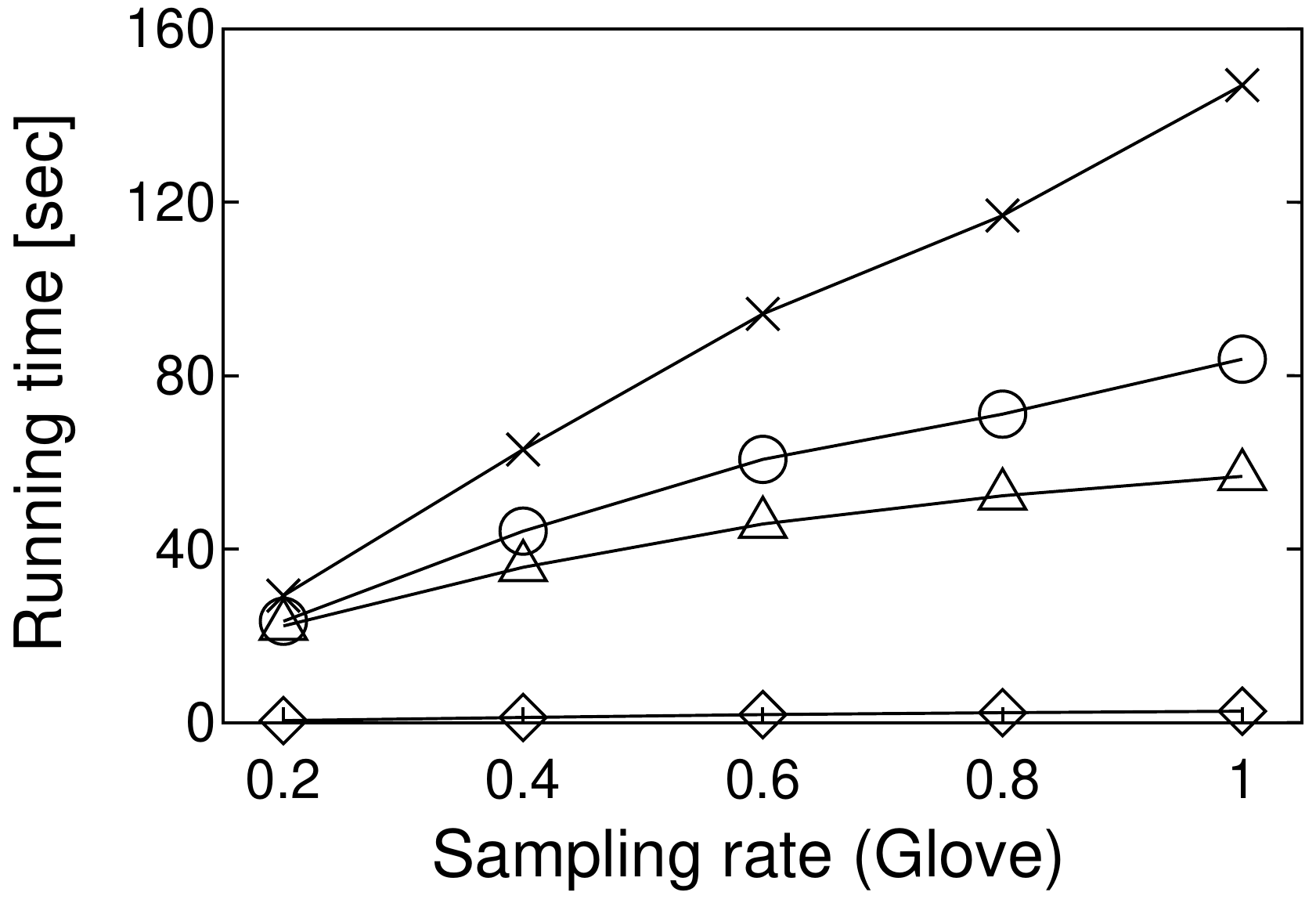}	\label{fig_glove_n}}
        \subfigure[HEPMASS]{%
		\includegraphics[width=0.235\linewidth]{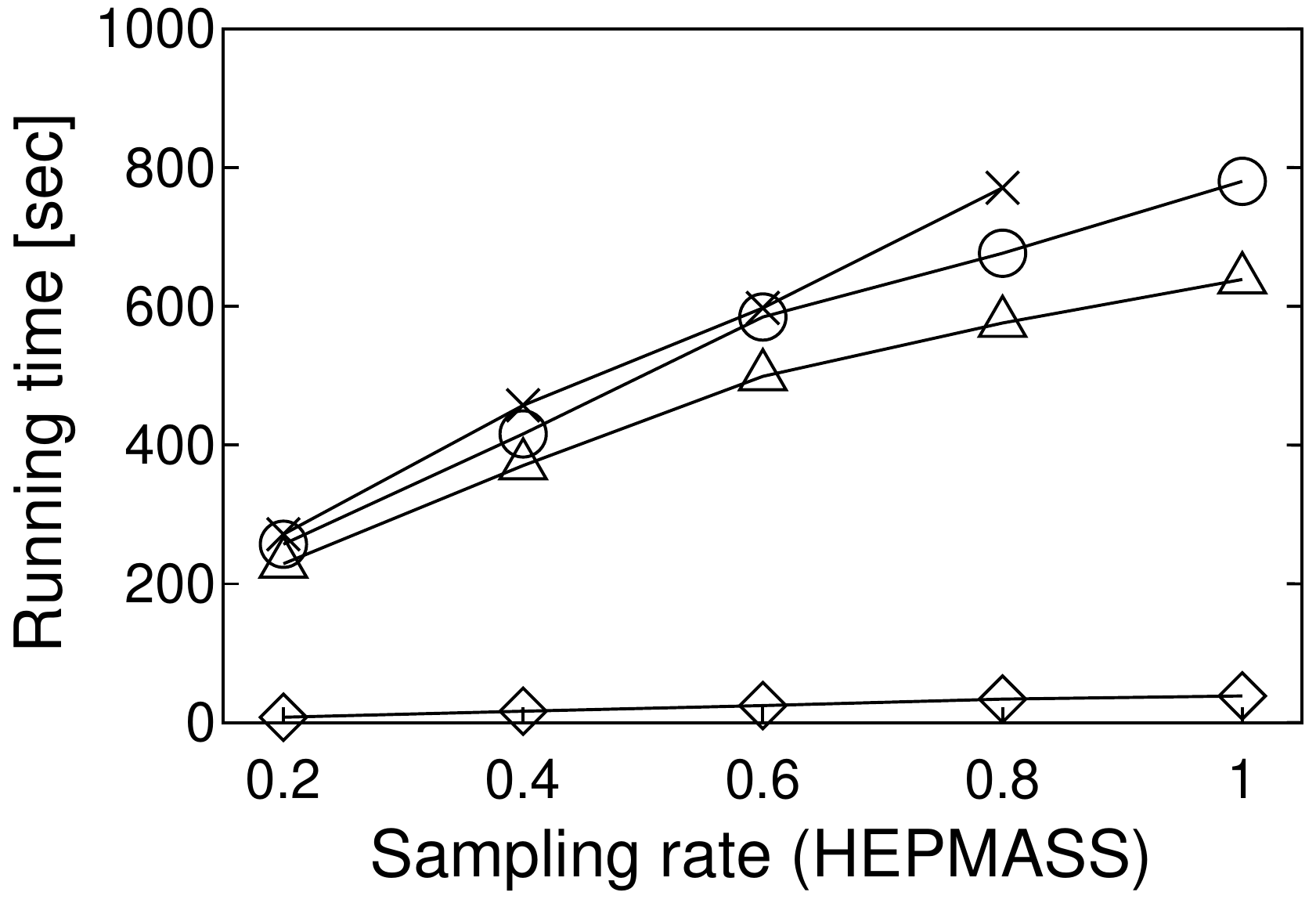}	\label{fig_hepmass_n}}
        \subfigure[MNIST]{%
		\includegraphics[width=0.235\linewidth]{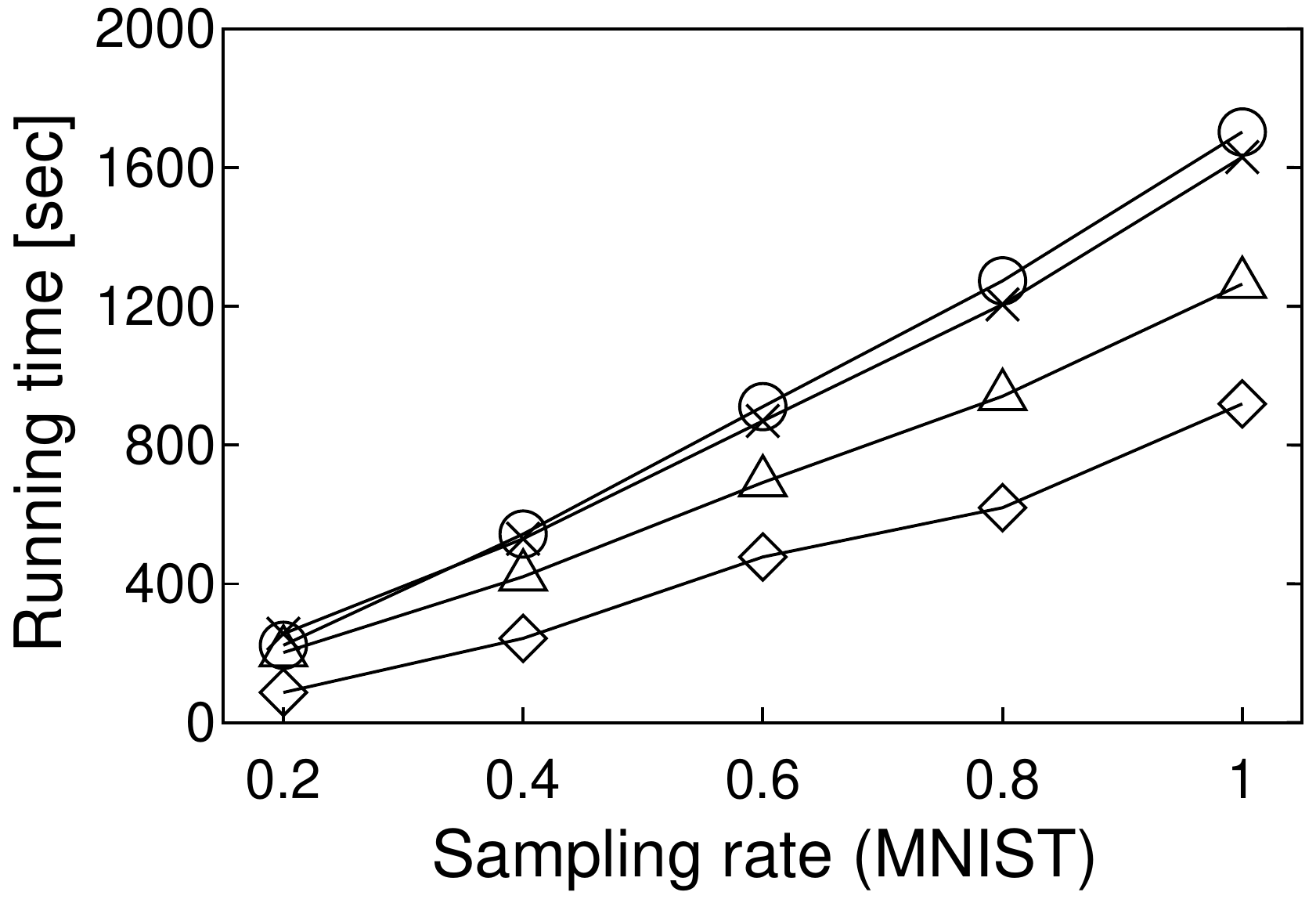}		\label{fig_mnist_n}}
        \subfigure[PAMAP2]{%
		\includegraphics[width=0.235\linewidth]{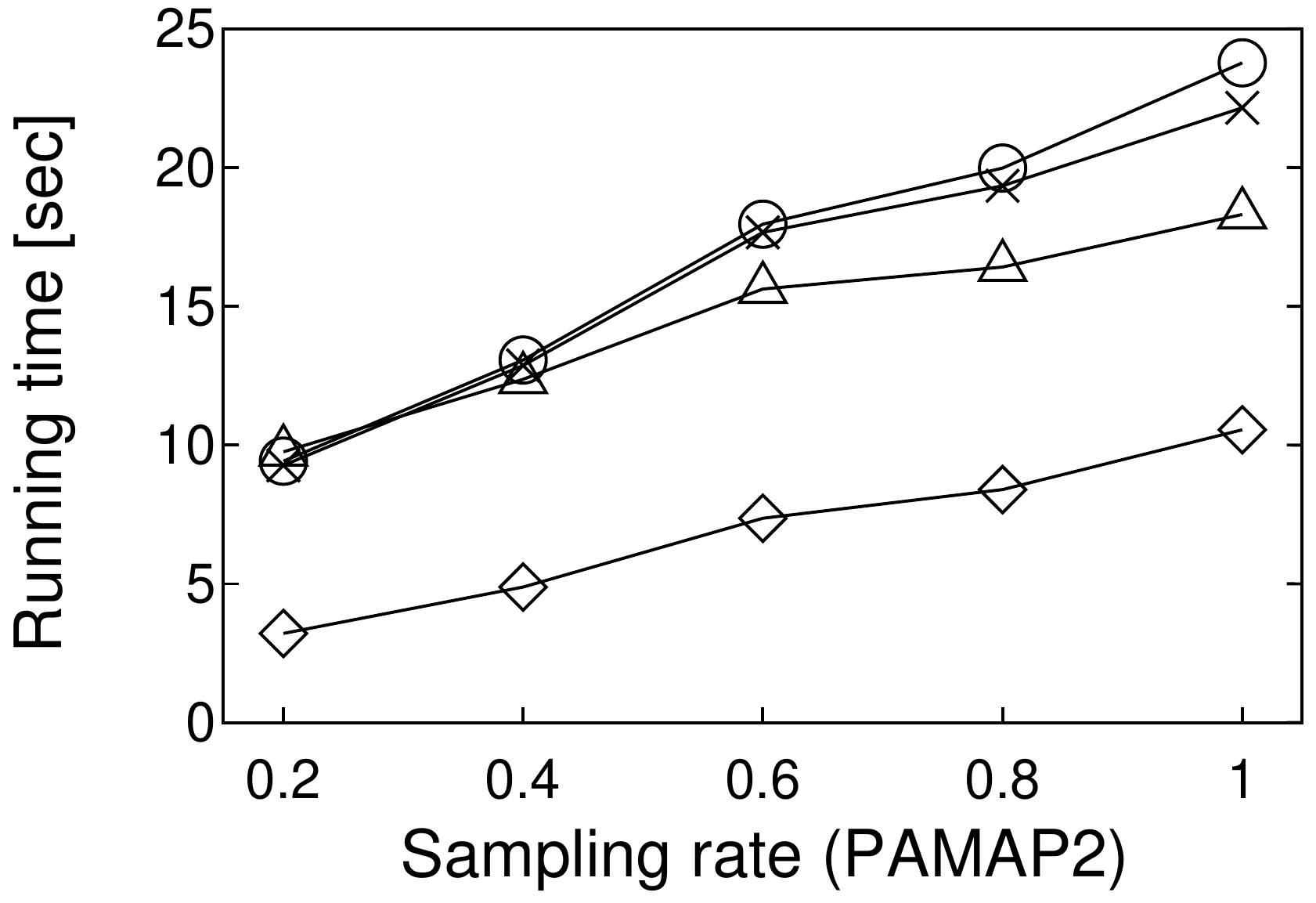}	\label{fig_pamap2_n}}
        \subfigure[SIFT]{%
		\includegraphics[width=0.235\linewidth]{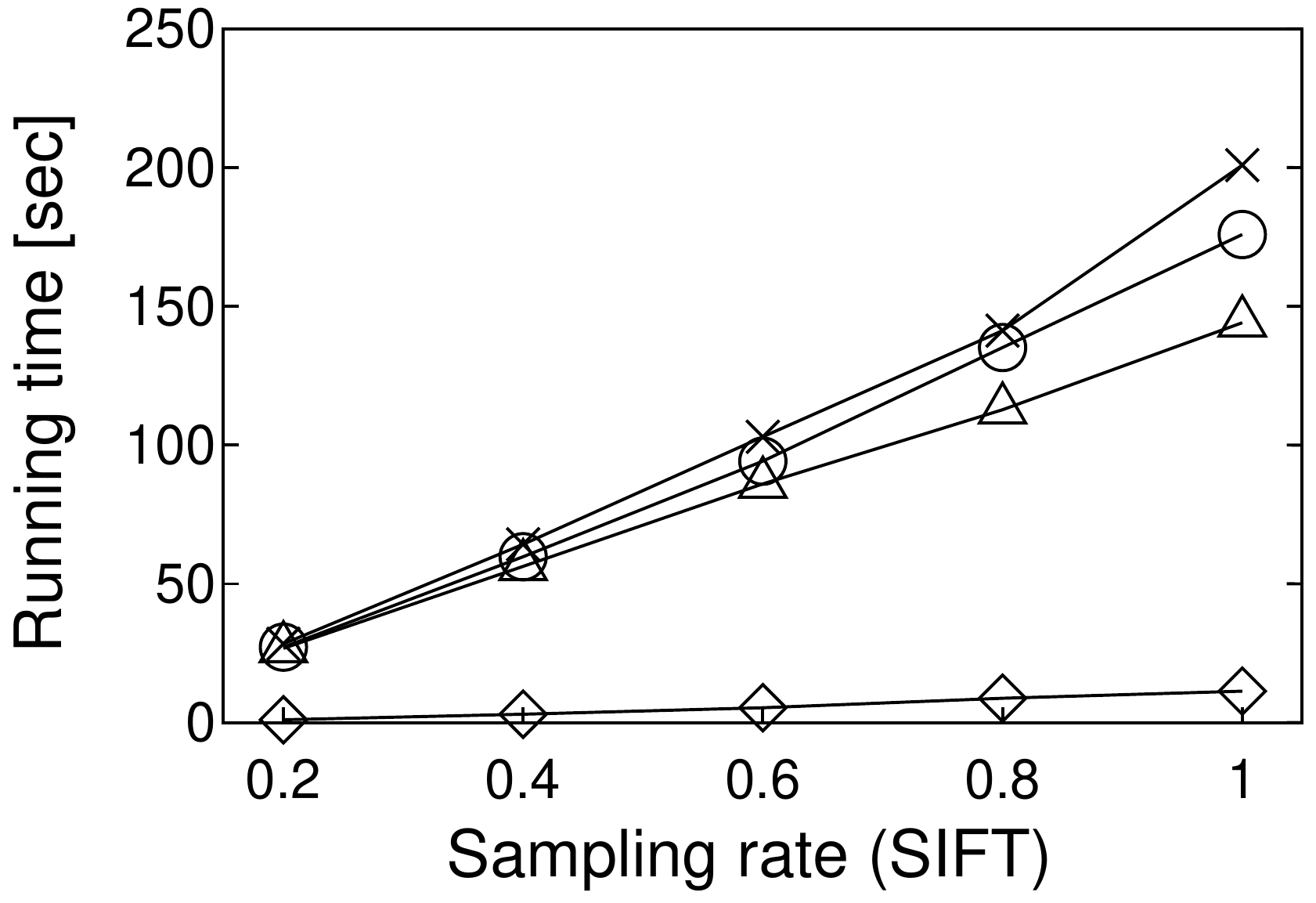}		\label{fig_sift_n}}
        \subfigure[Words]{%
		\includegraphics[width=0.235\linewidth]{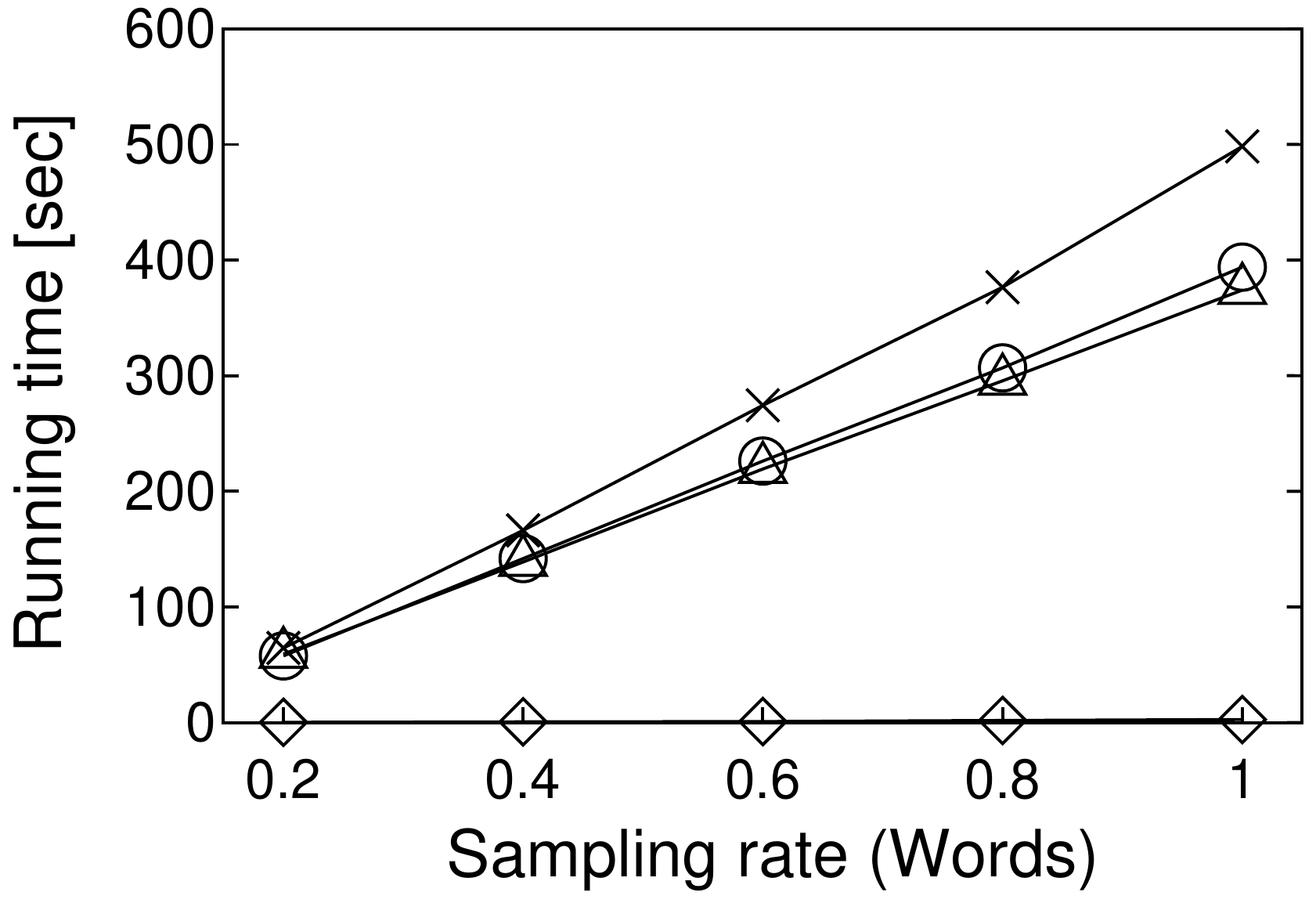}		\label{fig_words_n}}
        \vspace{-4.0mm}
        \caption{Impact of $n$}
        \label{figure_sample}
        \vspace{-2.0mm}
	\end{center}
\end{figure*}
\begin{figure*}[!t]
	\begin{center}
		\subfigure[Deep]{%
		\includegraphics[width=0.235\linewidth]{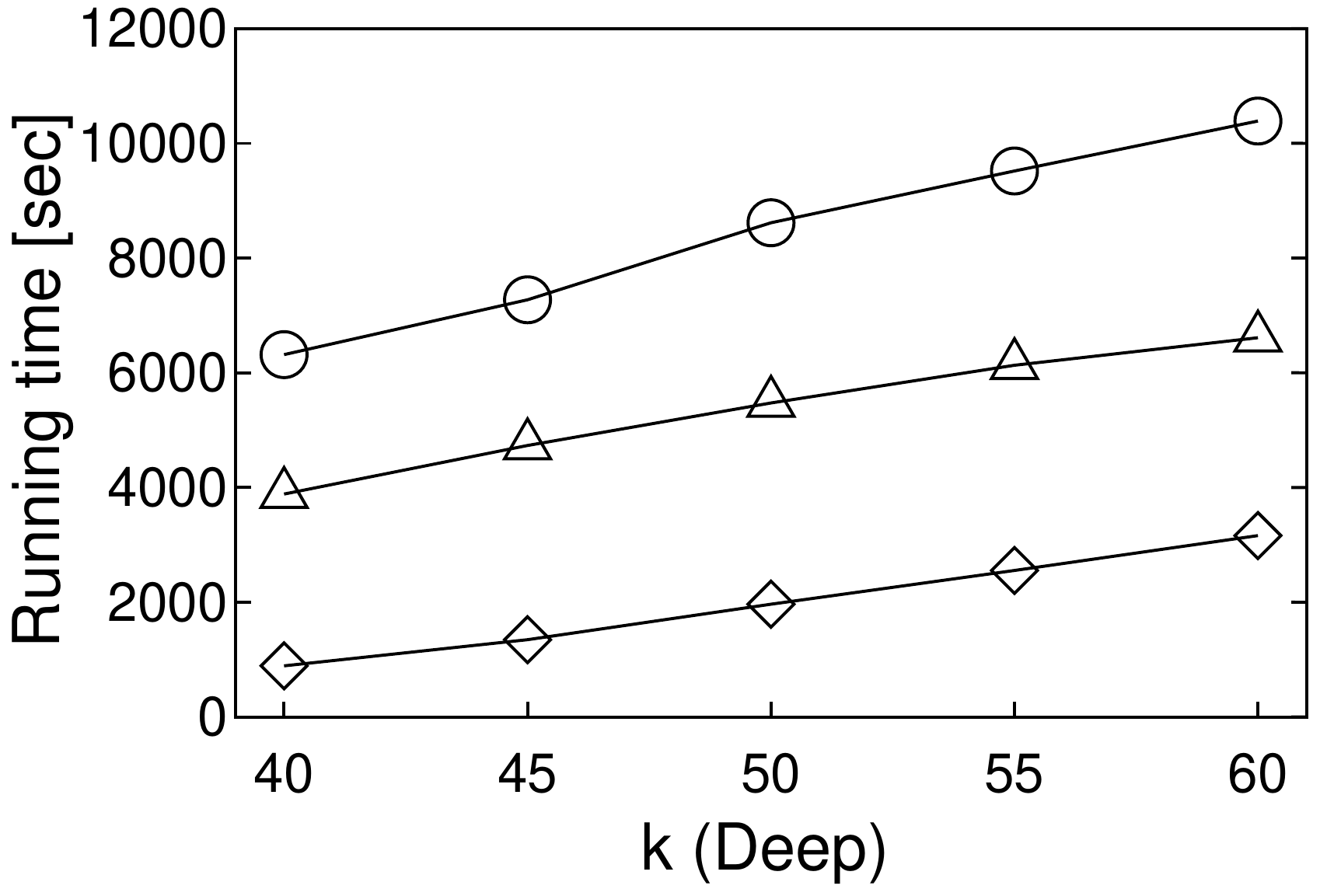}		\label{fig_deep_k}}
        \subfigure[Glove]{%
		\includegraphics[width=0.235\linewidth]{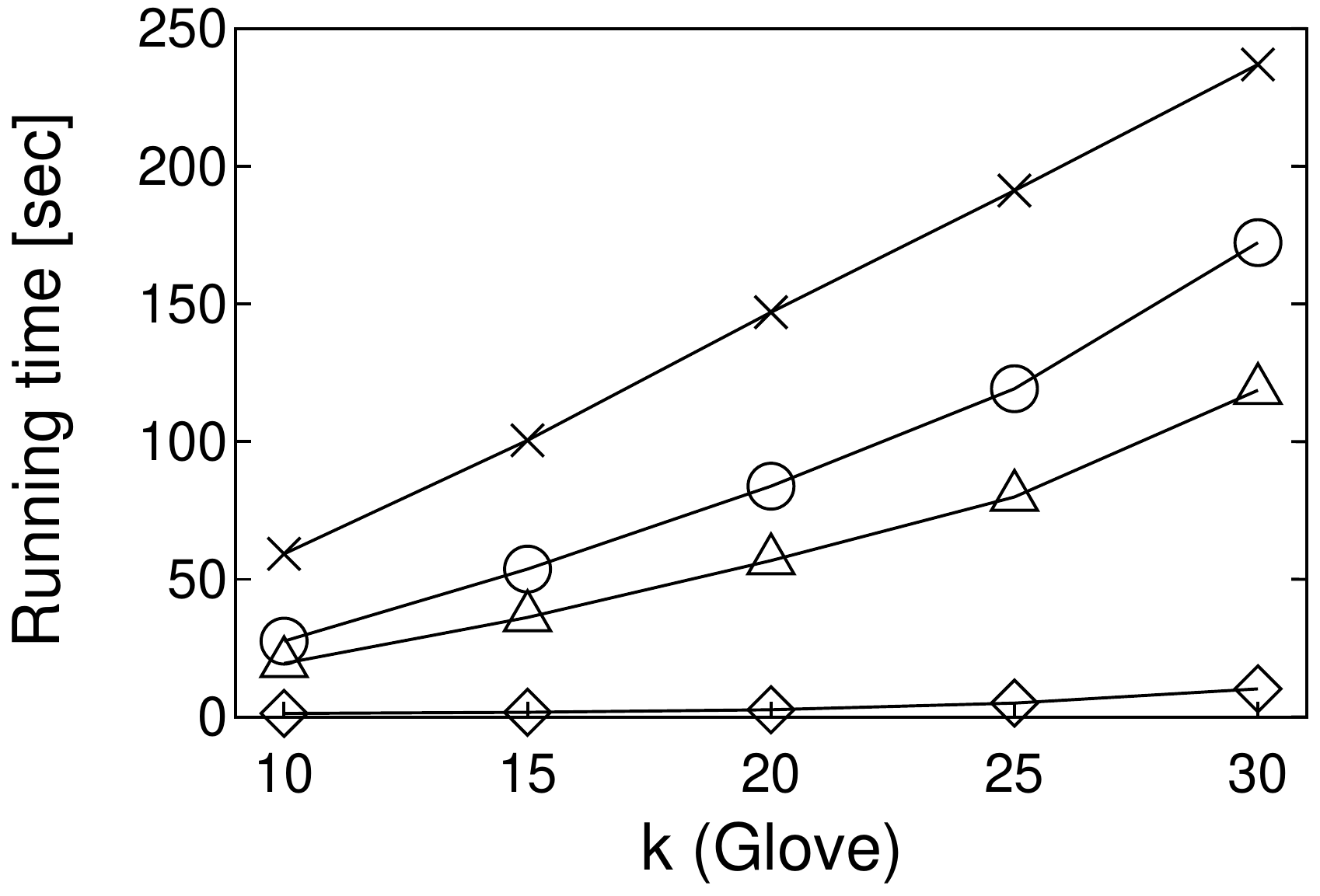}		\label{fig_glove_k}}
        \subfigure[HEPMASS]{%
		\includegraphics[width=0.235\linewidth]{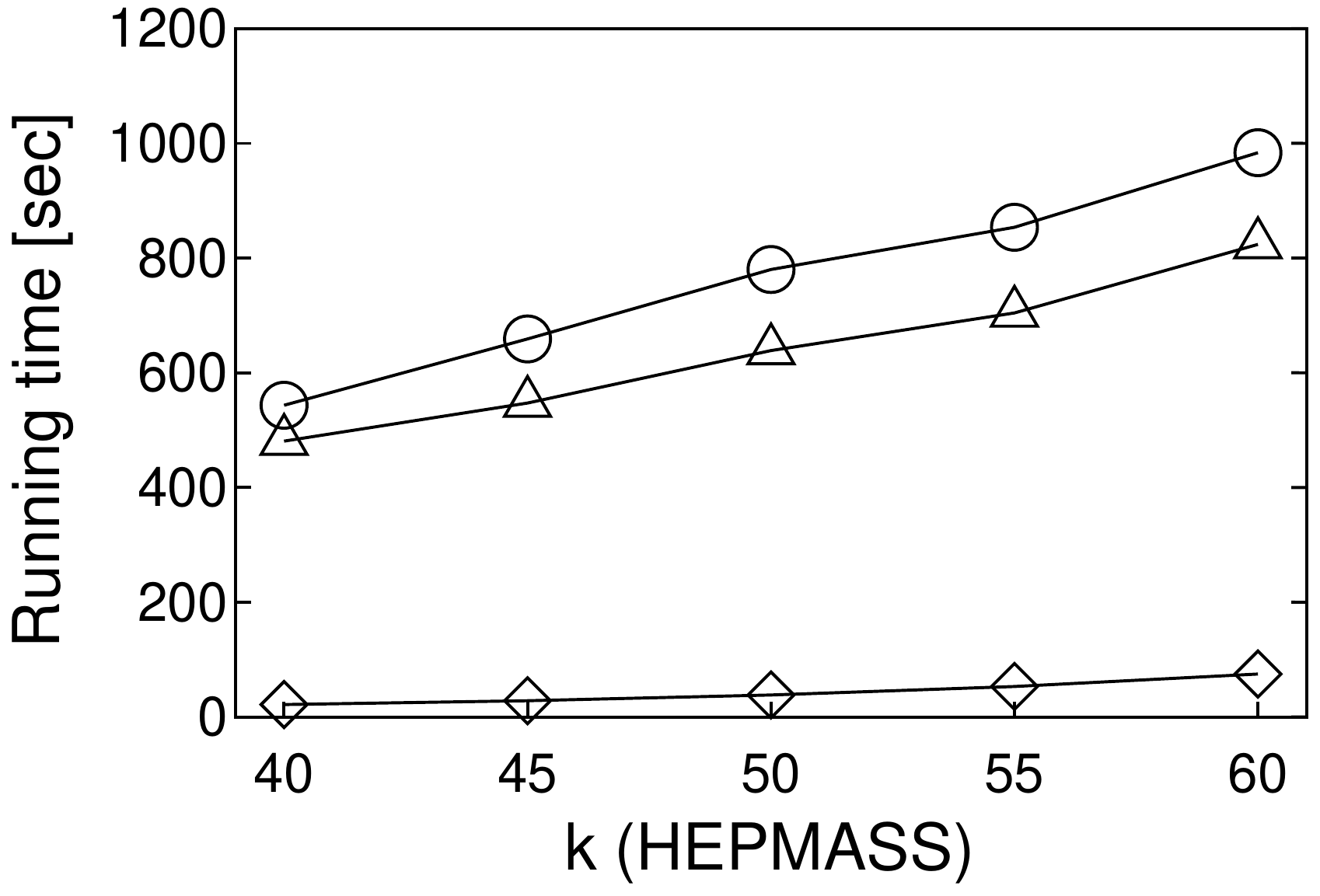}   \label{fig_hepmass_k}}
        \subfigure[MNIST]{%
		\includegraphics[width=0.235\linewidth]{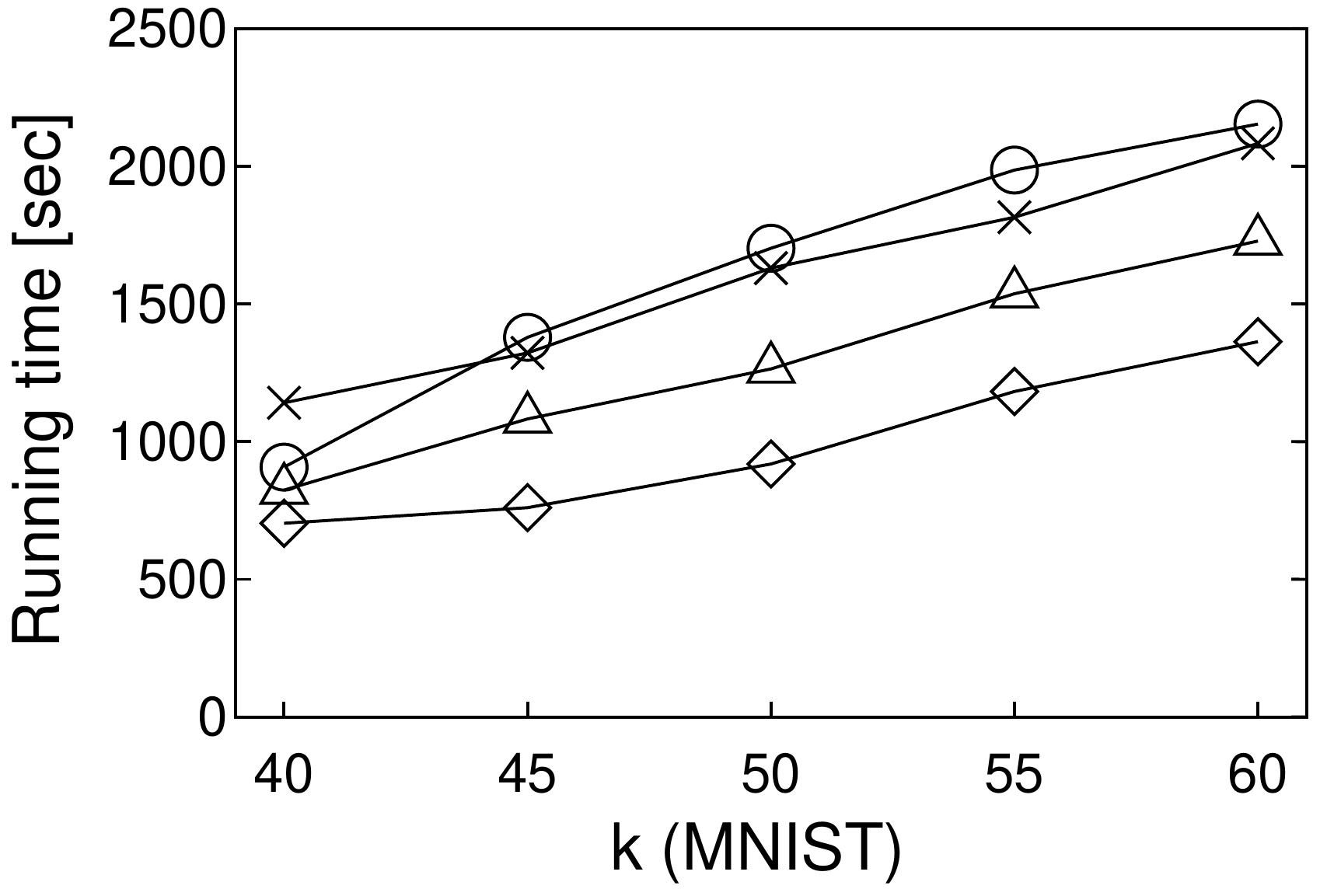}		\label{fig_mnist_k}}
        \subfigure[PAMAP2]{%
		\includegraphics[width=0.235\linewidth]{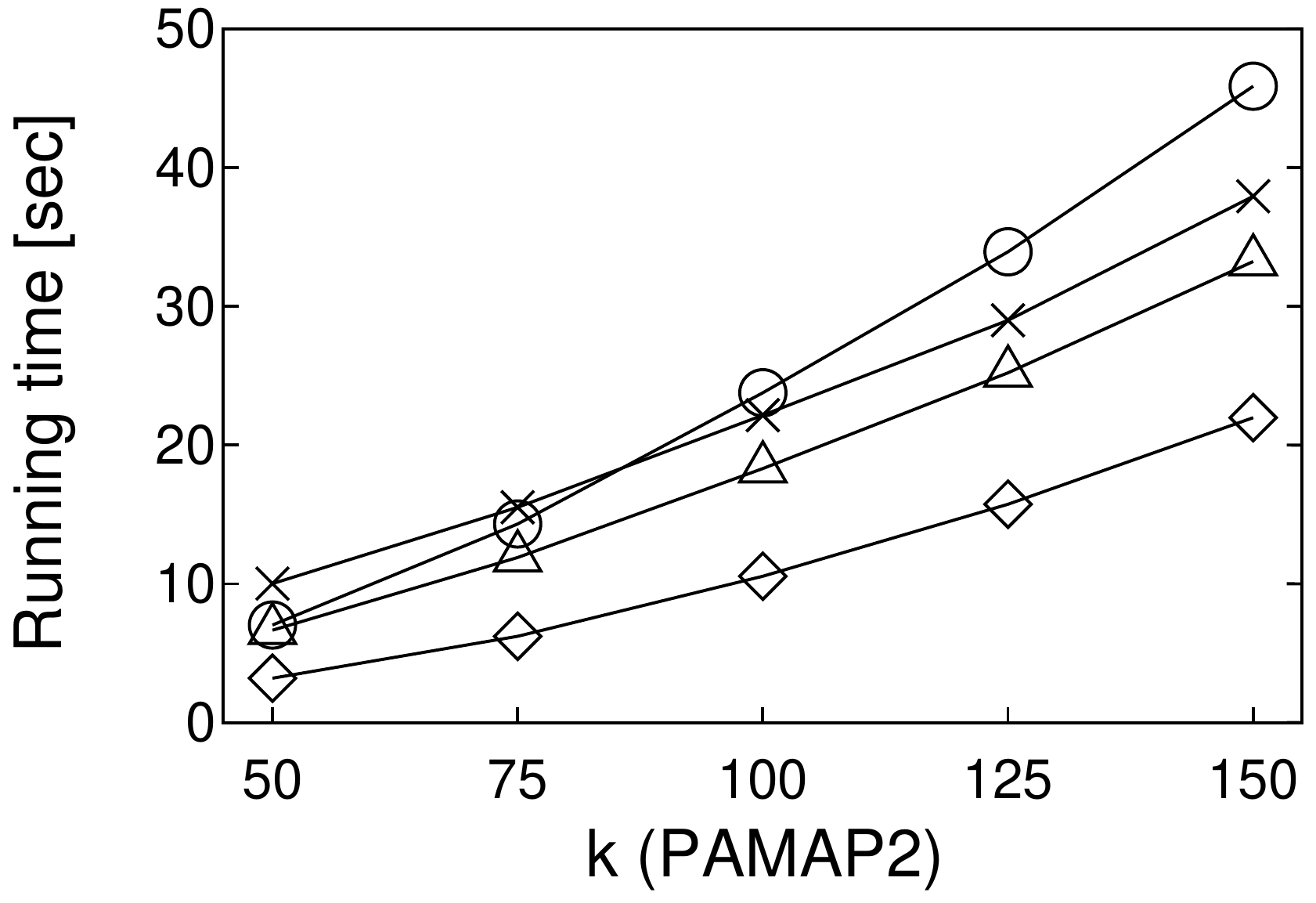}	\label{fig_pamap2_k}}
        \subfigure[SIFT]{%
		\includegraphics[width=0.235\linewidth]{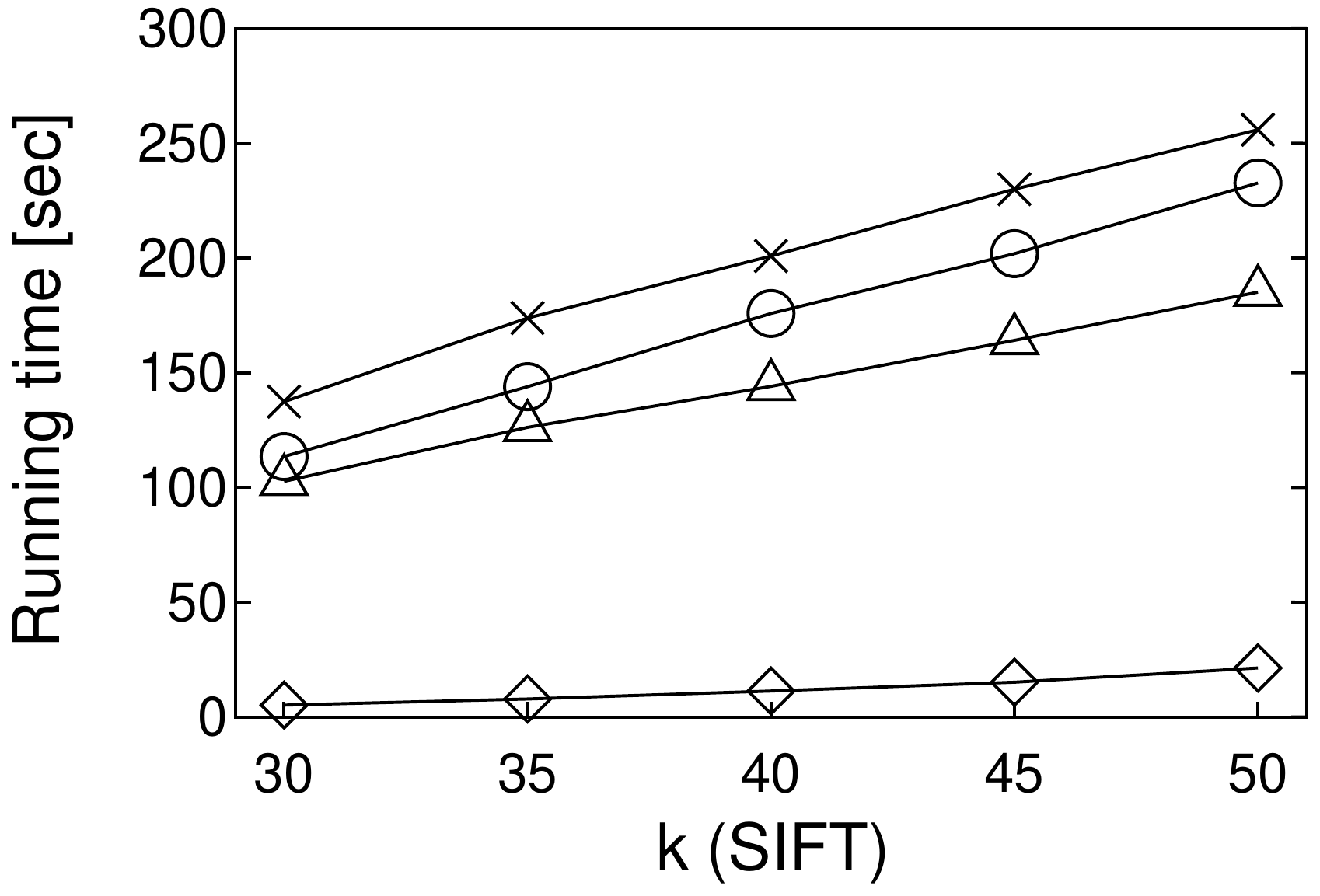}		\label{fig_sift_k}}
        \subfigure[Words]{%
		\includegraphics[width=0.235\linewidth]{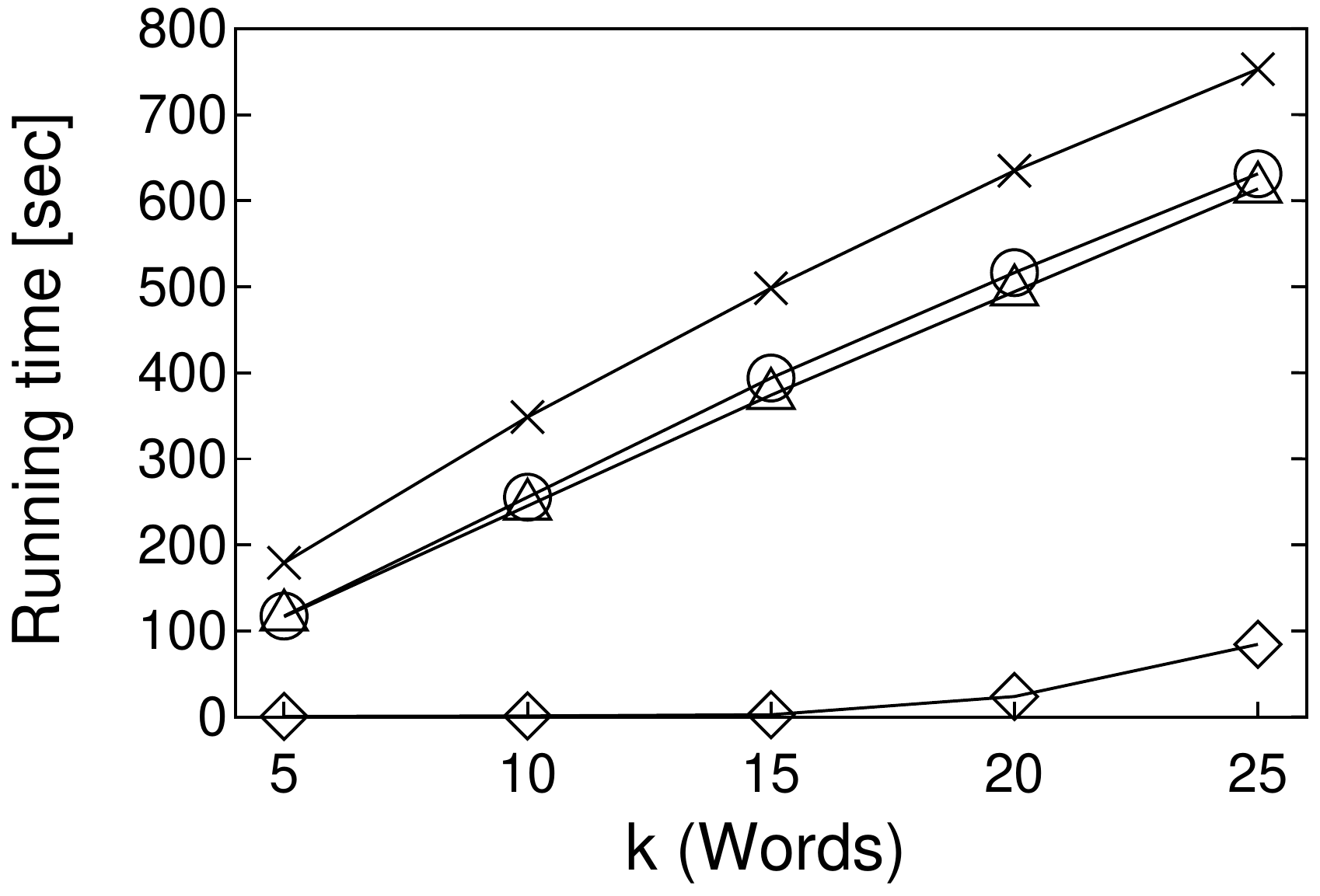}	\label{fig_words_k}}
        \vspace{-4.0mm}
        \caption{Impact of $k$}
        \label{figure_k}
        \vspace{-2.0mm}
	\end{center}
\end{figure*}

\vs
\noindent
\textbf{Varying $n$.}
Figure \ref{figure_sample} studies the scalability of each proximity graph in the same way as in Figure \ref{figure_sample_pre} (parameters were fixed as the default ones.).
Since Table \ref{table_time} confirms the superiority of our approach over the state-of-the-art, we omit the results of the state-of-the-art.

As the sampling rate increases, the running time of each proximity graph becomes longer.
This is reasonable, since both filtering and verification costs increase.
We have three observations.
The first one is that MRPG-basic keeps outperforming NSW and KGraph.
Second, MRPG significantly outperforms the other proximity graphs.
Last, MRPG and MRPG-basic scale better than the other proximity graphs, which confirms that pivot-based monotonic path creation provides their scalability.

In the case of Words, MRPG-basic and KGraph show similar performances.
We observed that outliers in Words have large dimensionality.
Because computing edit distance needs a quadratic cost to dimensionality, verification of outliers incurs a large computational cost.
For example, with the default parameters, MRPG-basic (KGraph) took 2.43 (0.73) and 371.65 (393.23) seconds for filtering and verification, respectively.
From the result in Table \ref{table_fp}, we see that false positives in Words are verified quickly (by early termination) and the verification of outliers dominates the most computational time.

\vs
\noindent
\textbf{Varying $k$.}
We investigate the influence of outlier ratio by varying $k$.
Figure \ref{figure_k} presents the results.
As $k$ increases, our approach needs to traverse more objects, rendering a larger filtering cost.
In addition, as $k$ increases, outlier ratio increases.
Our approach hence needs more verification cost when $k$ is large.

One difference between MRPG and the other proximity graphs is robustness to $k$, as MRPG(-basic) outperforms the other proximity graphs.
This is derived from \textsc{Connect-SubGraphs} and \textsc{Remove-Detours}, i.e., functions that make MRPG different from KGraph.
That is, the connectivity of the graph and the existence of monotonic paths (for similar objects) are important to exploit our algorithm.

\begin{figure*}[!t]
	\begin{center}
    \includegraphics[width=0.40\linewidth]{Figure/label.pdf}
    \vspace{-1.5mm}
    
		\subfigure[Deep]{%
		\includegraphics[width=0.235\linewidth]{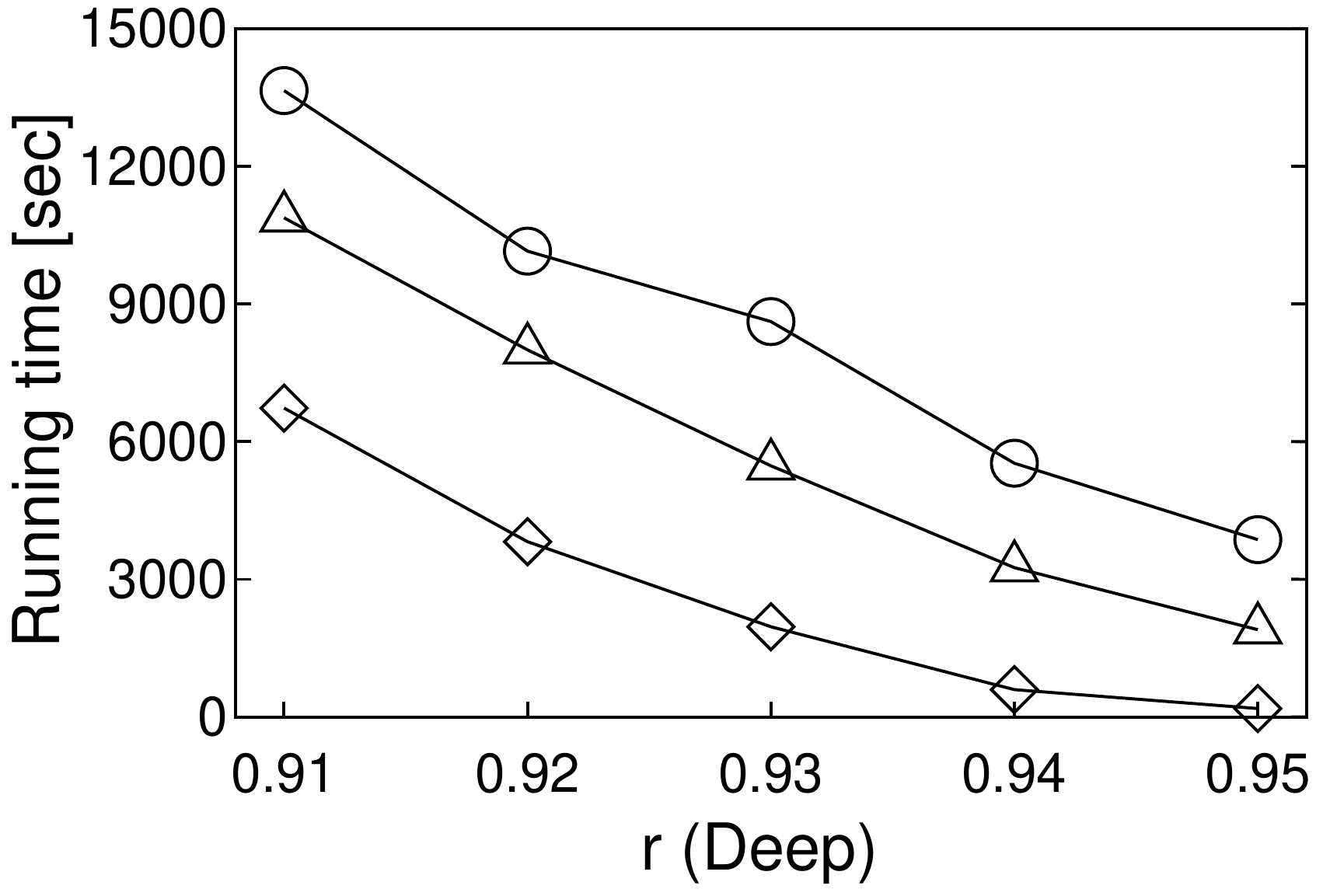}		\label{fig_deep_r}}
        \subfigure[Glove]{%
		\includegraphics[width=0.235\linewidth]{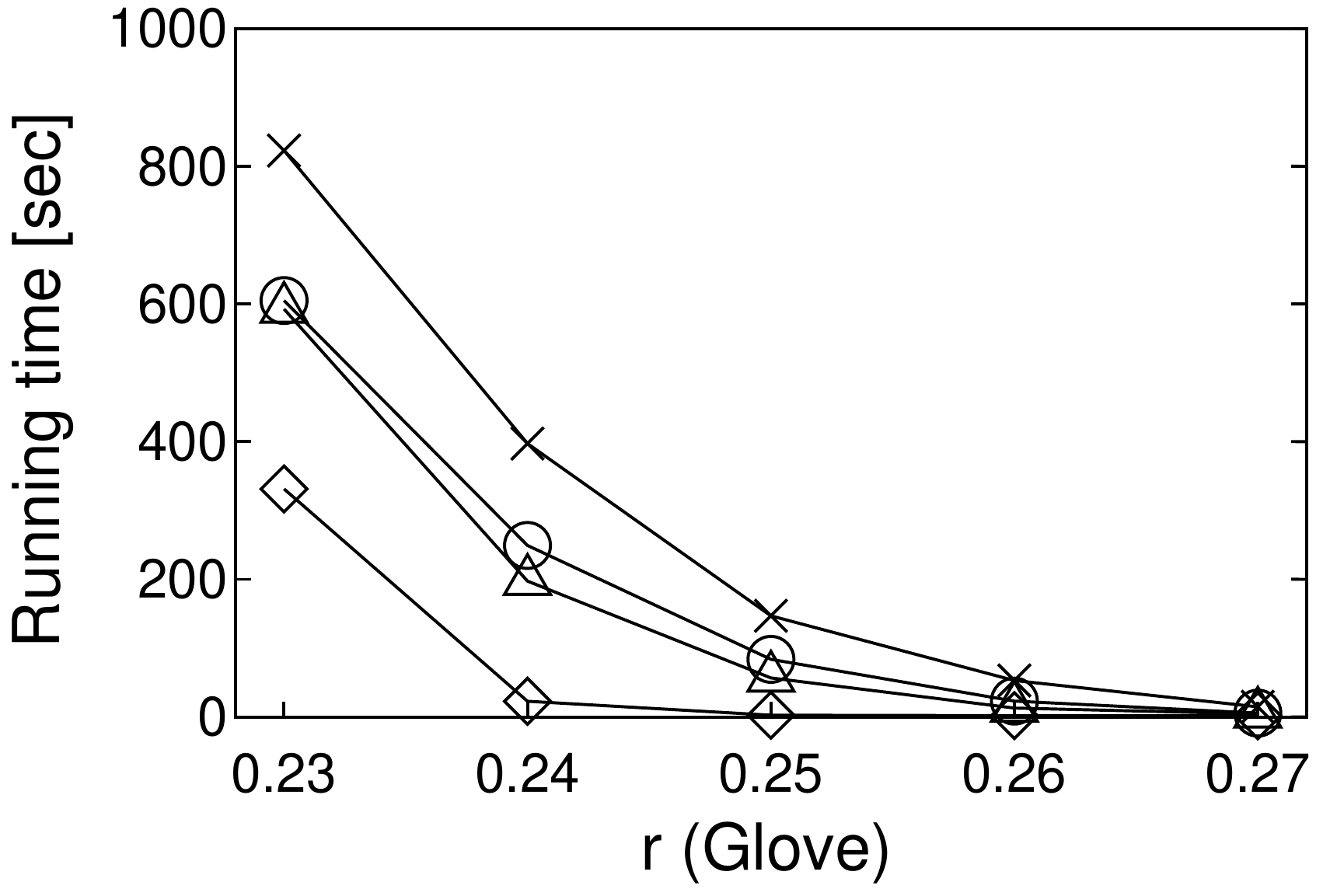}		\label{fig_glove_r}}
        \subfigure[HEPMASS]{%
		\includegraphics[width=0.235\linewidth]{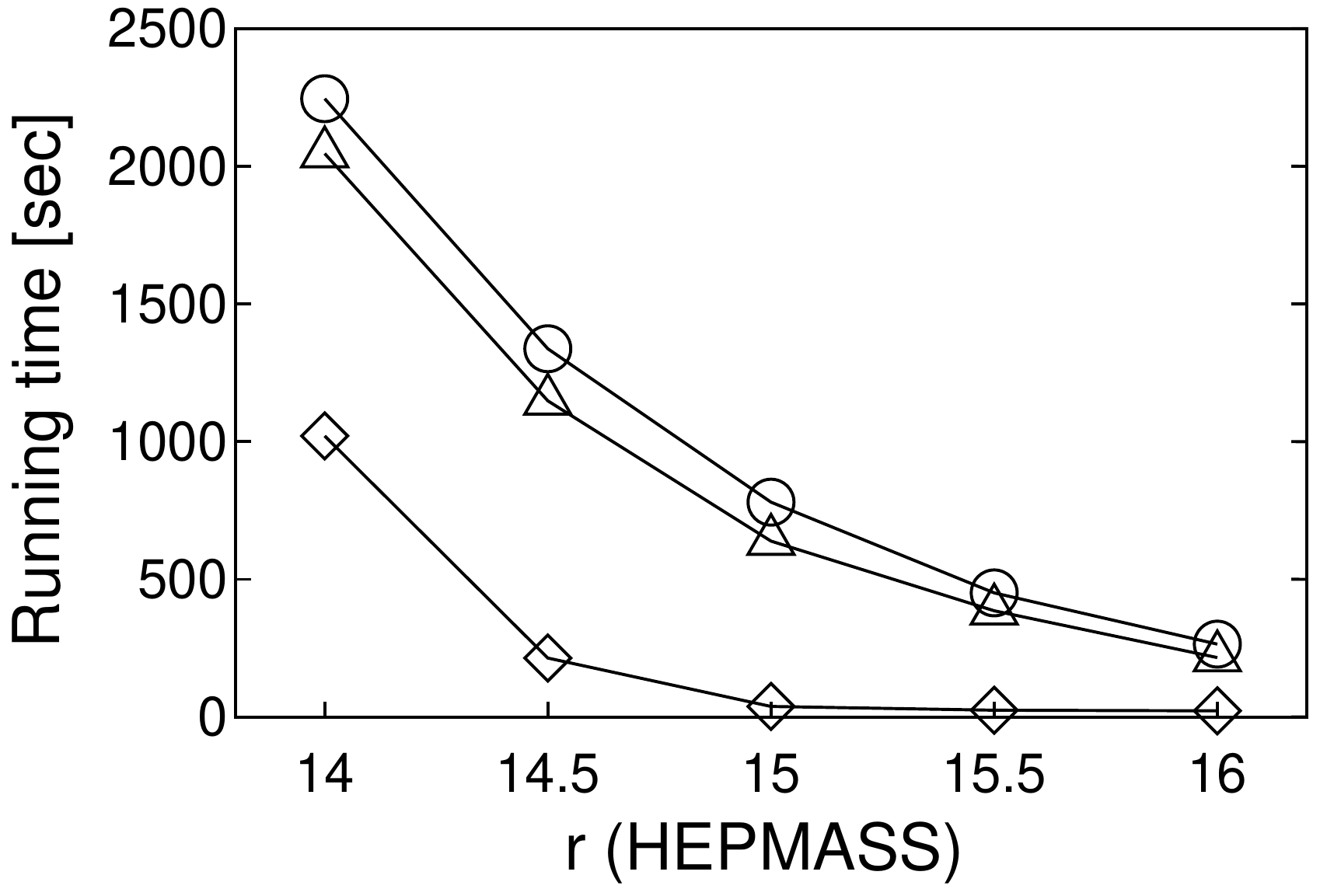}	\label{fig_hepmass_r}}
        \subfigure[MNIST]{%
		\includegraphics[width=0.235\linewidth]{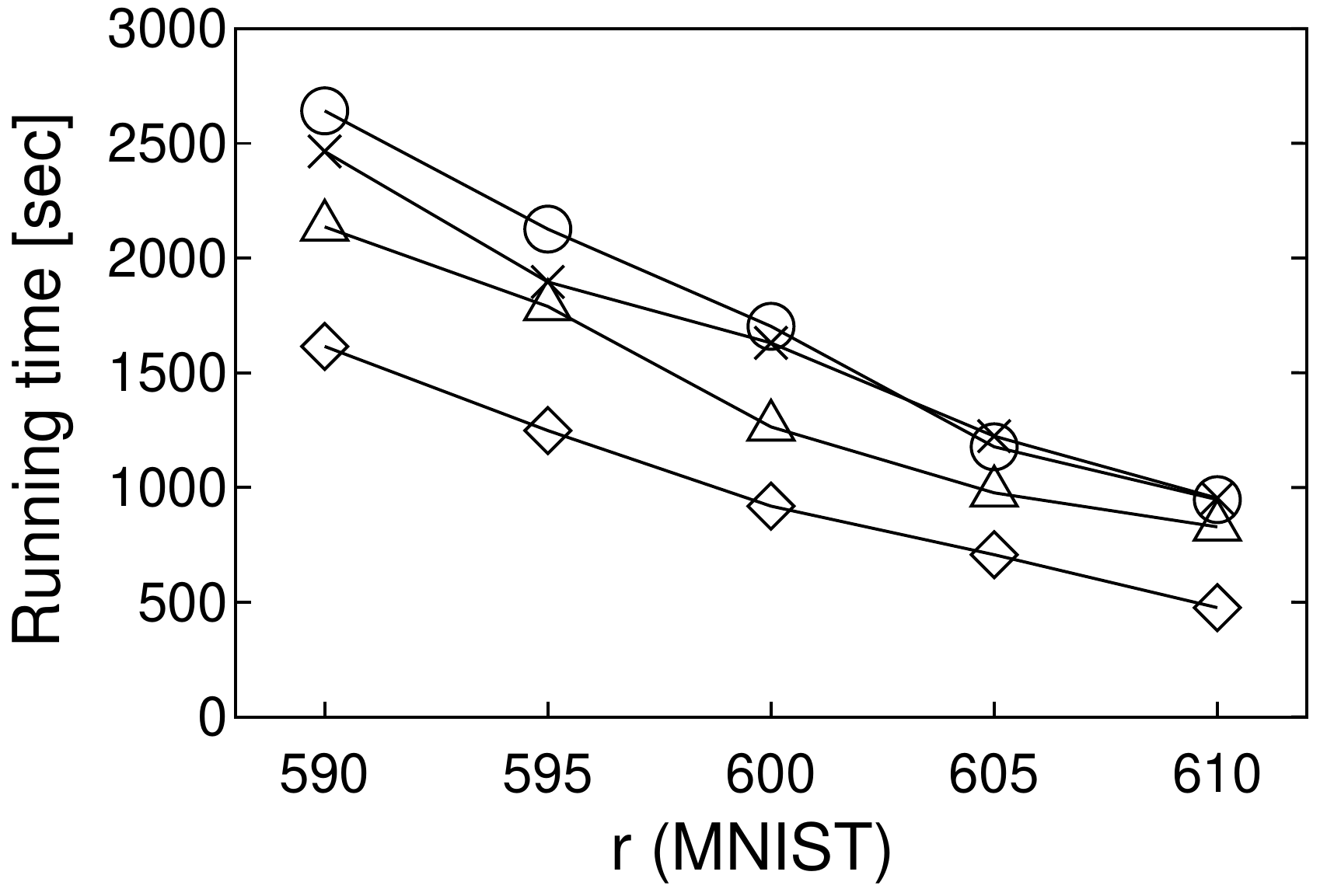}	    \label{fig_mnist_r}}
        \subfigure[PAMAP2]{%
		\includegraphics[width=0.235\linewidth]{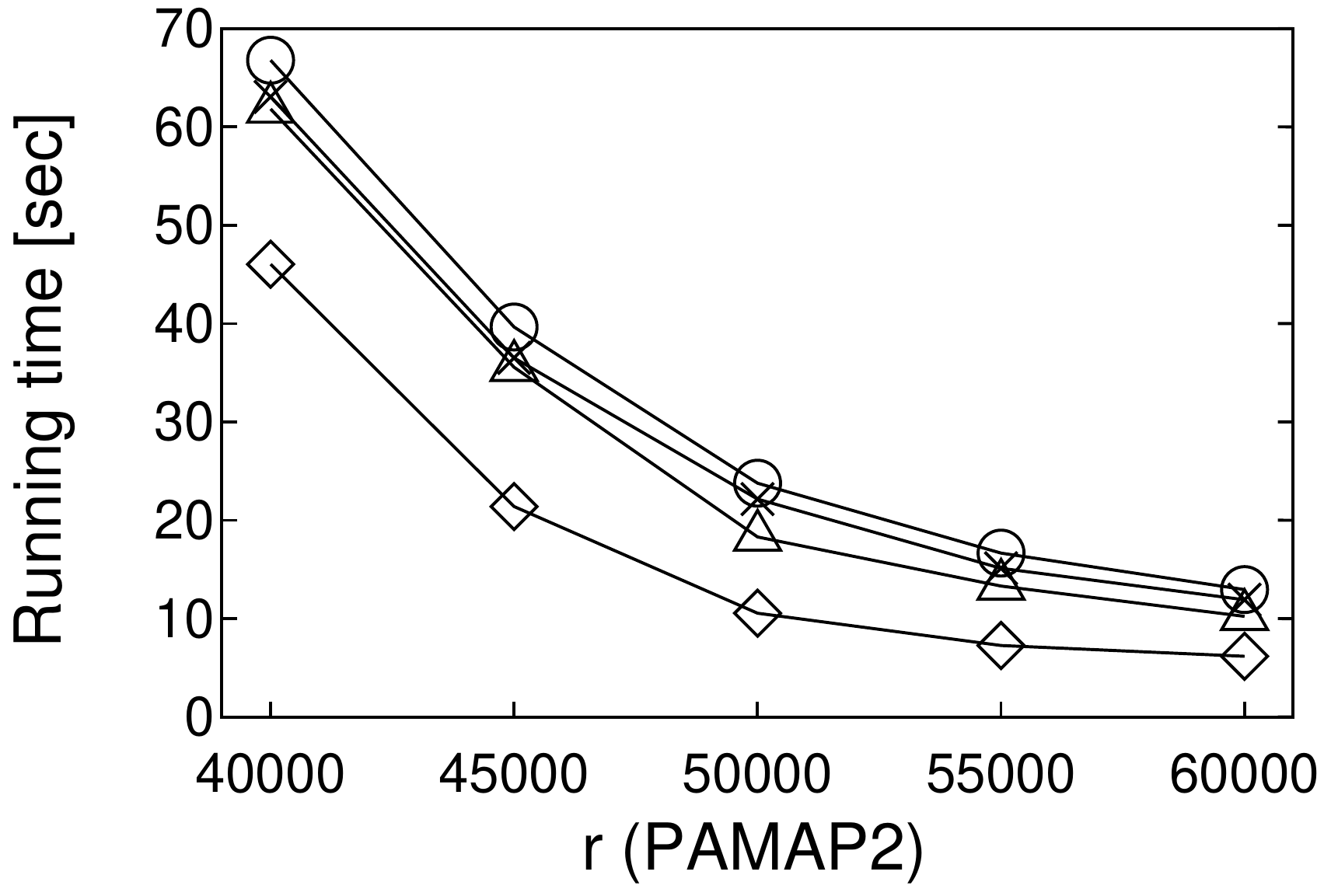}	\label{fig_pamap2_r}}
        \subfigure[SIFT]{%
		\includegraphics[width=0.235\linewidth]{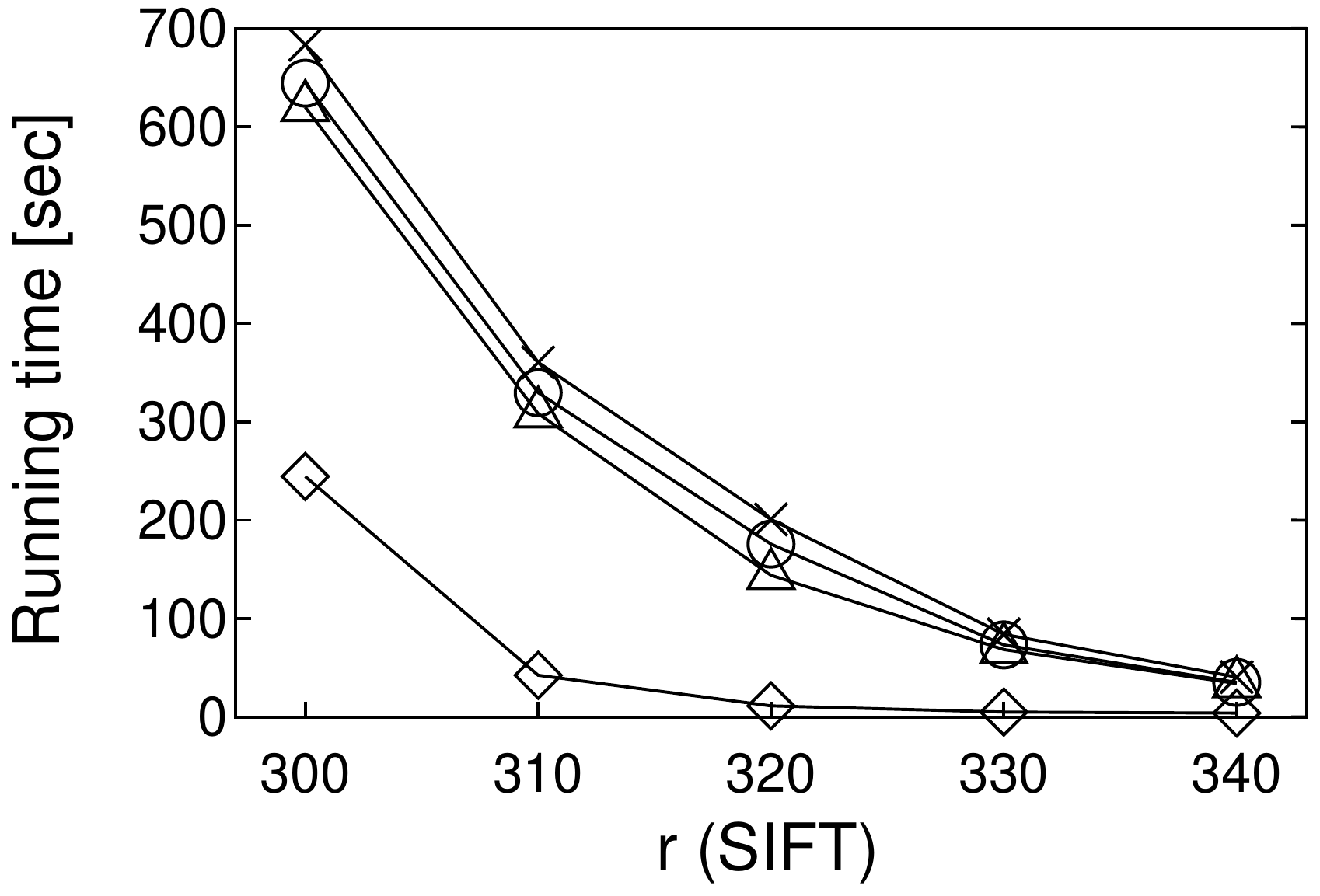}		\label{fig_sift_r}}
        \subfigure[Words]{%
		\includegraphics[width=0.235\linewidth]{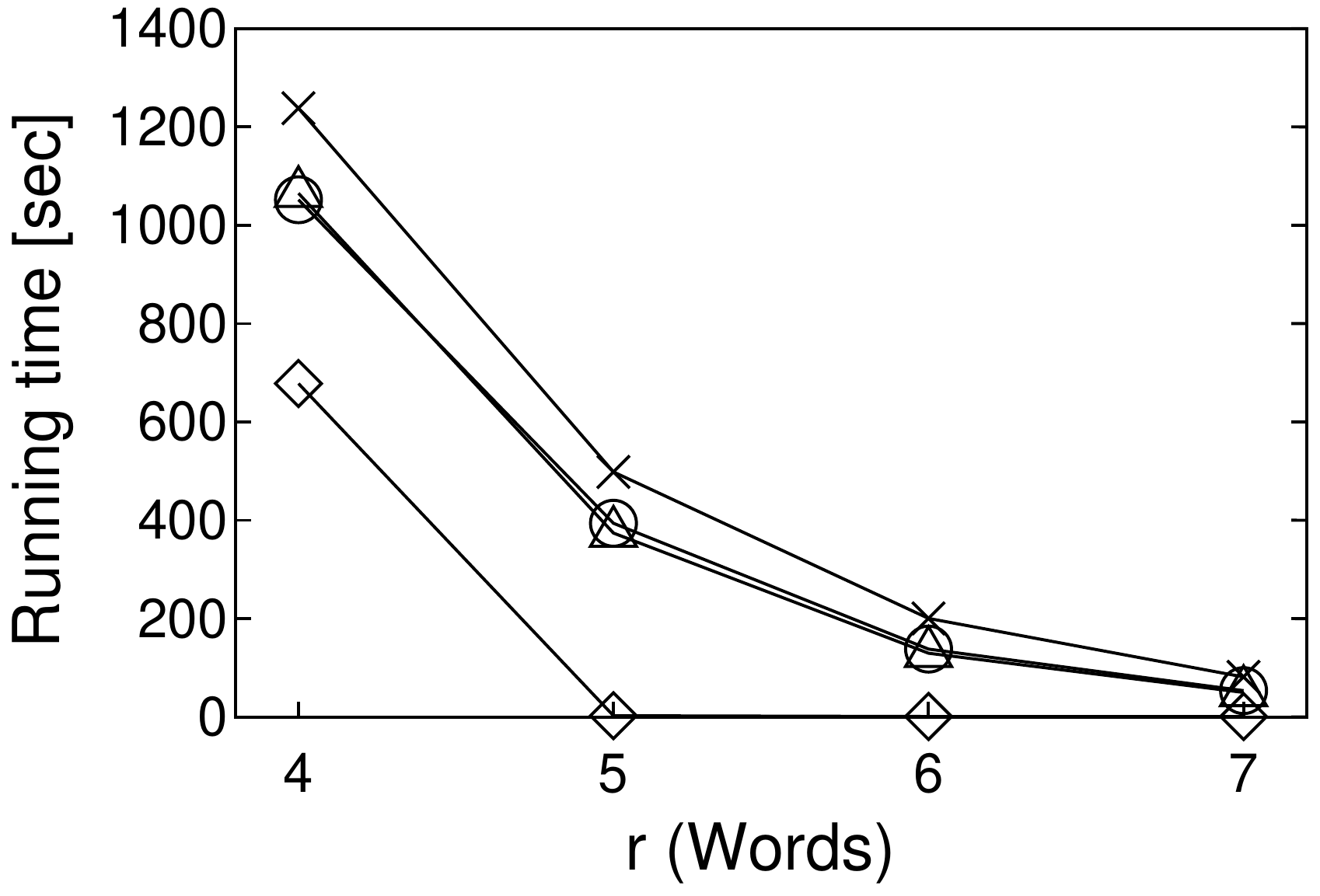}		\label{fig_words_r}}
        \vspace{-4.0mm}
        \caption{Impact of $r$}
        \label{figure_r}
        \vspace{-2.0mm}
	\end{center}
\end{figure*}

\vs
\noindent
\textbf{Varying $r$.}
The result of experiments with varying distance threshold $r$ is shown in Figure \ref{figure_r} ($k$ is fixed at the default value).
As $r$ increases, the outlier ratio decreases, and vice versa.
Similar to the results in Figure \ref{figure_k}, MRPG keeps outperforming KGraph and NSW both when outlier ratio is high and low.


\vs
\noindent
\textbf{Varying the number of threads.}
Last, we demonstrate that our approach is parallel-friendly.
Figure \ref{figure_t} shows the result on Glove, HEPMASS, PAMAP2, SIFT, and Words.
We see that our solution exploits the available threads and has linear scalability to the number of threads, for each proximity graph.
Also, the superiority among the proximity graphs does not change.

\section{Conclusion}	\label{section_conclusion}
In this paper, we addressed the problem of distance-based outlier detection in metric spaces and proposed a novel approach, namely proximity graph-based algorithm.
To exploit our DOD algorithm, we devised MRPG (Metric Randomized Proximity Graph), which improves reachability to neighbors and reduces the verification cost.
Our experiments on real datasets confirm that (i) our DOD algorithm is much faster than state-of-the-art and (ii) MRPG is superior to existing proximity graphs.

\section*{Acknowledgments}
This research is partially supported by JSPS Grant-in-Aid for Scientific Research (A) Grant Number 18H04095, JST CREST Grant Number J181401085, and JST PRESTO Grant Number JPMJPR1931.

\bibliographystyle{ACM-Reference-Format}
\bibliography{sigproc}


\begin{thebibliography}{38}


\ifx \showCODEN    \undefined \def \showCODEN     #1{\unskip}     \fi
\ifx \showDOI      \undefined \def \showDOI       #1{#1}\fi
\ifx \showISBNx    \undefined \def \showISBNx     #1{\unskip}     \fi
\ifx \showISBNxiii \undefined \def \showISBNxiii  #1{\unskip}     \fi
\ifx \showISSN     \undefined \def \showISSN      #1{\unskip}     \fi
\ifx \showLCCN     \undefined \def \showLCCN      #1{\unskip}     \fi
\ifx \shownote     \undefined \def \shownote      #1{#1}          \fi
\ifx \showarticletitle \undefined \def \showarticletitle #1{#1}   \fi
\ifx \showURL      \undefined \def \showURL       {\relax}        \fi
\providecommand\bibfield[2]{#2}
\providecommand\bibinfo[2]{#2}
\providecommand\natexlab[1]{#1}
\providecommand\showeprint[2][]{arXiv:#2}

\bibitem[\protect\citeauthoryear{Aggarwal}{Aggarwal}{2015}]%
        {aggarwal2015outlier}
\bibfield{author}{\bibinfo{person}{Charu~C Aggarwal}.}
  \bibinfo{year}{2015}\natexlab{}.
\newblock \showarticletitle{Outlier Analysis}. In
  \bibinfo{booktitle}{\emph{Data Mining}}. \bibinfo{pages}{237--263}.
\newblock


\bibitem[\protect\citeauthoryear{Amagata and Hara}{Amagata and Hara}{2021}]%
        {amagata2021dpc}
\bibfield{author}{\bibinfo{person}{Daichi Amagata} {and}
  \bibinfo{person}{Takahiro Hara}.} \bibinfo{year}{2021}\natexlab{}.
\newblock \showarticletitle{Fast Density-Peaks Clustering: Multicore-based
  Parallelization Approach}. In \bibinfo{booktitle}{\emph{SIGMOD}}.
  \bibinfo{pages}{49--61}.
\newblock


\bibitem[\protect\citeauthoryear{Amagata, Onizuka, and Hara}{Amagata
  et~al\mbox{.}}{2021}]%
        {amagata2021dod}
\bibfield{author}{\bibinfo{person}{Daichi Amagata}, \bibinfo{person}{Makoto
  Onizuka}, {and} \bibinfo{person}{Takahiro Hara}.}
  \bibinfo{year}{2021}\natexlab{}.
\newblock \showarticletitle{Fast and Exact Outlier Detection in Metric Spaces:
  A Proximity Graph-based Approach}. In \bibinfo{booktitle}{\emph{SIGMOD}}.
  \bibinfo{pages}{36--48}.
\newblock


\bibitem[\protect\citeauthoryear{Angiulli and Fassetti}{Angiulli and
  Fassetti}{2009}]%
        {angiulli2009dolphin}
\bibfield{author}{\bibinfo{person}{Fabrizio Angiulli} {and}
  \bibinfo{person}{Fabio Fassetti}.} \bibinfo{year}{2009}\natexlab{}.
\newblock \showarticletitle{Dolphin: An Efficient Algorithm for Mining
  Distance-based Outliers in Very Large Datasets}.
\newblock \bibinfo{journal}{\emph{ACM Transactions on Knowledge and Data
  Discovery}} \bibinfo{volume}{3}, \bibinfo{number}{1} (\bibinfo{year}{2009}),
  \bibinfo{pages}{4}.
\newblock


\bibitem[\protect\citeauthoryear{Arya and Mount}{Arya and Mount}{1993}]%
        {arya1993approximate}
\bibfield{author}{\bibinfo{person}{Sunil Arya} {and} \bibinfo{person}{David~M
  Mount}.} \bibinfo{year}{1993}\natexlab{}.
\newblock \showarticletitle{Approximate Nearest Neighbor Queries in Fixed
  Dimensions.}. In \bibinfo{booktitle}{\emph{SODA}}, Vol.~\bibinfo{volume}{93}.
  \bibinfo{pages}{271--280}.
\newblock


\bibitem[\protect\citeauthoryear{Babenko and Lempitsky}{Babenko and
  Lempitsky}{2016}]%
        {babenko2016efficient}
\bibfield{author}{\bibinfo{person}{Artem Babenko} {and} \bibinfo{person}{Victor
  Lempitsky}.} \bibinfo{year}{2016}\natexlab{}.
\newblock \showarticletitle{Efficient Indexing of Billion-scale Datasets of
  Deep Descriptors}. In \bibinfo{booktitle}{\emph{CVPR}}.
  \bibinfo{pages}{2055--2063}.
\newblock


\bibitem[\protect\citeauthoryear{Batista, Prati, and Monard}{Batista
  et~al\mbox{.}}{2004}]%
        {batista2004study}
\bibfield{author}{\bibinfo{person}{Gustavo~EAPA Batista},
  \bibinfo{person}{Ronaldo~C Prati}, {and} \bibinfo{person}{Maria~Carolina
  Monard}.} \bibinfo{year}{2004}\natexlab{}.
\newblock \showarticletitle{A Study of the Behavior of Several Methods for
  Balancing Machine Learning Training Data}.
\newblock \bibinfo{journal}{\emph{SIGKDD Explorations Newsletter}}
  \bibinfo{volume}{6}, \bibinfo{number}{1} (\bibinfo{year}{2004}),
  \bibinfo{pages}{20--29}.
\newblock


\bibitem[\protect\citeauthoryear{Bay and Schwabacher}{Bay and
  Schwabacher}{2003}]%
        {bay2003mining}
\bibfield{author}{\bibinfo{person}{Stephen~D Bay} {and} \bibinfo{person}{Mark
  Schwabacher}.} \bibinfo{year}{2003}\natexlab{}.
\newblock \showarticletitle{Mining Distance-based Outliers in Near Linear Time
  with Randomization and a Simple Pruning Rule}. In
  \bibinfo{booktitle}{\emph{KDD}}. \bibinfo{pages}{29--38}.
\newblock


\bibitem[\protect\citeauthoryear{Boguna, Krioukov, and Claffy}{Boguna
  et~al\mbox{.}}{2009}]%
        {boguna2009navigability}
\bibfield{author}{\bibinfo{person}{Marian Boguna}, \bibinfo{person}{Dmitri
  Krioukov}, {and} \bibinfo{person}{Kimberly~C Claffy}.}
  \bibinfo{year}{2009}\natexlab{}.
\newblock \showarticletitle{Navigability of Complex Networks}.
\newblock \bibinfo{journal}{\emph{Nature Physics}} \bibinfo{volume}{5},
  \bibinfo{number}{1} (\bibinfo{year}{2009}), \bibinfo{pages}{74}.
\newblock


\bibitem[\protect\citeauthoryear{Brin}{Brin}{1995}]%
        {brin1995near}
\bibfield{author}{\bibinfo{person}{Sergey Brin}.}
  \bibinfo{year}{1995}\natexlab{}.
\newblock \showarticletitle{Near Neighbor Search in Large Metric Spaces}. In
  \bibinfo{booktitle}{\emph{VLDB}}. \bibinfo{pages}{574--584}.
\newblock


\bibitem[\protect\citeauthoryear{Campos, Zimek, Sander, Campello,
  Micenkov{\'a}, Schubert, Assent, and Houle}{Campos et~al\mbox{.}}{2016}]%
        {campos2016evaluation}
\bibfield{author}{\bibinfo{person}{Guilherme~O Campos}, \bibinfo{person}{Arthur
  Zimek}, \bibinfo{person}{J{\"o}rg Sander}, \bibinfo{person}{Ricardo~JGB
  Campello}, \bibinfo{person}{Barbora Micenkov{\'a}}, \bibinfo{person}{Erich
  Schubert}, \bibinfo{person}{Ira Assent}, {and} \bibinfo{person}{Michael~E
  Houle}.} \bibinfo{year}{2016}\natexlab{}.
\newblock \showarticletitle{On the Evaluation of Unsupervised Outlier
  Detection: Measures, Datasets, and an Empirical Study}.
\newblock \bibinfo{journal}{\emph{Data Mining and Knowledge Discovery}}
  \bibinfo{volume}{30}, \bibinfo{number}{4} (\bibinfo{year}{2016}),
  \bibinfo{pages}{891--927}.
\newblock


\bibitem[\protect\citeauthoryear{Cao, Wang, and Rundensteiner}{Cao
  et~al\mbox{.}}{2016}]%
        {cao2016sharing}
\bibfield{author}{\bibinfo{person}{Lei Cao}, \bibinfo{person}{Jiayuan Wang},
  {and} \bibinfo{person}{Elke~A Rundensteiner}.}
  \bibinfo{year}{2016}\natexlab{}.
\newblock \showarticletitle{Sharing-aware Outlier Analytics over High-volume
  Data Streams}. In \bibinfo{booktitle}{\emph{SIGMOD}}.
  \bibinfo{pages}{527--540}.
\newblock


\bibitem[\protect\citeauthoryear{Chen, Gao, Zheng, Jensen, Yang, and Yang}{Chen
  et~al\mbox{.}}{2017}]%
        {chen2017pivot}
\bibfield{author}{\bibinfo{person}{Lu Chen}, \bibinfo{person}{Yunjun Gao},
  \bibinfo{person}{Baihua Zheng}, \bibinfo{person}{Christian~S Jensen},
  \bibinfo{person}{Hanyu Yang}, {and} \bibinfo{person}{Keyu Yang}.}
  \bibinfo{year}{2017}\natexlab{}.
\newblock \showarticletitle{Pivot-based Metric Indexing}.
\newblock \bibinfo{journal}{\emph{PVLDB}} \bibinfo{volume}{10},
  \bibinfo{number}{10} (\bibinfo{year}{2017}), \bibinfo{pages}{1058--1069}.
\newblock


\bibitem[\protect\citeauthoryear{Dearholt, Gonzales, and Kurup}{Dearholt
  et~al\mbox{.}}{1988}]%
        {dearholt1988monotonic}
\bibfield{author}{\bibinfo{person}{DW Dearholt}, \bibinfo{person}{N Gonzales},
  {and} \bibinfo{person}{G Kurup}.} \bibinfo{year}{1988}\natexlab{}.
\newblock \showarticletitle{Monotonic Search Networks for Computer Vision
  Databases}. In \bibinfo{booktitle}{\emph{ACSSC}}, Vol.~\bibinfo{volume}{2}.
  \bibinfo{pages}{548--553}.
\newblock


\bibitem[\protect\citeauthoryear{Dong, Moses, and Li}{Dong
  et~al\mbox{.}}{2011}]%
        {dong2011efficient}
\bibfield{author}{\bibinfo{person}{Wei Dong}, \bibinfo{person}{Charikar Moses},
  {and} \bibinfo{person}{Kai Li}.} \bibinfo{year}{2011}\natexlab{}.
\newblock \showarticletitle{Efficient k-nearest Neighbor Graph Construction for
  Generic Similarity Measures}. In \bibinfo{booktitle}{\emph{WWW}}.
  \bibinfo{pages}{577--586}.
\newblock


\bibitem[\protect\citeauthoryear{Ester, Kriegel, Sander, Xu,
  et~al\mbox{.}}{Ester et~al\mbox{.}}{1996}]%
        {ester1996density}
\bibfield{author}{\bibinfo{person}{Martin Ester}, \bibinfo{person}{Hans-Peter
  Kriegel}, \bibinfo{person}{J{\"o}rg Sander}, \bibinfo{person}{Xiaowei Xu},
  {et~al\mbox{.}}} \bibinfo{year}{1996}\natexlab{}.
\newblock \showarticletitle{A density-based algorithm for discovering clusters
  in large spatial databases with noise.}. In \bibinfo{booktitle}{\emph{KDD}}.
  \bibinfo{pages}{226--231}.
\newblock


\bibitem[\protect\citeauthoryear{Fu, Xiang, Wang, and Cai}{Fu
  et~al\mbox{.}}{2019}]%
        {fu2019fast}
\bibfield{author}{\bibinfo{person}{Cong Fu}, \bibinfo{person}{Chao Xiang},
  \bibinfo{person}{Changxu Wang}, {and} \bibinfo{person}{Deng Cai}.}
  \bibinfo{year}{2019}\natexlab{}.
\newblock \showarticletitle{Fast Approximate Nearest Neighbor Search with the
  Navigating Spreading-out Graph}.
\newblock \bibinfo{journal}{\emph{PVLDB}} \bibinfo{volume}{12},
  \bibinfo{number}{5} (\bibinfo{year}{2019}), \bibinfo{pages}{461--474}.
\newblock


\bibitem[\protect\citeauthoryear{Harwood and Drummond}{Harwood and
  Drummond}{2016}]%
        {harwood2016fanng}
\bibfield{author}{\bibinfo{person}{Ben Harwood} {and} \bibinfo{person}{Tom
  Drummond}.} \bibinfo{year}{2016}\natexlab{}.
\newblock \showarticletitle{FANNG: Fast Approximate Nearest Neighbour Graphs}.
  In \bibinfo{booktitle}{\emph{CVPR}}. \bibinfo{pages}{5713--5722}.
\newblock


\bibitem[\protect\citeauthoryear{Hodge and Austin}{Hodge and Austin}{2004}]%
        {hodge2004survey}
\bibfield{author}{\bibinfo{person}{Victoria Hodge} {and} \bibinfo{person}{Jim
  Austin}.} \bibinfo{year}{2004}\natexlab{}.
\newblock \showarticletitle{A Survey of Outlier Detection Methodologies}.
\newblock \bibinfo{journal}{\emph{Artificial Intelligence Review}}
  \bibinfo{volume}{22}, \bibinfo{number}{2} (\bibinfo{year}{2004}),
  \bibinfo{pages}{85--126}.
\newblock


\bibitem[\protect\citeauthoryear{Ilyas and Chu}{Ilyas and Chu}{2019}]%
        {ilyas2019data}
\bibfield{author}{\bibinfo{person}{Ihab~F Ilyas} {and} \bibinfo{person}{Xu
  Chu}.} \bibinfo{year}{2019}\natexlab{}.
\newblock \bibinfo{booktitle}{\emph{Data Cleaning}}.
\newblock


\bibitem[\protect\citeauthoryear{Knorr and Ng}{Knorr and Ng}{1998}]%
        {knorr1998algorithms}
\bibfield{author}{\bibinfo{person}{Edwin~M Knorr} {and}
  \bibinfo{person}{Raymond~T Ng}.} \bibinfo{year}{1998}\natexlab{}.
\newblock \showarticletitle{Algorithms for Mining Distance-based Outliers in
  Large Datasets}. In \bibinfo{booktitle}{\emph{VLDB}},
  Vol.~\bibinfo{volume}{98}. \bibinfo{pages}{392--403}.
\newblock


\bibitem[\protect\citeauthoryear{Kontaki, Gounaris, Papadopoulos, Tsichlas, and
  Manolopoulos}{Kontaki et~al\mbox{.}}{2011}]%
        {kontaki2011continuous}
\bibfield{author}{\bibinfo{person}{Maria Kontaki}, \bibinfo{person}{Anastasios
  Gounaris}, \bibinfo{person}{Apostolos~N Papadopoulos},
  \bibinfo{person}{Kostas Tsichlas}, {and} \bibinfo{person}{Yannis
  Manolopoulos}.} \bibinfo{year}{2011}\natexlab{}.
\newblock \showarticletitle{Continuous Monitoring of Distance-based Outliers
  over Data Streams}. In \bibinfo{booktitle}{\emph{ICDE}}.
  \bibinfo{pages}{135--146}.
\newblock


\bibitem[\protect\citeauthoryear{Larson, Mahendran, Lee, Kummerfeld, Hill,
  Laurenzano, Hauswald, Tang, and Mars}{Larson et~al\mbox{.}}{2019}]%
        {larson2019outlier}
\bibfield{author}{\bibinfo{person}{Stefan Larson}, \bibinfo{person}{Anish
  Mahendran}, \bibinfo{person}{Andrew Lee}, \bibinfo{person}{Jonathan~K
  Kummerfeld}, \bibinfo{person}{Parker Hill}, \bibinfo{person}{Michael~A
  Laurenzano}, \bibinfo{person}{Johann Hauswald}, \bibinfo{person}{Lingjia
  Tang}, {and} \bibinfo{person}{Jason Mars}.} \bibinfo{year}{2019}\natexlab{}.
\newblock \showarticletitle{Outlier Detection for Improved Data Quality and
  Diversity in Dialog Systems}. In \bibinfo{booktitle}{\emph{NAACL-HLT}}.
  \bibinfo{pages}{517--527}.
\newblock


\bibitem[\protect\citeauthoryear{Lerman and Maunu}{Lerman and Maunu}{2018}]%
        {lerman2018overview}
\bibfield{author}{\bibinfo{person}{Gilad Lerman} {and} \bibinfo{person}{Tyler
  Maunu}.} \bibinfo{year}{2018}\natexlab{}.
\newblock \showarticletitle{An Overview of Robust Subspace Recovery}.
\newblock \bibinfo{journal}{\emph{Proc. IEEE}} \bibinfo{volume}{106},
  \bibinfo{number}{8} (\bibinfo{year}{2018}), \bibinfo{pages}{1380--1410}.
\newblock


\bibitem[\protect\citeauthoryear{Li, Zhang, Sun, Wang, Li, Zhang, and Lin}{Li
  et~al\mbox{.}}{2019}]%
        {li2019approximate}
\bibfield{author}{\bibinfo{person}{Wen Li}, \bibinfo{person}{Ying Zhang},
  \bibinfo{person}{Yifang Sun}, \bibinfo{person}{Wei Wang},
  \bibinfo{person}{Mingjie Li}, \bibinfo{person}{Wenjie Zhang}, {and}
  \bibinfo{person}{Xuemin Lin}.} \bibinfo{year}{2019}\natexlab{}.
\newblock \showarticletitle{Approximate Nearest Neighbor Search on High
  Dimensional Data -- Experiments, Analyses, and Improvement}.
\newblock \bibinfo{journal}{\emph{IEEE Transactions on Knowledge and Data
  Engineering}} (\bibinfo{year}{2019}).
\newblock


\bibitem[\protect\citeauthoryear{Malkov, Ponomarenko, Logvinov, and
  Krylov}{Malkov et~al\mbox{.}}{2014}]%
        {malkov2014approximate}
\bibfield{author}{\bibinfo{person}{Yury Malkov}, \bibinfo{person}{Alexander
  Ponomarenko}, \bibinfo{person}{Andrey Logvinov}, {and}
  \bibinfo{person}{Vladimir Krylov}.} \bibinfo{year}{2014}\natexlab{}.
\newblock \showarticletitle{Approximate Nearest Neighbor Algorithm based on
  Navigable Small World Graphs}.
\newblock \bibinfo{journal}{\emph{Information Systems}}  \bibinfo{volume}{45}
  (\bibinfo{year}{2014}), \bibinfo{pages}{61--68}.
\newblock


\bibitem[\protect\citeauthoryear{Malkov and Yashunin}{Malkov and
  Yashunin}{2020}]%
        {malkov2018efficient}
\bibfield{author}{\bibinfo{person}{Yu~A Malkov} {and} \bibinfo{person}{DA
  Yashunin}.} \bibinfo{year}{2020}\natexlab{}.
\newblock \showarticletitle{Efficient and Robust Approximate Nearest Neighbor
  Search Using Hierarchical Navigable Small World Graphs}.
\newblock \bibinfo{journal}{\emph{IEEE Transactions on Pattern Analysis and
  Machine Intelligence}} \bibinfo{volume}{42}, \bibinfo{number}{4}
  (\bibinfo{year}{2020}), \bibinfo{pages}{824--836}.
\newblock


\bibitem[\protect\citeauthoryear{Pennington, Socher, and Manning}{Pennington
  et~al\mbox{.}}{2014}]%
        {pennington2014glove}
\bibfield{author}{\bibinfo{person}{Jeffrey Pennington},
  \bibinfo{person}{Richard Socher}, {and} \bibinfo{person}{Christopher~D
  Manning}.} \bibinfo{year}{2014}\natexlab{}.
\newblock \showarticletitle{Glove: Global Vectors for Word Representation}. In
  \bibinfo{booktitle}{\emph{EMNLP}}. \bibinfo{pages}{1532--1543}.
\newblock


\bibitem[\protect\citeauthoryear{Perdacher, Plant, and B{\"o}hm}{Perdacher
  et~al\mbox{.}}{2019}]%
        {perdacher2019cache}
\bibfield{author}{\bibinfo{person}{Martin Perdacher}, \bibinfo{person}{Claudia
  Plant}, {and} \bibinfo{person}{Christian B{\"o}hm}.}
  \bibinfo{year}{2019}\natexlab{}.
\newblock \showarticletitle{Cache-oblivious High-performance Similarity Join}.
  In \bibinfo{booktitle}{\emph{SIGMOD}}. \bibinfo{pages}{87--104}.
\newblock


\bibitem[\protect\citeauthoryear{Tao, Xiao, and Zhou}{Tao
  et~al\mbox{.}}{2006}]%
        {tao2006mining}
\bibfield{author}{\bibinfo{person}{Yufei Tao}, \bibinfo{person}{Xiaokui Xiao},
  {and} \bibinfo{person}{Shuigeng Zhou}.} \bibinfo{year}{2006}\natexlab{}.
\newblock \showarticletitle{Mining Distance-based Outliers from Large Databases
  in any Metric Space}. In \bibinfo{booktitle}{\emph{KDD}}.
  \bibinfo{pages}{394--403}.
\newblock


\bibitem[\protect\citeauthoryear{Tran, Fan, and Shahabi}{Tran
  et~al\mbox{.}}{2016}]%
        {tran2016distance}
\bibfield{author}{\bibinfo{person}{Luan Tran}, \bibinfo{person}{Liyue Fan},
  {and} \bibinfo{person}{Cyrus Shahabi}.} \bibinfo{year}{2016}\natexlab{}.
\newblock \showarticletitle{Distance-based Outlier Detection inData Streams}.
\newblock \bibinfo{journal}{\emph{PVLDB}} \bibinfo{volume}{9},
  \bibinfo{number}{12} (\bibinfo{year}{2016}), \bibinfo{pages}{1089--1100}.
\newblock


\bibitem[\protect\citeauthoryear{Tran, Mun, and Shahabi}{Tran
  et~al\mbox{.}}{2020}]%
        {tran2020real}
\bibfield{author}{\bibinfo{person}{Luan Tran}, \bibinfo{person}{Min~Y Mun},
  {and} \bibinfo{person}{Cyrus Shahabi}.} \bibinfo{year}{2020}\natexlab{}.
\newblock \showarticletitle{Real-time Distance-based Outlier Detection in Data
  Streams}.
\newblock \bibinfo{journal}{\emph{PVLDB}} \bibinfo{volume}{14},
  \bibinfo{number}{2} (\bibinfo{year}{2020}), \bibinfo{pages}{141--153}.
\newblock


\bibitem[\protect\citeauthoryear{Wang, Bah, and Hammad}{Wang
  et~al\mbox{.}}{2019}]%
        {wang2019progress}
\bibfield{author}{\bibinfo{person}{Hongzhi Wang},
  \bibinfo{person}{Mohamed~Jaward Bah}, {and} \bibinfo{person}{Mohamed
  Hammad}.} \bibinfo{year}{2019}\natexlab{}.
\newblock \showarticletitle{Progress in Outlier Detection Techniques: A
  Survey}.
\newblock \bibinfo{journal}{\emph{IEEE Access}}  \bibinfo{volume}{7}
  (\bibinfo{year}{2019}), \bibinfo{pages}{107964--108000}.
\newblock


\bibitem[\protect\citeauthoryear{Wang, Liu, Ma, Bailey, Zha, Song, and
  Xia}{Wang et~al\mbox{.}}{2018}]%
        {wang2018iterative}
\bibfield{author}{\bibinfo{person}{Yisen Wang}, \bibinfo{person}{Weiyang Liu},
  \bibinfo{person}{Xingjun Ma}, \bibinfo{person}{James Bailey},
  \bibinfo{person}{Hongyuan Zha}, \bibinfo{person}{Le Song}, {and}
  \bibinfo{person}{Shu-Tao Xia}.} \bibinfo{year}{2018}\natexlab{}.
\newblock \showarticletitle{Iterative Learning with Open-set Noisy Labels}. In
  \bibinfo{booktitle}{\emph{CVPR}}. \bibinfo{pages}{8688--8696}.
\newblock


\bibitem[\protect\citeauthoryear{Yianilos}{Yianilos}{1993}]%
        {yianilos1993data}
\bibfield{author}{\bibinfo{person}{Peter~N Yianilos}.}
  \bibinfo{year}{1993}\natexlab{}.
\newblock \showarticletitle{Data Structures and Algorithms for Nearest Neighbor
  Search in General Metric Spaces}. In \bibinfo{booktitle}{\emph{SODA}}.
  \bibinfo{pages}{311--21}.
\newblock


\bibitem[\protect\citeauthoryear{Yoon, Lee, and Lee}{Yoon
  et~al\mbox{.}}{2019}]%
        {yoon2019nets}
\bibfield{author}{\bibinfo{person}{Susik Yoon}, \bibinfo{person}{Jae-Gil Lee},
  {and} \bibinfo{person}{Byung~Suk Lee}.} \bibinfo{year}{2019}\natexlab{}.
\newblock \showarticletitle{NETS: Extremely Fast Outlier Detection from a Data
  Stream via Set-based Processing}.
\newblock \bibinfo{journal}{\emph{PVLDB}} \bibinfo{volume}{12},
  \bibinfo{number}{11} (\bibinfo{year}{2019}), \bibinfo{pages}{1303--1315}.
\newblock


\bibitem[\protect\citeauthoryear{Zhang, Chen, Ooi, Tan, and Zhang}{Zhang
  et~al\mbox{.}}{2015}]%
        {zhang2015memory}
\bibfield{author}{\bibinfo{person}{Hao Zhang}, \bibinfo{person}{Gang Chen},
  \bibinfo{person}{Beng~Chin Ooi}, \bibinfo{person}{Kian-Lee Tan}, {and}
  \bibinfo{person}{Meihui Zhang}.} \bibinfo{year}{2015}\natexlab{}.
\newblock \showarticletitle{In-memory Big Data Management and Processing: A
  Survey}.
\newblock \bibinfo{journal}{\emph{IEEE Transactions on Knowledge and Data
  Engineering}} \bibinfo{volume}{27}, \bibinfo{number}{7}
  (\bibinfo{year}{2015}), \bibinfo{pages}{1920--1948}.
\newblock


\bibitem[\protect\citeauthoryear{Zois, Tsotras, and Najjar}{Zois
  et~al\mbox{.}}{2019}]%
        {zois19efficient}
\bibfield{author}{\bibinfo{person}{Vasileios Zois}, \bibinfo{person}{Vassilis~J
  Tsotras}, {and} \bibinfo{person}{Walid~A Najjar}.}
  \bibinfo{year}{2019}\natexlab{}.
\newblock \showarticletitle{Efficient Main-Memory Top-K Selection for Multicore
  Architectures}.
\newblock \bibinfo{journal}{\emph{PVLDB}} \bibinfo{volume}{13},
  \bibinfo{number}{2} (\bibinfo{year}{2019}), \bibinfo{pages}{114--127}.
\newblock


\end{thebibliography}

\appendix

\section{Proofs}

\subsection{Proof of \textsc{Lemma} \ref{lemma_false-negative}.}
Given $p_{1}$, \textsc{Greedy-Counting} computes the distance between $p_{1}$ and $p_{2}$, only when $p_{2}$ is visited for the first time.
Let $p_{3}$ be a neighbor of $p_{1}$.
A proximity graph does not guarantee to have a path from $p_{1}$ to $p_{3}$ that can be traversed by \textsc{Greedy-Counting}.
Therefore, the count (the number of neighbors of $p_{1}$) returned by \textsc{Greedy-Counting} is always $k$ or less.
As outliers have less than $k$ neighbors, \textsc{Greedy-Counting} does not filter them.	\wsq

\subsection{Proof of \textsc{Theorem} \ref{theorem_online-time}}
Let $\rho$ be the average number of accessed objects until \textsc{Greedy-Counting} terminates, for an object.
The filtering phase incurs $O(\rho n)$ time.
The verification phase incurs $O((f + t)n)$ time.
Because $\rho$ is small and often $\rho = O(k)$, we have $\rho \ll f + t$.
Algorithm \ref{algo_framework} hence requires $O((f + t)n)$ time.   \wsq

\subsection{Proof of \textsc{Theorem} \ref{theorem_nndescent}}
The initial operation needs $O(nK\log K)$ time, because each object needs $O(K\log K)$ time to sort the random objects.
To update the current AKNNs of an object $p$, \textsc{NNDescent} accesses the similar object list of $p'$, where $p'$ is one of the current AKNNs or reverse AKNNs of $p$.
The amortized size of the similar object list of $p'$ is $O(K)$, thus updating the current AKNN of $p$ needs $O(K) \times K \times O(\log K) = O(K^2\log K)$ time.
That is, the second operation needs $O(nK^2\log K)$ time.
This is conducted iteratively, and the number of iterations is almost constant \cite{dong2011efficient} (and can be fixed).
The time complexity of \textsc{NNDescent} is then $O(nK^2\log K)$.	\wsq

\subsection{Proof of \textsc{Lemma} \ref{lemma_init}}
Building a VP-tree needs $O(n\log n)$ time \cite{brin1995near}.
The number of leaf nodes, which are the left nodes of their parents, is $O(2^{\log n}) = O(n)$.
This is because VP-tree is a balanced tree, i.e., its height is $O(\log n)$, and it has $2^{\log n}$ leaf nodes.
Since each leaf node contains $O(K)$ objects, updating AKNNs of them needs $O(K^2\log K)$ time.
These facts conclude that this lemma holds.	\wsq

\subsection{Proof of \textsc{Lemma} \ref{lemma_nndescent-plus}}
Lemma \ref{lemma_init} proves that the first procedure of \textsc{NNDescent+} needs $O(nK^2\log K)$ time.
The second procedure of \textsc{NNDescent+} is essentially the same as that of \textsc{NNDescent}.
Therefore, it needs $O(nK^2\log K)$ time.
Exact $K$-NN retrieval, the last procedure of \textsc{NNDescent+}, needs $O(n(K + \log n))$ time.
In total, the time complexity is $O(nK^2\log K)$.	\wsq

\subsection{Proof of \textsc{Lemma} \ref{lemma_connect-subgraph}}
The reverse AKNN phase incurs $O(nK)$ time, since we scan all links.
In the BFS with ANN phase, BFS needs $O(nK)$ time, as each object checks its links.
The number of disjoint sub-graphs is at most $O(\frac{n}{K})$, since each object has at least $K$ links.
Our ANN search incurs only $O(K)$ time, and lines \ref{algo_connect_ann_b}--\ref{algo_connect_second_e} of Algorithm \ref{algo_connect} need at most $O(\frac{n}{K}) \times O(K) = O(n)$ time.
We now see that both the first and second phases need $O(nK)$ time, which proves this lemma.	\wsq

\subsection{Proof of \textsc{Theorem} \ref{theorem_msg-time}}
BFS and sorting incur $O(nK)$ and $O(n\log n)$ time, respectively.
We do these operations for each object, and $s \leq n$.
Hence this theorem holds.	\wsq

\subsection{Proof of \textsc{Lemma} \ref{lemma_remove-detours}}
The proof of Theorem \ref{theorem_msg-time} suggests that 3-hop BFS needs $O(K^3\log K^3)$ $=$ $O(K^3\log K)$ time for a target object.
Similarly, 2-hop BFS needs $O(K^2\log K)$ time, and this is done $O(K)$ times.
We hence need $O(K^3\log K) + O(K) \times O(K^2\log K) = O(K^3\log K)$ time for computing objects with no monotonic path from the target object.
Since $|P'| = O(\frac{n}{K})$, lines \ref{algo_remove_b}--\ref{algo_remove_2hop_e} of Algorithm \ref{algo_remove} need $O(\frac{n}{K}) \times O(K^3\log K) = O(nK^2\log K)$ time.
The size of $A_{i}$ is $O(K^2)$, so lines \ref{algo_remove_add_b}--\ref{algo_remove_add_e} of Algorithm \ref{algo_remove} need $O(nK)$ time.
Now we see that the lemma holds.	\wsq

\subsection{Proof of \textsc{Theorem} \ref{theorem_mrpg-space}}
The AKNN graph built by \textsc{NNDescent+} has $O(nK)$ links, and then links are added in \textsc{Connect-SubGraphs} and \textsc{Remove-Detours}.
In \textsc{Connect-SubGraphs}, we add at most $O(\frac{n}{K})$ links, since the number of disjoint sub-graphs is at most $O(\frac{n}{K})$.
On the other hand, in \textsc{Remove-Detours}, we add at most $O(\frac{n}{K}) \times O(K^2) = O(nK)$ links.
As $\frac{n}{K} \ll nK$, a MRPG has at most $O(nK)$ links.	\wsq

\end{document}